\documentclass[prd,aps,showpacs,floats,floatfix,superscriptaddress,nofootinbib]{revtex4}

\usepackage{graphicx}
\usepackage{amssymb}

\usepackage{longtable}

\def\laq{\raise 0.4ex\hbox{$<$}\kern -0.8em\lower 0.62ex\hbox{$\sim$}}
\def\gaq{\raise 0.4ex\hbox{$>$}\kern -0.7em\lower 0.62ex\hbox{$\sim$}}

\newcommand{\beq}{\begin{equation}}
\newcommand{\eeq}{\end{equation}}
\newcommand{\bea}{\begin{eqnarray}} 
\newcommand{\eea}{\end{eqnarray}}
\newcommand{\ba}{\begin{array}}
\newcommand{\ea}{\end{array}}

\newcommand{\eqref}[1]{(\ref{#1})}

\newcommand{\mytextrm}[1]{{}}

\newlength{\sizeonefig}
\newlength{\sizetwofig}
\newlength{\sizeonefigb}
\newlength{\sizetwofigb}
\begin{document}

\title{Inspiral, merger and ring-down of equal-mass black-hole binaries}

\author{Alessandra Buonanno} 

\affiliation{Department of Physics, University of Maryland, College Park, Maryland 20742 }

\author{Gregory B.\ Cook} 

\affiliation{Department of Physics, Wake Forest University, Winston-Salem, North-Carolina, 27109}

\author{Frans Pretorius}

\affiliation{Department of Physics, University of Alberta, Edmonton, AB T6G 2G7 Canada}
\affiliation{Canadian Institute for Advanced Research, Cosmology and Gravity Program}

\begin{abstract}
We investigate the dynamics and gravitational-wave (GW) emission in the
binary merger of equal-mass black holes as obtained from numerical relativity simulations. 
The simulations were performed with an evolution code based on generalized harmonic 
coordinates developed by Pretorius, and used quasi-equilibrium initial data sets 
constructed by Cook and Pfeiffer. Results from the evolution of three sets of initial data
are explored in detail, corresponding to different initial separations
of the black holes, and exhibit between $2\mbox{--}8$ GW cycles 
before coalescence. We find that to a good approximation the inspiral phase of
the evolution is quasi-circular, followed by a ``blurred, quasi-circular
plunge'' lasting for about $1\mbox{--} 1.5$ GW cycles. After this plunge
the GW frequency decouples from the orbital 
frequency, and we define this time to be the start of the merger 
phase. Roughly $10\mbox{--} 15 m$
separates the time between the beginning of the
merger phase and when we are able to extract quasi-normal
ring-down modes from gravitational waves emitted by the newly formed black hole. This 
suggests that the merger lasts for a correspondingly short amount
of time, approximately $0.5\mbox{--}0.75$ of a full GW cycle. 
We present first-order comparisons between analytical models
of the various stages of the merger and the numerical results---more detailed and 
accurate comparisons will need to await numerical simulations
with higher accuracy, better control of systemic errors (including
coordinate artifacts), and initial configurations where the binaries 
are further separated. During the inspiral,
we find that if the orbital {\em phase} is well modeled, the leading order
Newtonian quadrupole formula is able to match both the amplitude 
and phase of the numerical GW quite accurately until close to the point of merger. 
We provide comparisons between the numerical results 
and analytical predictions based on the adiabatic post-Newtonian (PN) and 
non-adiabatic resummed-PN models (effective-one-body and Pad\'e models). 
For all models considered, 3PN and 3.5PN orders 
match the {\em inspiral} numerical data the best. 
From the ring-down portion of the GW we extract the fundamental
quasi-normal mode and several of the overtones.
Finally, we estimate the optimal signal-to-noise ratio for typical binaries detectable  
by GW experiments. We find that when the merger and ring-down phases are included, 
binaries with total mass larger than $40 M_\odot$ (sources for ground-based detectors)
are brought in band and can be detected with signal-to-noise up to $\approx 15$ at 100 Mpc, whereas 
for binaries with total mass larger than $2 \times 10^6 M_\odot$ (sources 
for space-based detectors) the SNR can be  $\approx 10^4$ at 1Gpc. 
\end{abstract}


\date{\today}

\maketitle

\section{Introduction}
\label{sec0}

With gravitational-wave (GW) detectors operating~\cite{LIGO,GEO,TAMA} or under 
commissioning~\cite{VIRGO}, it is more and more desirable to improve the theoretical 
predictions of the GW signals. Compact binaries composed of black holes (BH) and/or neutron stars 
are among the most promising candidates for the first detection. 

The last year has been marked by breakthroughs in numerical 
relativity (NR) with several independent groups being able to simulate
binary black hole coalescence through the last stages of inspiral ($2\mbox{--}4$ orbits), merger,
ring-down and sufficiently long afterwards
to extract the emitted GW signal~\cite{FP,Bakeretal1,CLMZ,HLS,S}. 
In Ref.~\cite{FP} the first stable evolution
of such an entire merger process was presented. A couple of the key elements responsible
for this success where the use of a formulation of the field
equations based on a generalization of harmonic 
coordinates~\cite{friedrich,garfinkle,szilagyi_et_al} and the
addition of constraint damping terms to the equations~\cite{gundlach_et_al,lambda_ref}. 
Similar techniques have been successfully
incorporated into other efforts since~\cite{lindblom_et_al,scheel_et_al,szilagyi_nfnr}.
Some months afterwords, two groups~\cite{CLMZ,Bakeretal1} independently presented 
modifications of the BSSN (or NOK)~\cite{nok,bs,sn} formulation of the field
equations that allowed them to simulate complete merger events. Among the
key modifications were gauge conditions allowing the BHs 
to move through the computational domain in a so-called puncture evolution~\cite{bruegmann};
these methods have also been successfully reproduced by 
other groups since~\cite{HLS,S,bruegmann_nfnr,marronetti_nfnr,pollney_nfnr}.

In this paper, after analyzing the main features of the dynamics and waveforms obtained 
from the NR simulations, we present a preliminary comparison between 
the numerical and analytical waveforms. The analytical model
we use for the inspiral phase is the post-Newtonian (PN) 
approximation~\cite{35PNnospin,LB} in the {\it adiabatic} limit~\footnote{The natural 
adiabatic parameter during the inspiral phase is $\frac{\dot{\omega}}{\omega^2} = {\cal O}\left 
(\frac{v^5}{c^5} \right )$.}~\cite{LB}, whereas for the inspiral--(plunge)--merger--ring-down phases we 
consider the PN {\it non-adiabatic} and resummed models, such as the 
effective-one-body (EOB)~\cite{BD1,BD2,BD3,DJS,TD,BCD} and Pad\'e resummations~\cite{DIS98}. 
Due to the limited resolution, initial eccentricity, relatively close initial configurations and possible coordinate artifacts, 
it is difficult to claim very high accuracy when comparing with analytical models.
Thus, we shall refer to those comparisons as first-order comparisons. We will {\it only} 
consider a few dynamical quantities of the analytical models characterizing the binary evolution, notably gauge-invariant 
quantities such as the orbital and wave frequencies, and the GW phase, all 
as measured by an observer at infinity. When comparing, we will assume that the numerical 
and analytical waveforms refer to equal-mass binaries, but we apply a 
{\it fitting} procedure to obtain the best-match 
time of coalescence and the spin variables. 
We will use the confrontation with analytical models as an interesting diagnostic 
of the numerical results. 
When simulations that are more accurate and begin closer to an inspiralling circular binary 
become available we will be able to do more stringent tests of analytical models, 
compare all dynamical quantities expressed in the same gauge, 
and use those results to discriminate between models.

Certainly, the most intriguing and long-awaited result of the numerical simulation is the transition 
inspiral--merger--ring-down. Is it a strongly non-linear phase? 
How much energy and angular momentum is released? Over how many GW cycles does it occur? 
How spread in frequency is the signal power spectrum?
Answers to these questions are relevant from a theoretical point of view, e.g., 
to study general relativity in the strongly coupled regime, and also from an 
observational point of view, e.g., to build faithful templates to detect GW
waves and test GR with GW experiments. In this paper we shall start 
to scratch the surface of this problem. We pinpoint several interesting features 
of the inspiral to ringdown transition
that need to be investigated more quantitatively in the future when 
more accurate simulations become available. 
  
Quasi-normal modes (QNM) (or ring-down modes) of Schwarzschild and Kerr 
BHs were predicted a long time ago~\cite{CVV,Press,Davis,QNR}. Their associated 
signal can be described analytically in terms of damped sinusoids. By fitting to the numerical 
waveforms we extract the dominant QNMs of the final Kerr BH---i.e. the fundamental mode and several
of the overtones---and try to make connections to the
previous dynamical phase. 

Finally, we discuss the impact of the merger and ring-down phases on 
the detectability of GWs emitted by equal-mass binaries for ground-based and 
space-based detectors, and compare those results with predictions from 
analytical models.   

This paper is organized as follows. In Sec.~\ref{sec1} we review the initial-data sets 
of Cook and Pfeiffer~\cite{CP} used in the numerical simulations. In Sec.~\ref{sec2} we present 
and discuss the results of NR simulations of binary BH mergers obtained 
with the generalized-harmonic-gauge code of Pretorius~\cite{FP}. In Sec.~\ref{sec3} we 
provide a first-order comparison between numerical and analytical results for the last stages 
of inspiral. In Secs.~\ref{sec4} and \ref{secmerger} we analyze the ring-down and merger phases as  
predicted by the numerical simulations. In Sec.~\ref{secEOB} we present a first-order 
comparison between the numerical results and the EOB predictions for 
inspiral, plunge, merger and ring-down. In Sec.~\ref{sec5} we evaluate the Fourier transform of 
the waveforms and discuss how the inclusion of the merger and ring-down phases will increase  
the optimal signal-to-noise ratio of ground-based and space-based detectors. 
Section~\ref{sec:discussion} contains our main conclusions and a 
discussion on how to make more robust comparisons with analytical models in 
future NR simulations. 

Some material we defer to appendices. The majority
of the comparisons and analysis of gravitational waveforms focus
on the dominant quadrupole multipole moment of the wave; in Appendix~\ref{appendix_mult}
we briefly describe the sub-dominant multipole moments extracted 
from the waves. Appendix~\ref{appendix_comp} compares the energy
and angular momentum flux of the numerically extracted GW with analytical
models of the fluxes. In Appendix~\ref{appendix_extract} we describe some possible
artifacts induced by extracting the GW a finite distance from the
source in the simulations. Appendix~\ref{appendix_tables} contains
tables of fitting coefficients from the QNM ring-down fits. We have
analyzed three sets of BBH evolutions---some figures from the case
with the closest initial separation are contained in Appendix~\ref{appendix_d13}
to simplify the main text.

\section{Initial data}
\label{sec1}

The evolutions presented in this paper begin with initial data that
has been prepared using the methods developed by Cook and
Pfeiffer~\cite{CP,CCGP,PKST,CPID}.  This approach incorporates the
extended conformal thin-sandwich (CTS) decomposition~\cite{PY,CTS}, the
``Komar mass method'' for locating circular orbits~\cite{GGB1,GGB2},
and quasiequilibrium boundary conditions on BH excision
surfaces~\cite{C2,CP}.  The data we have used are designed to represent
an equal-mass binary BH configuration in which the binary is in
quasiequilibrium with the holes in a nearly circular orbit and where
the spins of the individual holes correspond to ``corotation''.
Within the CTS approach, the conformal metric, the trace of the
extrinsic curvature, and their time derivatives must be freely
specified.  The time derivatives of the conformal metric and the trace of the
extrinsic curvature are chosen to vanish.  This time derivative is
taken along an approximate helical Killing vector which defines the
notion of quasiequilibrium.  The initial data is constructed on a
``maximal slice'' which fixes the trace of the extrinsic curvature to
zero.  Finally, the conformal metric is chosen to be flat.

The initial data produced by this procedure do a very good job of
representing the desired astrophysical situation of a pair of BHs 
nearing the point of coalescence.  However, two important
approximations have been made in the construction of these
initial-data sets.  When the initial data are evolved, these
approximations will affect the subsequent dynamics and the
GW that is produced.  The first approximation is
that the initial data are ``conformally flat''.  The choice of a flat
conformal metric is known to introduce small errors in representing
both individual spinning BHs and binary systems.  As part of this
error, some amount of unphysical gravitational radiation is included
in the initial data.  The second approximation is in placing the
binary in a circular orbit.  This approximation is motivated by the
fact that for large enough separation, the time scale for radial
motion due to radiation reaction is {\em large} compared to the
orbital period.  However, this approximation results in BHs having
little, if any, initial radial momentum.  For sufficiently small
separations, this is clearly not ``astrophysically correct.''

Until now, the best way to estimate the quality of BH binary initial
data has been to compare them against the results obtained by PN 
methods.  Comparisons of gauge-invariant
quantities such as the total energy and angular momentum of the system
or the orbital angular velocity, all measured at infinity, are in good
agreement with adiabatic sequences of circular orbits as determined by
third-order PN (3PN) calculations~\cite{LB} and their EOB 
resummed extension~\cite{BD1,DJS} (see Figs.~10--18 in
Ref.~\cite{CP}, Figs.~3--5 in Ref.~\cite{DGG} and Figs.~3--5 in
Ref.~\cite{ICO} and discussion around them). There are, of course,
differences between the numerical initial data and the PN/EOB models
and these differences increase as the binary separation decreases and
the system becomes more relativistic.  Among these differences, there
is also some evidence that the initial orbit incorporates a small
eccentricity~\cite{BIW}.

Adiabatic sequences of BH binaries exhibit an inner-most
stable circular orbit (ISCO) defined by a turning point in the total
{\it conserved} energy.  For the numerical initial data, the quasiequilibrium
approximation becomes less accurate as the binary separation decreases
and we would expect the approximation to be rather poor at the ISCO.
PN methods restricted to circular orbits suffer similar problems as they
approach the ISCO.  The PN/EOB and numerical initial-data circular-orbit 
models are in reasonable agreement up to the ISCO, but it is
difficult to ascertain the accuracy of either in this limit.

The comparisons done in Refs.~\cite{DGG,ICO,CP} between numerical
initial data and PN/EOB models are limited by the fact that they use
adiabatic circular orbits.  Essentially, these comparisons lend
strength to the belief that the conformal-flatness approximation is
not causing significant problems with the non-radiative aspects of the
initial data.  Clearly, they cannot shed any light on the effects of
the circular-orbit approximation.  Some insight on this issue has been
obtained by performing full dynamical evolutions of the PN/EOB
equations of motion for equal-mass binaries~\cite{BD2,MM}.  These
studies show that neglecting the radial momentum, at both the initial
time and throughout the evolution as done in adiabatic circular
orbits, results in a phase error in the waveforms (see e.g., Fig.~5 in
Ref.~\cite{BD2}). Neglecting the radial momentum at the initial time 
also introduces eccentricity into the dynamics (see e.g., Fig.~4 in
Ref.~\cite{MM}).

\section{Numerical relativity results and their diagnostics}
\label{sec2}

\subsection{Initial Data for Generalized Harmonic Evolution}

The corotating quasi-circular BH inspiral data discussed in the previous section were evolved
using a numerical code based on a generalized harmonic (GH) decomposition of the
field equations, as described in detail in Ref.~\cite{FP1,FP,FP3}. 
As supplied by Pfeiffer~\cite{CPID,PKST},
the initial data is given in terms of standard 
$3+1$ or ADM (Arnowitt-Deser-Misner) variables~\cite{L,ADM,CB,Y}, namely
the lapse function $\alpha$, shift vector $\beta^i$, spatial metric $h_{ij}$ and extrinsic curvature {$K_{ij}$}:
\begin{eqnarray}
ds^2 &=& -\alpha^2 dt^2 + h_{ij} \left(dx^i + \beta^idt\right)\left(dx^j + \beta^jdt\right), \label{adm}\\
K_{ij}&=& - h_i{}^{l} h_j{}^{m} \nabla_m n_l.\label{kdef}
\end{eqnarray}
In the above $n_\mu=-\alpha \nabla_\mu t$ is the unit time-like vector normal to $t={\rm const.}$ hypersurfaces, and we use units where $G=c=1$.
The GH code directly integrates the 4-metric elements $g_{\mu\nu}$
\begin{equation}
ds^2=g_{\mu\nu} dx^\mu dx^\nu,
\end{equation}
and therefore needs the values of $g_{\mu\nu}$ and $\partial g_{\mu\nu}/\partial t$ at $t=0$ as initial
conditions. The initial data (\ref{adm}), (\ref{kdef}) provides most of what is required to construct
$g_{\mu\nu}|_{t=0},\partial g_{\mu\nu}/\partial t|_{t=0}$; what must still be specified
are the components of the gauge encoded in $\partial \alpha/\partial t|_{t=0}$ and $\partial \beta^i/\partial t|_{t=0}$. We choose
the time derivatives of the lapse and shift such that the slice is spacetime harmonic
at $t=0$:
\begin{eqnarray}
\partial_{\gamma} \alpha \ n^\gamma &=& -\alpha K\\
\partial_\gamma \beta^i \ n^\gamma &=& \alpha\bar{\Gamma}^i_{jk} h^{jk} - \partial_j \alpha \ h^{i j},
\end{eqnarray}
where $K$ is the trace of the extrinsic curvature, and
$\bar{\Gamma}^i_{jk}$ are the Christoffel symbols of the spatial metric $h_{ij}$.

\subsection{Characterization of the Waveform}
\label{sec_num_extract}

Gravitational wave information is obtained by computing the Weyl
scalar $\Psi_4$, which has the asymptotic property of being equal to
the outgoing radiation if the complex null tetrad is chosen correctly.
To be explicit, we define a spherical coordinate system centered on
the center of mass of the binary with orthonormal bases
$(\hat{r},\hat\theta,\hat\phi)$.  The coordinates are chosen so
that the azimuthal axis is aligned with the orbital angular momentum
and the binary orbits in the direction of increasing azimuthal
coordinate.

To define our complex null tetrad, we use the time-like unit vector normal
to a given hypersurface $\hat{n}$ and the radial unit vector $\hat{r}$
to define an ingoing ($\vec{k}$) and outgoing null vector ($\vec{\ell}$) by
\begin{eqnarray}
  \vec{k} &\equiv& \frac1{\sqrt{2}}(\hat{n} + \hat{r}), \\
  \vec{\ell} &\equiv& \frac1{\sqrt{2}}(\hat{n} - \hat{r}).
\end{eqnarray}
We define the complex null vector $\vec{m}$ by
\begin{equation}
  \vec{m} \equiv \frac1{\sqrt{2}}(\hat\phi - i\hat\theta).
\end{equation}
In terms of this tetrad, we define $\Psi_4$ as
\begin{equation}
\label{eq:Psi4_Weyl}
  \Psi_4 \equiv C_{\alpha\beta\gamma\delta}\ell^\alpha(m^\beta)^*\ell^\gamma(m^\delta)^*,
\end{equation}
where $C_{\alpha\beta\gamma\delta}$ is the Weyl tensor and $*$ denotes complex conjugation.

To relate $\Psi_4$ to the GWs, we note that in transverse-traceless (TT) gauge,
\begin{eqnarray}
  \frac14(\ddot{h}^{TT}_{\hat\theta\hat\theta} 
      - \ddot{h}^{TT}_{\hat\phi\hat\phi}) &=&
      -R_{\hat{n}\hat\theta\hat{n}\hat\theta} =
      -R_{\hat{n}\hat\phi\hat{r}\hat\phi} =
      -R_{\hat{r}\hat\theta\hat{r}\hat\theta} =
      R_{\hat{n}\hat\phi\hat{n}\hat\phi} =
      R_{\hat{n}\hat\theta\hat{r}\hat\theta} =
      R_{\hat{r}\hat\phi\hat{r}\hat\phi}, \\
  \frac12\ddot{h}^{TT}_{\hat\theta\hat\phi} &=&
      -R_{\hat{n}\hat\theta\hat{n}\hat\phi} =
      -R_{\hat{r}\hat\theta\hat{r}\hat\phi} =
      R_{\hat{n}\hat\theta\hat{r}\hat\phi} =
      R_{\hat{r}\hat\theta\hat{t}\hat\phi}.
\end{eqnarray}
Following convention, we take the $h_+$ and $h_\times$ polarizations
of the GW to be given by
\begin{eqnarray}
  \ddot{h}_+ &=& \frac12(\ddot{h}^{TT}_{\hat\theta\hat\theta} 
      - \ddot{h}^{TT}_{\hat\phi\hat\phi}), \\
\ddot{h}_\times &=& \ddot{h}^{TT}_{\hat\theta\hat\phi}.
\end{eqnarray}
We find, then, that in vacuum regions of the spacetime,
\begin{equation}
\label{eq:Psi4_ddh_defn}
  \Psi_4 = \ddot{h}_+ - i\ddot{h}_\times.
\end{equation}

It is most convenient to deal with $\Psi_4$ in terms of its harmonic
decomposition.  Given the definition of $\Psi_4$ in
Eq.~(\ref{eq:Psi4_Weyl}) and the fact that $\vec{m}^*$ carries a
spin-weight of $-1$, it is appropriate to decompose $\Psi_4$ in terms
of spin-weight $-2$ spherical harmonics $_{-2}Y_{\ell
m}(\theta,\phi)$.  There is some freedom in the definition of the
spin-weighted spherical harmonics.  To be explicit, we defined the
general spin-weighted spherical harmonics by
\begin{equation}
  _sY_{\ell m}(\theta,\phi) \equiv
  (-1)^s\sqrt{\frac{2\ell+1}{4\pi}}d^\ell_{m(-s)}(\theta)e^{im\phi},
\end{equation}
where $d^\ell_{ms}$ is the Wigner $d$-function 
\begin{equation}
   d^\ell_{ms}(\theta) \equiv
   \sum_{t=C_1}^{C_2}\frac{(-1)^t\sqrt{(\ell+m)!(\ell-m)!(\ell+s)!(\ell-s)!}}
      {(\ell+m-t)!(\ell-s-t)!t!(t+s-m)!}
      (\cos\theta/2)^{2\ell+m-s-2t}(\sin\theta/2)^{2t+s-m},
\end{equation}
and where $C_1= \max(0,m-s)$ and $C_2=\min(\ell+m,\ell-s)$. 

Finally, for convenience, we always decompose the dimensionless 
Weyl scalar $rM\Psi_4$ where $M=m_1+m_2$ is the mass of the
{\em initial} binary system with $m_1$ and $m_2$ the irreducible~\cite{CR} 
masses of the individual BHs, and $r$ is the generalized harmonic 
radial coordinate.  We then define
\begin{equation}\label{eq:psi4Ylmdef}
  rM\Psi_4(t,\vec{r}) = 
\sum_{\ell m}{}_{-\!2}C_{\ell m}(t,r){}_{-\!2}Y_{\ell m}(\theta,\phi).
\end{equation}
The complex mode amplitudes $_{-\!2}C_{\ell m}(t,r)$, extracted at
a fixed generalized harmonic coordinate radius $r$, contain the full
information about the gravitational waveforms as a time-series.

In the numerical code the four orthonormal vectors 
$(\hat{n},\hat{r},\hat{\theta},\hat{\phi})$ used to construct the null
tetrad are computed as follows. The spacetime is evolved using
Cartesian coordinates $x,y,z$ with time $t$, and we use the standard transformation
to define the spherical coordinates:
\begin{eqnarray}
x&=&r\cos(\phi)\sin(\theta)\,, \\
y&=&r\sin(\phi)\sin(\theta)\,, \\
z&=&r\cos(\theta)\,,
\end{eqnarray} 
$\hat{n}$ is the time-like unit vector normal to $t={\rm const.}$ surfaces,
and $\hat{r}$ is the unit space-like vector pointing in the
direction $(\partial/\partial r)^a$. In the limit $r\rightarrow\infty$ 
the time coordinate $t$ coincides with $TT$ time. $\hat{\theta}$ is 
computed by making $(\partial/\partial \theta)^a$ orthonormal
to $(\hat{n},\hat{r})$ using a Gramm-Schmidt process, and then
$\hat{\phi}$ is calculated by making $(\partial/\partial \phi)^a$
orthonormal to ($\hat{n},\hat{r},\hat{\theta}$). All norms are
computed with the full spacetime metric $g_{\mu\nu}$. The Weyl scalar
$\Psi_4$ is evaluated over the entire numerical grid (i.e. at
all $x,y,z$ mesh points) at regular intervals in time.
We then interpolate $\Psi_4(t,x,y,z)$ to a set of coordinate
spheres at several ``extraction radii'' $r_i$,
with a uniform distribution of points in $(\theta,\phi)$~\footnote{$33\times65$
points in $\theta\in0..\pi,\phi\in0..2\pi$ for this set of simulations}.
All the waveform related data from the simulations presented
here are taken from such samplings of $\Psi_4(t,r=r_i,\theta,\phi)$,
and we have used $r_i=12.5m,25m,37.5m$ and $50m$. The plots and comparisons
shown in the main part of the paper use $r_i=50m$, while in 
Appendix \ref{appendix_extract} we discuss the trends that are seen in $\Psi_4$
as a function of $r_i$. To summarize the results of the 
Appendix, prior to merger the extraction radii at $37.5m$ and $50m$ appear to
be well within the ``wave-zone'', and thus gives a decent representation of
the waveform. Specifically, the coordinate propagation speed of the wave from 
one extraction sphere to the next is very close to unity, and the 
structure of the
wave, normalized by $r$, is similar at the two extraction points. 
Interestingly though, at later times during the simulation,
apparently in coincidence with the strongest wave emission around
the merger part of the evolution, the gauge seems to change slightly
in the extraction zone. The coordinate speed drops by several percent,
and the amplitude of the normalized waveform also decreases as
the wave moves outward. The effect is more pronounced for binaries
that are initially further separated. As not all metric information was saved
when the simulations were run we cannot describe what underlying
properties of the metric are responsible for the change
in propagation behavior, though for the purposes of this paper the
effect is sufficiently small that we do not believe it will
alter any of the primary conclusions. A couple of exceptions are
in the estimates of the total energy and angular momentum 
radiated---these calculations involve double and triple time
integrations of squares of the wave, and so significantly
amplify even small systematic errors in the waveform. This is,
we believe, the cause of the increasing over-estimate of these
quantities as the initial orbital separation increases, as shown
later in Table~\ref{tab_res}. Future work will attempt
to address these gauge-related issues.

\subsection{Numerical results and a discussion of errors}
\label{sec_num_res_and_errors}

We have evolved 3 sets of initial data, labeled by $d=13,16$ and $19$ in Ref.~\cite{CP};
the initial orbital parameters are summarized in Table \ref{tab_idparams}~\footnote{Note that 
when writing $S_{1,2}/m_{1,2}^2$ we mean by $m_{1,2}$ the irreducible mass $m_{1,2}^{\rm irr}$. When later 
on we compare with PN models, we should first express the rest masses appearing in the 
PN formula in terms of $m_{1,2}^{\rm irr}$ and the spin variable $S_{1,2}$. However, 
since we deal here with very small spins, the error in not doing so is very small.}.
Each initial data set was evolved using three different grid resolutions, summarized
in Table~\ref{tab_res}. Most of the results presented in this paper are from the highest resolution
simulations, with the lower resolution runs providing error estimates
via the Richardson expansion.
During evolution the same temporal source-function evolution equations were used as with
the scalar-field-collapse binaries described in Ref.~\cite{FP1,FP}, and the (covariant)
spatial source-functions were kept equal to zero.  We have quite extensively
tested this code to make sure we are solving the Einstein equations, and convergence
tests of the residual of the Einstein equation for similar simulations were
presented in Ref.~\cite{FP3}.

\begin{table}
{\small
\begin{tabular}[t]{| l l || c | c | c | c | c |}
\hline
 & ``d'' & $M_{\rm ADM}/M$ & $M\omega$ & $J_{\rm ADM}/M^2$ & $\ell/M$ & $S_1/m_1^2$\\
\hline
\hline
 & 13 & 0.986 & 0.0562 & 0.875 & 7.96 & 0.107\\
 & 16 & 0.988 & 0.0416 & 0.911 & 9.77 & 0.0802\\
 & 19 & 0.989 & 0.0325 & 0.951 & 11.5 & 0.0629\\
\hline
\end{tabular}
}
\caption{Several parameters describing the initial data~\cite{CP} (left to right): the ADM mass of the
spacetime, the initial angular velocity of each BH, the ADM angular momentum,
the initial proper separation between the holes, and their initial spins. The units
have been scaled to $M$, the sum of initial AH masses.
}
\label{tab_idparams}
\end{table}
\begin{table}
{\small
\begin{tabular}[t]{| l l || c | c | c |}
\hline
 & ``Resolution'' & wave-zone res.& orbital-zone res.& BH res.\\
\hline
\hline
 & h     & $1.6  M$ & $0.20 M$  & $ 0.048 M$ \\
 & 3/4 h & $1.2  M$ & $0.15 M$  & $ 0.036 M$ \\
 & 1/2 h & $0.82 M$ & $0.10 M$  & $ 0.024 M$ \\
\hline
\end{tabular}
}
\caption{The three sets of characteristic spatial resolutions used in the
simulation discussed here, where each resolution
is labeled relative to the coarsest resolution $h$.
The grid is adaptive with a total of 8 levels of refinement,
and the coordinate system is compactified. The wave zone
is defined to be at $r=50 M$, the orbital zone within about $r=10 M$
and the BH zone is within $2-3M$ of each AH.
A Courant-Friedrichs-Lewy (CFL) factor
of $0.2$ was used in all cases.
}
\label{tab_res}
\end{table}
\begin{table}
{
\begin{tabular}[t]{| l || c | c | c | }
\hline
                                  & d=13             & d=16             & d=19 \\
\hline
\hline
$M_f/M$                        & $0.950\pm 0.005$ & $0.954\pm 0.005$ & $0.952 \pm 0.005$\\
$E_{\rm GW}/M$                 & $0.036\pm 0.004$ & $0.043\pm 0.004$ & $0.052  \pm 0.004$\\
$a_f/M_f$                      & $0.71 \pm 0.02$  & $0.71 \pm 0.02$  & $0.70 \pm 0.02$\\
$J^z_f/M^2$                    & $0.64 \pm 0.02$  & $0.65 \pm 0.02$  & $0.63 \pm 0.02$\\
$J^z_{\rm GW}/M^2$             & $0.23 \pm 0.02$  & $0.31 \pm 0.02$  & $0.42 \pm 0.02$ \\
number of orbits               & $1.47\pm 0.10$   & $2.47\pm 0.09$   & $4.39 \pm 0.18$ \\
$t_{\rm CAH}/M$                & $109 \pm 4$      & $228 \pm 16$     & $529  \pm 22$   \\
$(t_{\rm peak}-t_{\rm CAH})/M$ & $\approx9 $      & $\approx9 $      & $\approx9   $   \\
$(t_{\rm dec}-t_{\rm CAH})/M$  & $\approx-11 $     & $\approx-11$      & $\approx-9  $   \\
initial eccentricity $e_1$     & --- & --- & $0.018 \pm 0.003$\\
initial eccentricity $e_2$     & --- & --- & $0.012 (+ 0.014, - 0.012)$\\
Max. GW amp. error             & 8\%              & 9\%              & 8\%             \\
(Max. GW phase error)/$(2\pi)$ & 0.08             & 0.7              & 1               \\
(Max.``shifted'' GW phase 
error)/$(2\pi)$ & 0.04         & 0.06             & 0.05            \\
\hline
\end{tabular}
}
%
%
\caption{A summary of simulation results. From top to bottom: (i) the mass $M_f$ of the final
BH estimated from AH properties, (ii) the energy $E_{\rm GW}$ emitted in 
GWs extracted using the Newman-Penrose scalar $\Psi_4$ at a coordinate radius of $r=50M$
from the origin, (iii) the Kerr spin parameter $a_f$ and corresponding non-zero z-component $J^z$ of the 
angular-momentum vector of the final BH (from AH properties), (iv) 
an estimate $J^z_{\rm GW}$ of the angular-momentum radiated in 
GWs (from $\Psi_4$), (v) the number of orbits in coordinate space before
a common AH, at coordinate time $t=t_{\rm CAH}$, is first detected, 
(vi) an estimate $t=t_{\rm peak}$ of when the GW amplitude reaches its peak,
(vii) an estimate $t=t_{\rm dec}$ of when the GW frequency decouples from the orbital frequency,  
(viii) for $d=19$ two estimates of the eccentricity in the initial data calculated using
(\ref{e1_def}) ($e_1$) and (\ref{e2_def}) ($e_2$), (ix) the estimated {\em maximum} error in the
amplitude of the extracted GWs (occurring near the peak of emission),
(x) an estimate of the cumulative phase error in the wave from $t=0$ until merger, and
(xi) the phase error after {\em maquillage}.
The estimated uncertainties and errors do {\em not} include possible systematic
errors; see the discussion in Secs.~\ref{sec_num_extract}, \ref{sec_num_res_and_errors} 
and Appendix~\ref{appendix_extract} for more information. In all plots in the paper where we
have shifted time by either $t_{\rm CAH}$ or $t_{\rm peak}$, any quantity shown that was
measured from the waveform at $r=50M$  is also shifted by the
propagation time $\delta t$ of the wave to $r=50M$. We have estimated $\delta t$ 
by finding the shift resulting in a best-fit between 
orbital frequency measured using the GW versus AH orbital motion, 
as shown in Fig.~\ref{Fig:OmegaNQC22d16-19}; specifically, we get
$\delta t = 65M,68M$ and $70M$ for $d=13,16$ and $19$ respectively.}
\label{tab_simnums}
\end{table}
 
\begin{figure}
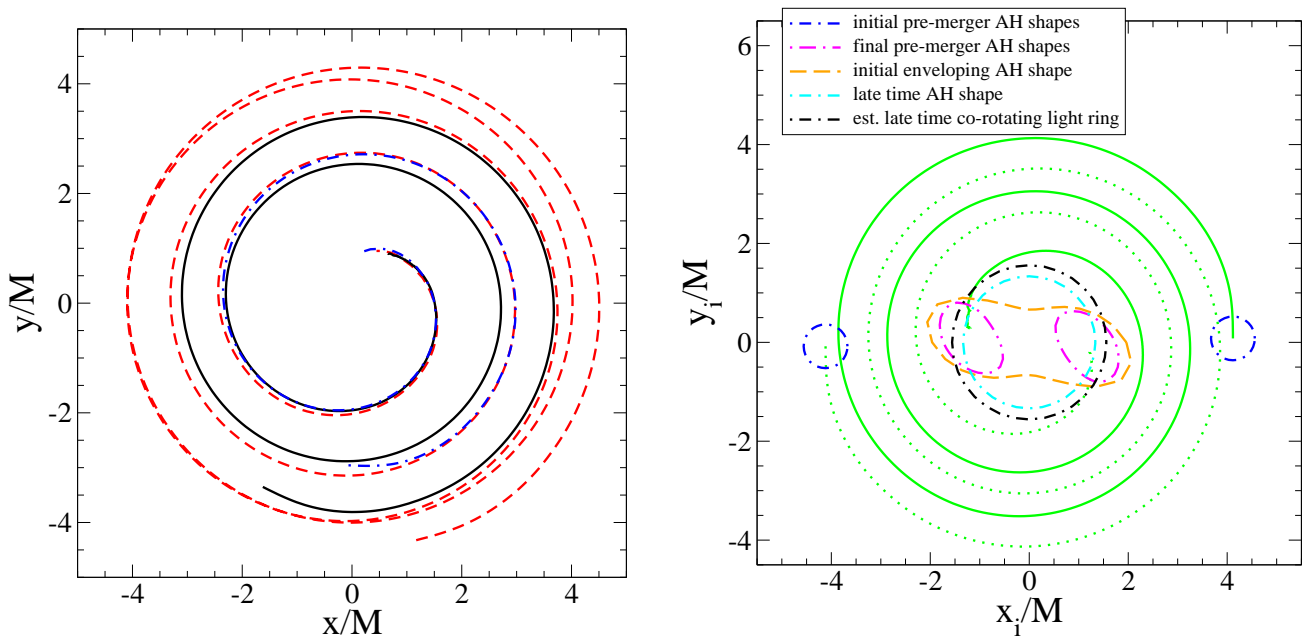

\includegraphics[width=3.25in,clip]{orbits}
\hspace{0.5cm}
\includegraphics[width=3.25in,clip]{orbit}
\caption{(left) The orbital motion of one BH from each of the three 
cases: the dot-dashed (blue) line is $d=13$, the solid (black) $d=16$ and the dashed (red) 
$d=19$. The position of the BH 
is defined as the center of its AH, and the curve ends once an encompassing
horizon is found. The eccentricity present in the initial data is particularly
evident for the $d=19$ case,
though in part this is due to numerical error---see Fig.~\ref{d19_coord_e}. To aid in the comparison
each trajectory was rotated by a constant phase so that they coincide at a coordinate
separation of $3M$. (right) The orbital motion of the BH's from the d=16 simulation,
showing the coordinate shapes of the AHs at several key moments. Also shown
is the location of the co-rotating light ring of the final BH---see the discussion
in Sec.~\ref{secmerger}.
\label{orbits}}
\end{figure}

Table~\ref{tab_simnums} lists some key information obtained from each merger simulation.
Figure~\ref{orbits} shows the orbital motion prior to merger. The uncertainties
and error estimates listed in the table were calculated using an assumed
Richardson expansion, as discussed in \cite{FP3}. At least three simulations are needed
to verify that one is in the convergent regime in general, and for the properties
listed in the table we do see close to second order convergence for most
properties --- a couple of anomalous cases (the phase error and merger time for the $d=19$
case) are discussed further below. Assuming
one is in the convergent regime, results from two simulations can then be used to estimate
the truncation error. This estimate will {\em not} account for systematic
errors, and here we list several potential sources of such error.
Some quantities are in principle susceptible to gauge or coordinate effects, including
the apparent horizons (AHs), and therefore properties measured using them;
orbital parameters such as angular frequency and eccentricity
deduced from the positions of the AHs; the finite GW 
extraction radius and nature of the coordinates at the extraction
surface (see Sec.\ref{sec_num_extract} and Appendix\ref{appendix_extract}); and the choice of tetrad 
used to calculate $\Psi_4$~\cite{NBRE,NBBBP,BBLP}.
Certain post-processing operations have a truncation error associated
with them and we have not estimated the magnitude of these.  For example, in all cases
$\Psi_4$ was sampled on a mesh of size $33\times65$ points in $\theta\times\phi$
at the extraction radius at a given time. We suspect that many of these systematic
errors are small, though eventually the accuracy of the simulations will
be improved (through a combination of high-order methods and higher resolution),
and then it may become important to quantify and eliminate these additional 
uncertainties.

In a waveform, the error of most significance to GW detection is an
error in the phase. The first generation of numerical binary BH 
merger simulations \cite{FP,Bakeretal1,CLMZ,Bakeretal2,CLZ,HLS,FP3,CLZ2}, with the 
notable exception of the Caltech/Cornell effort \cite{CalCor}, 
all suffer from rather significant cumulative phase errors in the 
inspiral portion of the waveform for the longer duration merger events. 
In Ref.~\cite{Bakeretal2} it was argued that the dominant portion
of the phase could be factored out as a constant phase shift within the wave,
resulting in a ``universal'' merger result for the class of initial conditions considered.
Henceforth, we shall denote by {\em maquillage} the operation of phase shifting waveforms 
to make them agree at a specified point in time, improving their coincidence (appeareance) 
over a larger interval of time.
For certain applications this maquillage waveform is the relevant one.
For example, in a matched-filter search the initial phase of the waveform is an
extrinsic parameter~\cite{DIS98} and is irrelevant for detectability of the signal. Also,
when comparing waveforms from different initial-data sets the waveforms need to
be aligned in some manner for a meaningful comparison, and so again a constant
phase difference, whatever the source, is largely irrelevant. However, for parameter
estimation in an inspiral search where a hybrid PN/numerical template is used,
the association of a numerical merger/ring-down to a PN inspiral waveform
will be very sensitive to all phase errors. In particular, an uncertainty 
in the overall phase evolution prior to merger in the numerical waveform 
is directly related to an uncertainty in the merger time for the given
initial conditions, and this will translate to an uncertainty in the PN
binary parameters identified with the match.

\begin{figure}
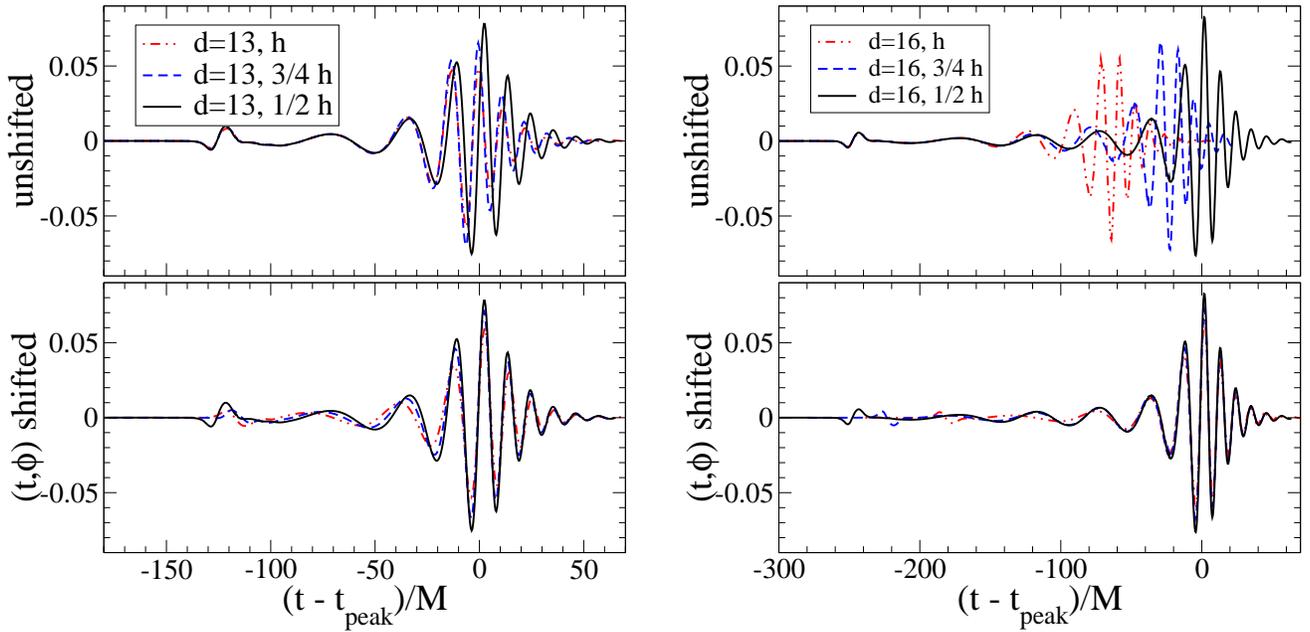

\begin{center}
\includegraphics[width=3.25in,clip]{d13_wave_c22}
\hspace{0.5cm}
\includegraphics[width=3.25in,clip]{d16_wave_c22}
\caption{
The ${\rm Re}[_{-2}C_{22}]$ component of the $d=13$ and
$d=16$ waveforms, unshifted (top) and shifted (bottom). All resolutions
are shown to demonstrate the size of numerical errors in the simulation,
and data such as this was used to compute the errors for waveform
quantities listed in Table \ref{tab_res}.
\label{d13-16_wave}}
\end{center}
\end{figure}

\begin{figure}
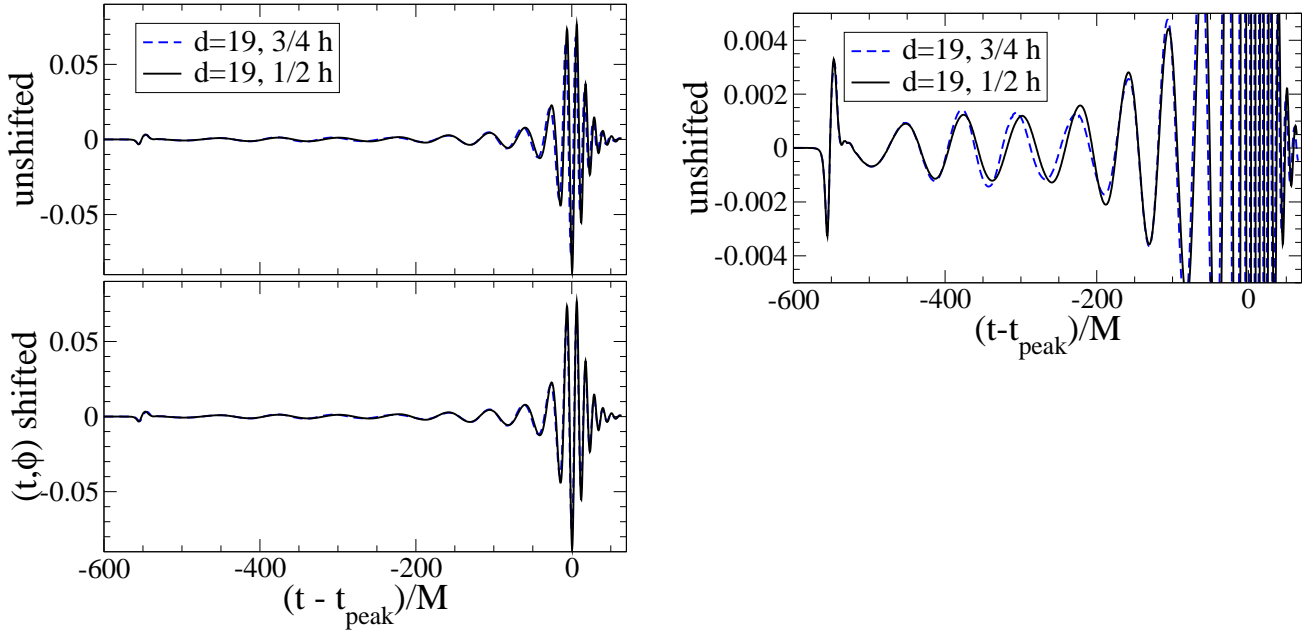

\begin{center}
\includegraphics[width=3.25in,clip]{d19_wave_c22}
\hspace{0.5cm}
\includegraphics[width=3.25in,clip]{d19_wave_c22_zoom}
\caption{In the left panel we show the
the ${\rm Re}[_{-2}C_{22}]$ component of the $d=19$
waveform, unshifted (top) and shifted (bottom).
The lower resolution data for the $d=19$ case was accidentally deleted. 
In the right panel we zoom in for a close-up of the inspiral part of the 
unshifted waveform. From these two results alone it would appear as if
the $d=19$ simulations have anomalously good convergence behavior (compare
to Fig. \ref{d13-16_wave}). However, this is not the
case---refer to the discussion in Sec.~\ref{sec_num_res_and_errors}, and 
see Figs.~\ref{d19_am} and 
\ref{d19_coord_e} for other estimators of the convergence of the solution.
\label{d19_wave}}
\end{center}
\end{figure}

\begin{figure}
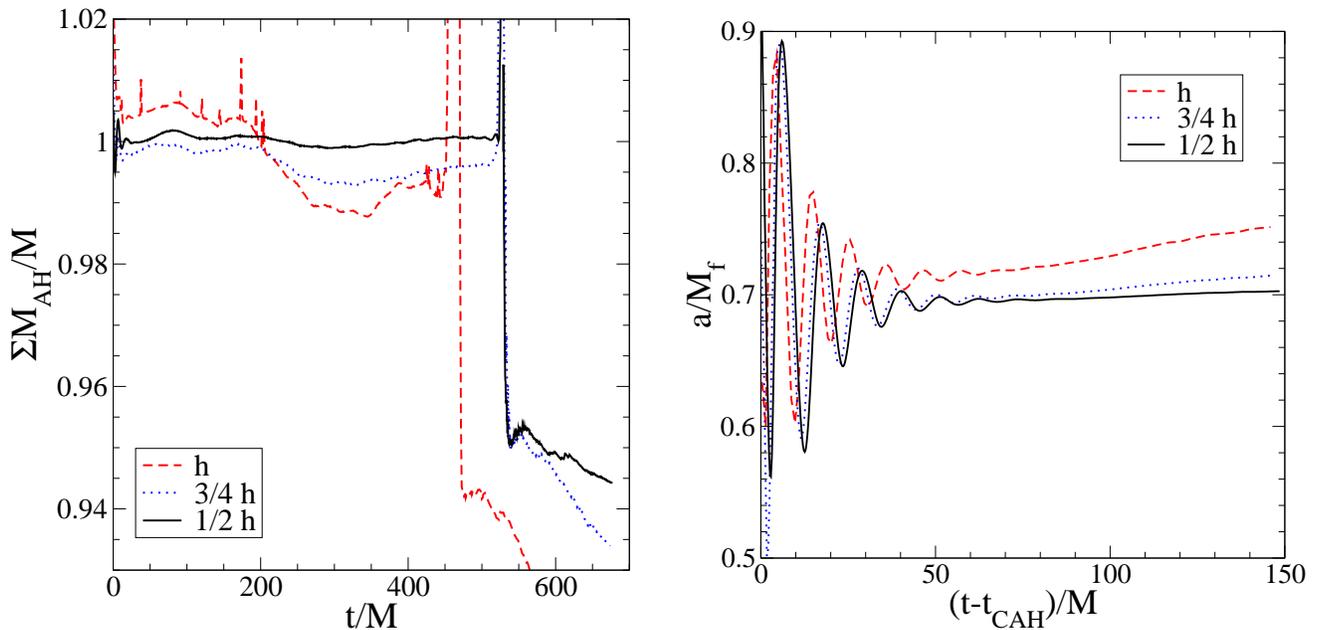

\begin{center}
\includegraphics[width=3.25in,clip]{M_qe_19}
\hspace{0.5cm}
\includegraphics[width=3.25in,clip]{a_qe_19}
\caption{Sum of AH masses (left panel), and the Kerr angular momentum parameter
of the final BH (right panel) for the $d=19$ simulations. The angular momentum was estimated
using the ratio of polar to equatorial proper circumference of the 
horizon~\cite{brandt_seidel}; the
dynamical horizons estimate~\cite{ashtekar_krishnan}
gives similar results modulo the oscillations about
the mean. Except near the time of merger the sum of AH masses in the spacetime 
should be conserved, and similarly at
late times for the Kerr parameter. As resolution increases we see the expected trends
in these quantities. Note that the "jaggies" in the AH mass estimate is a reflection
of AH finder problems in the code, and not irregularities in the underlying 
solution~\cite{FP3}.
\label{d19_am}}
\end{center}
\end{figure}

\begin{figure}
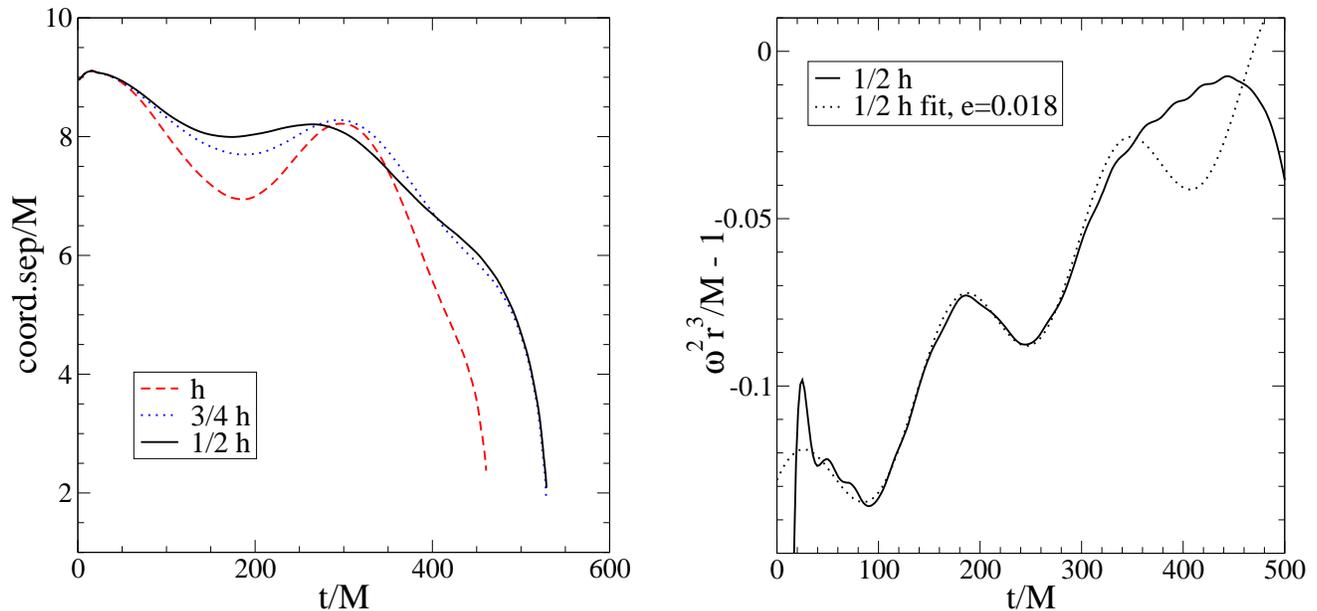

\begin{center}
\includegraphics[width=3.25in,clip]{coord_sep_qe_19}
\hspace{0.5cm}
\includegraphics[width=3.25in,clip]{circ_test_qe_19}
\caption{In the left panel we show the coordinate separation 
of the BHs as a function of time for the $d=19$ simulations. 
This plot highlights the eccentricity within the orbit and it also reflects 
the phasing behavior in the waveform for the $3/4 h$ and $1/2 h$ 
cases---see Fig. \ref{d19_wave}. In the right panel we estimate the eccentricity for the $d=19$ case:
shown is a plot of the left hand side of Eq.~(\ref{e1_def}) together with 
a fit of the form $a_0+a_1 t + e\,\cos(a_2 t + a_3)$ to the early time behavior of this function.
We estimate the eccentricity to be the amplitude of the sinusoidal part of the
fitting function.
\label{d19_coord_e}}
\end{center}
\end{figure}

For the reasons just outlined, in Table \ref{tab_simnums} we give {\em two} 
estimates of the phase errors in the waveforms. The first is the cumulative error in the
phase directly measured from the waveform, and the second is the cumulative
error after the maquillage. Specifically, in the latter case we shift all
the waveforms in time so that the peak amplitudes (corresponding to the peak of 
the energy radiated) occur at t=0, and then
apply a constant rotation in the complex plane of the waveform to give 
optimal overlap with a reference waveform (typically the highest resolution
result). Example waveforms before and after the shift for the three
resolutions are given in Figs.~\ref{d13-16_wave}, \ref{d19_wave}.
Unfortunately, the lowest resolution waveform data
for the $d=19$ case was accidentally deleted, and so only the medium and higher resolution
results are shown. The two $d=19$ cases are unusual in that the phase difference
between them does not grow monotonically with time; rather the phase difference
initially grows then decreases so that by merger time the difference 
is close to zero. We are not entirely sure why the error in the phase evolution behaves so
in this case. One possibility is that for this longer integration time
we are too far from the convergent regime to see the trends in phase evolution
observed for the shorter d=13 and d=16 runs. However, other estimators
of convergence, including AH properties and orbital
motion as shown in Figs.~\ref{d19_am}, \ref{d19_coord_e} which do include
data from the lowest resolution suggest the $d=19$ case {\em is} in 
the convergent regime. Another possibility is that
the numerical error in the phase has a periodic time component whose frequency is
sufficiently low that the d=13 and 16 runs do not show it. Regardless,
that the $1/2 h$ and $3/4 h$ d=19 cases merge at almost the same time makes it
impossible to use them to estimate the error in merger time and phase.
Therefore, the errors quoted for these two numbers in Table \ref{tab_simnums} for d=19 were
estimated using the difference in merger time between the $h$ and $1/2 h$ d=19
runs, with the estimated phase error being the error in the merger time
divided by the waveform period at ring-down multiplied by $2\pi$.

A final note regarding the phase difference in maquillage waveforms:
for two orbits that are quite similar, either because of similar initial conditions
or the same initial conditions but different numerical truncation error, 
the time/phase shifting can be performed at {\em any} time during the common
inspiral/merger phase. The phase difference will then by construction be identically zero
at the time of the match, and slowly drift as one moves away from the matching time.
The choice of matching at the peak of the wave amplitude effectively {\it minimizes}
the net phase error as this is where the wave frequency is highest.


\subsection{Diagnostic of the orbital evolution}
\label{sec2.2}

The initial orbits of the $d=19$ case displayed in Fig.~\ref{orbits}
are clearly neither circular nor a smooth adiabatic inspiral.  It is
natural to refer to such orbits as being eccentric.  However,
describing orbits as ``eccentric'' when radiative effects are strong
can be problematic.  The notion of eccentricity is precise in
Newtonian physics, where the eccentricity is one of two parameters
needed to describe a general, bound elliptic orbit.  In general
relativity, even when considering only the conservative dynamics,
binaries do not follow closed elliptic orbits.  When the dissipative
effects of gravitational radiation are strong, it becomes even more
difficult to define the concept of eccentricity.

The initial data we use starts with essentially no radial momentum.
If radiative dissipation is neglected, such orbits can be circular or
eccentric depending on the magnitude of the orbital angular velocity.
But, because of radiation reaction, initial data with no radial
momenum cannot represent a binary on a smooth quasi-circular inspiral.
In fact, such an orbit must have some effective eccentricity.  

We will use two methods to attempt to calculate this eccentricity in the $d=19$ case. Neither of these
methods work well for the $d=13$ and $d=16$ cases as they do not exhibit enough orbital 
motion prior to merger. The first method uses the following relationship that holds for 
an orbit with eccentricity $e$, orbital angular frequency $\omega$ and separation 
$r$ in Newtonian theory:
\beq
\label{e1_def}
\omega^2(t)\,r^3(t)/M - 1 = e \cos(\phi(t)).
\eeq
Lower order general relativistic corrections (in particular perihelion precession) will change
the argument to the cosine function, though the amplitude remains
$e$. The right panel in Fig.~\ref{d19_coord_e} shows the LHS of Eq.~(\ref{e1_def}) for the
$d=19$ simulation, with $\omega$ and $r$ calculated from the coordinate 
motion of the BHs. A certain amount of eccentricity is due to numerical error, though 
the trend in the curves of Fig.~\ref{d19_coord_e} as resolution increases indicates that 
some amount of eccentricity does come from the initial data. 
The numerical data does not follow Eq.~(\ref{e1_def}) too closely, though at 
early times there are clear oscillations about a line, and we will use 
the amplitude of these oscillations to define
$e$. For a fitting function we use $a_0+a_1 t + e\,\cos(a_2 t + a_3)$, and 
guided by Eq.~(\ref{e1_def}), we define the amplitude of the oscillation 
to be the eccentricity. For the $1/2 h$ run the fit gives $e=0.018\pm0.003$, with the 
uncertainty calculated using the $1/2 h$ and $3/4 h$ data and
assumed second order convergence.

For a second estimate of the eccentricity we use another Newtonian
definition
given in Ref.~\cite{MW}:
\beq
\label{e2_def}
e = \frac{\sqrt{\omega_p} - \sqrt{\omega_a}}{\sqrt{\omega_p} + \sqrt{\omega_a}},
\eeq
where $\omega_p$ is the frequency $\omega$ at a local maxima, and
$\omega_a$ is the frequency at the following local minima. Using this
definition, and the data for the $d=19$ case shown in
Fig.~\ref{Fig:OmegaNQC22d16-19}, we get $e=0.012$ ($0.029,0.068$)
for the $1/2 h$ ($3/4 h, h$) resolution runs. The rather large
differences in the values calculated using the different resolutions
means that the corresponding uncertainty in $e$ calculated using
Eq.~(\ref{e2_def}) is also large: $\pm 0.014$ (for the values quoted
in Table \ref{tab_simnums} we restricted $e\geq0$).

As mentioned in Sec.~\ref{sec1}, Refs.~\cite{MW,BIW} 
have shown that 3PN estimates of {\em eccentric}
orbits suggest the quasicircular initial data being used has some
intrinsic eccentricity.  From Fig.~2 of Ref.~\cite{BIW}, we find a 3PN
estimate of $e\sim0.01$ for the $d=19$ case.  We note that this is
remarkably close to the eccentricity estimate obtained via
Eq.~(\ref{e2_def}).  However, despite this coincidence, we should be
cautious in attributing the ``eccentricity'' observed in the orbit of
the $d=19$ case to a non-vanishing eccentricity in the initial data.
It is important to remember that the initial data are constructed to
have vanishing radial velocity.  As shown by Miller~\cite{MM},
initially circular orbits clearly lead to the kind of effective
eccentric behavior seen in our numerical evolutions.  A comparison of
Fig.~4 of Ref.~\cite{MM} with Fig.~\ref{d19_coord_e} of this paper
also shows striking similarity.  It is clear that the initial data,
through a combination of a vanishing initial radial velocity and
possibly non-vanishing initial eccentricity, results in an evolution
that exhibits some undesired eccentric behavior.  However, it is not
yet possible to determine which, if either, effect dominates.

\begin{figure}
\includegraphics[width=3.25in,clip]{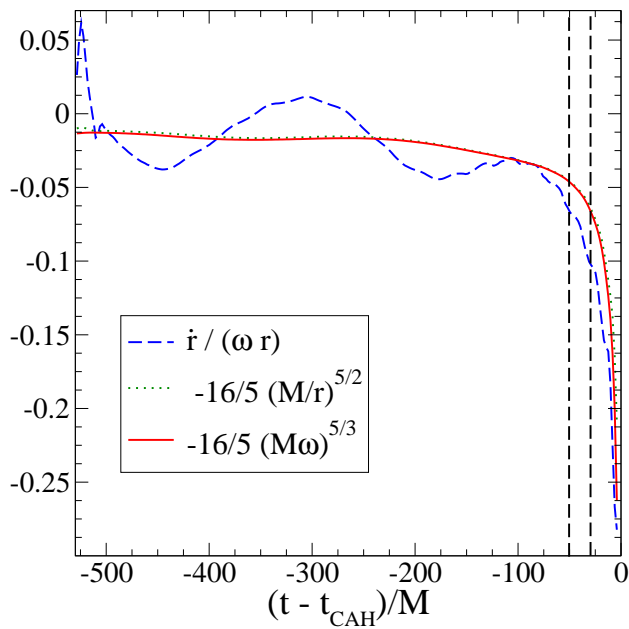}
\caption{We plot the ratio between the radial velocity and the tangential 
velocity as evaluated in the numerical evolution of the $d=19$ case (dashed curve),  
and compare it with the Newtonian predictions (dot and continuous curves) obtained using the 
quadrupole formula. The two dashed vertical lines span the region in which a 
dynamical ISCO could be present.  
\label{d19_vel}}
\end{figure}
Despite the presence of the eccentricity, the orbital motion on {\it average} 
is quasi-circular. By this we mean that throughout the evolution the radial 
velocity is smaller than the tangential velocity. 
At leading order, the quadrupole formula predicts for the radial velocity 
$\dot{r} = -16/5\,(M/r)^3$ and for the ratio between the radial and tangential 
velocity $\dot{r}/(\omega\,r) = -2/3\, \dot{\omega}/{\omega^2} = 
-16/5\,(M/r)^{5/2} = -16/5\,(M\omega)^{5/3}$, where we use $\omega^2\,r^3/M=1$. 
In Fig.~\ref{d19_vel} we show how the above relations are satisfied by the numerical simulations. 
For simplicity we only consider the high-resolution run $d=19$. Quite interestingly, the curves 
$-16/5\,(M/r)^{5/2}$ or $-16/5\,(M\omega)^{5/3}$ average the behavior of $\dot{r}/(\omega\,r)$ during the 
inspiral part and converge to it at later times. Between $30\mbox{--} 50 M$ before the 
formation of the CAH, we notice an abrupt change in the behavior of the ratio 
between the radial velocity and the tangential velocity, which suggests the presence 
of a blurred dynamical ISCO with subsequent plunge~\cite{BD2}. Even during the plunge, the radial velocity is 
still much smaller than the tangential velocity, reaching the value of 
$20\%$ only at the end of the plunge. 
This result is a further confirmation that the numerical, equal-mass dynamics is quasi-circular until 
the end, as predicted by the EOB approach~\cite{BD2}.

\section{The inspiral}
\label{sec3}

The analysis in Sec.~\ref{sec2.2} has shown that, despite the
presence of an initial eccentricity, the dynamics is quasi-circular.
If the dynamics is sufficiently quasi-circular, then it should be
possible to model the inspiral waveform and frequency using Newtonian
and PN methods together with the quadrupole formula.  In this section,
we will compare the numerical waveforms to the expected results from
Newtonian theory and PN theory~\cite{LB} assuming an adiabatic
inspiral.  In a subsequent section (Sec.~\ref{secEOB}) we will also
consider the non-adiabatic EOB model~\cite{BD1,BD2,DJS,BD3} and Pad\'e 
approximants~\cite{DIS98}. Our
analysis should be considered as a first-order attempt to assess the
closeness of analytical and numerical results.  More rigorous
comparisons will be tackled in the future when numerical simulations
start with initial conditions that more accurately model a binary
on an adiabatic path closer to that describe by PN methods, 
and with simulations that have smaller or better understood systematic errors.
\footnote{Note that when comparing with analytical models we assume 
that the binary total mass $M$, introduced in Sec.~\ref{sec2} as
the sum of the irreducible~\cite{CR} BH masses computed from the AH, coincides 
with the rest masses appearing in the PN waveforms, and is
constant. In a numerical evolution,
the mass estimated from the AH can change during evolution. In these simulations
we believe most of this change is due to numerical error, though in principle
part of it could be accretion of gravitational energy. Also, given that
the AH is a coordinate dependent object part of the change could be
gauge-related, though this is unlikely. Regardless of the source, for the highest
resolution simulations the change in $M$ is relatively small.
For example, for the $d=19$ case, we find
that the maximum drift in $M$ is less that about $0.2\%$ before CAH
formation. After the CAH, the final AH mass drops by about $1\%$ in the last $100M$. 
Those variations are within or smaller than other errors present 
in the numerical simulation.}

In addition to examining the full waveforms, it is useful to focus
attention on the angular frequency of the waves and the underlying orbital
motion.  From the evolved data, there are several methods for
determining the orbital angular frequency.  The most direct measure is
obtained by tracking the coordinate locations of the centers of the 
AHs of each individual BH. We label this measure
of the frequency by $\omega_c$.  Because it is based directly on
generalized harmonic coordinate values, this measure of $\omega$ is
susceptible to gauge effects.  A second method for determining
$\omega$ is to track the phase of the maximum of $\Psi_4$ in the
equatorial plane as it intersects the extraction surface at $r=50M$.
We denote the orbital angular frequency determined by this method by
$\omega_\lambda$. Since the angular resolution at which $\Psi_4$ is
sampled is coarser than the temporal resolution, we use spatial interpolation
to find the phase $\phi_{\rm max}(t)$ of the maximum at time $t$, then
smooth the curve in $t$ before computing $\omega_\lambda=d\phi_{\rm max}/dt$.
As a third method, we note that if a complex signal
$f(t)$ has a dominant frequency and it is circular polarized, 
then that frequency is given by ${\rm
Im}[\dot f/f]$, where the dot ($\dot{\ }$) denotes a time derivative.
In terms of the mode amplitudes $_{-2}C_{\ell m}(t,r)$, the dominant 
circular polarized frequency can be estimated by  
\begin{equation}\label{eq:omega_Dm}
\omega_{\rm Dm} = -\frac{1}{m} {\rm Im} 
\left [\frac{{_{-2}\dot{C}_{\ell m}}}
             {_{-2}C_{\ell m}}\right ],
\end{equation}
where we note that $m$ in this equation is the azimuthal index and 
should not be confused with the total mass. These different 
definitions of the frequency are summarized in Table~\ref{tab_frequency}.
\begin{table}
{\small
\begin{tabular}[t]{|c|c | c |}
\hline
Symbol & Type & Computed  \\\hline\hline
$\omega_c$ & orbital frequency & from AH centers (see Sec.~IIB)\\
$\omega_\lambda$ & $_{-2}C_{22}$ frequency & by tracking wave peak (see Sec.~IIB)\\ 
$\omega_{\rm Dm}$ & $_{-2}C_{lm}$ dominant (circular-polarized) frequency & from Eq.~(\ref{eq:omega_Dm}) \\
$\omega_{\rm NQC}$ & $_{-2}C_{22}$ Newtonian quasi-circular frequency & from Eq.~(\ref{eq:omegafromPsi4})\\
\hline
\end{tabular}
}
\caption{We summarize several frequency variables used in the text.} 
\label{tab_frequency}
\end{table}

Figure~\ref{Fig:OmegaNQC22d16-19} compares the orbital angular
velocity $M\,\omega(t)$ obtained by these approaches, for the initial
separations $d=16$ and $19$.  First, note that various frequencies
have been appropriately shifted in time to account for the wave
propagation time to the extraction sphere.  The initial numerical
waveform is dominated by spurious radiation associated with the
initial-data, and $\omega_\lambda$ and $\omega_{{\rm D}2}$ are quite
noisy at early times.  Though it is somewhat difficult to see in
these plots, $\omega_\lambda$ and $\omega_{{\rm D}2}$ also extend to
earlier times than does $\omega_c$.  This is a manifestation of
the fact that $\omega_\lambda$ and $\omega_{{\rm D}2}$ are obtained
from information at the extraction sphere at $r=50M$.  Also small and
difficult to see, we note that there are unexpected deviations at
early times in $\omega_c$.  We find that all three measures of
$\omega$ agree quite well except near the beginning of the evolution
and near the end of the inspiral $30$--$40M$ before the time of the
peak in $|\Psi_4|$.

The aberrant behavior of $\omega$ at early times is primarily due to 
the use of conformally-flat initial data and, consequently, to a
lack of physically realistic initial radiative modes.  This strongly
affects $\omega_\lambda$ and $\omega_{D2}$.  The small anomalous
behavior in $\omega_c$ at early times might also be caused by
artifacts in the initial data, and note that imposing spacetime
harmonic coordinates at the initial time does create some
``artificial'' coordinate dynamics.  Each of these methods for
measuring $\omega$ are based directly on generalized harmonic
coordinate values, and are susceptible to gauge effects.  In
particular, the coordinate position of an AH is certainly not a
gauge-invariant quantity, and given that the BHs are in the
strong-field region of the spacetime there is no {\em a~priori}
reason to expect the coordinate locations to have any simple mapping
to what one may describe as the physical orbit.  It is therefore
somewhat surprising how well these ``almost-harmonic'' coordinates
describe the orbit---more examples of this are given in the next section.

\subsection{Newtonian quadrupole approximation}
\label{sec:Newt_Quadrupole}
Now consider a {\em Newtonian} binary in a circular orbit with orbital
angular frequency $\omega$.  For a binary with reduced mass $\mu=m_1m_2/M$
and mass ratio $\nu=\mu/M$, the standard quadrupole formula yields
\begin{equation}\label{eq:NewtQuadPsi4}
r M \Psi_4 = 32\sqrt{\frac{\pi}5}\nu(M\omega)^{8/3}\left[
e^{-2i(\omega t - \phi_0)}{}_{-\!2}Y_{22} 
+ e^{2i(\omega t - \phi_0)}{}_{-\!2}Y_{2-\!2} \right]
\end{equation}
where $\phi_0$ fixes the initial phase of the orbit and assuming
right-handed rotation about the positive $z$-axis.  If we replace
$\omega t$ by the accumulated phase of the orbit
\begin{equation}\label{eq:phifromomega}
\phi(t) = \int_0^t{\omega(t^\prime)dt^\prime},
\end{equation}
then we find that we can approximate $\ell=2$ modes of the inspiral waveform by
\begin{equation}\label{eq:circ_orb_22}
_{-\!2}C_{2\pm\!2}(t) = 32\sqrt{\frac{\pi}5}\,\nu\,[M\omega(t)]^{8/3}\,e^{\mp2i(\phi(t) - \phi_0)}.
\end{equation}
Note that we have assumed an adiabatic inspiral and have replaced the
constant orbital angular frequency $\omega$ of the circular orbit with
a time-dependent orbital angular frequency $\omega(t)$.  We refer to
the result in Eq.~(\ref{eq:circ_orb_22}) as the Newtonian quadrupole
circular orbit (NQC) approximation.

\begin{figure}
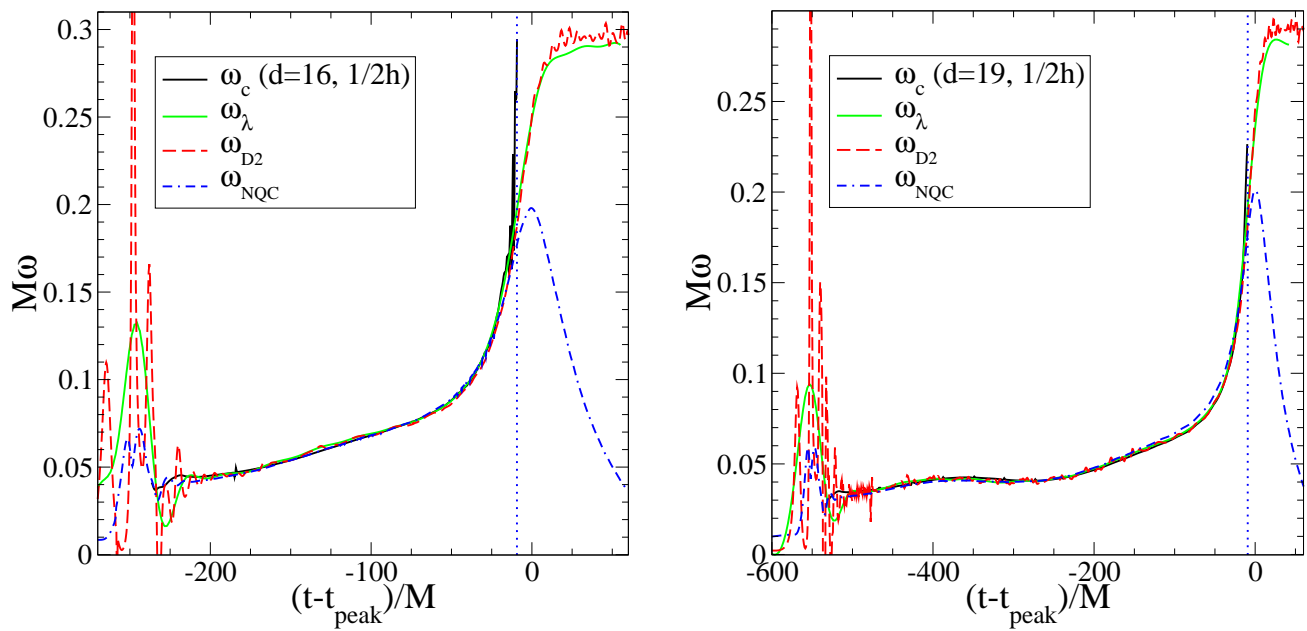

\begin{center}
\includegraphics[width=3.25in,clip]{OmegaNQC22d16}
\hspace{0.5cm}
\includegraphics[width=3.25in,clip]{OmegaNQC22d19}
\caption{\label{Fig:OmegaNQC22d16-19} Orbital angular frequencies
for the $d=16$ and $d=19$ cases evaluated using several different
methods.  The solid (black) line labeled $\omega_c$ displays $M\omega$
as determined from the coordinate locations of the center of each
BH's AH.  The solid (green) line labeled
$\omega_\lambda$ displays $M\omega$ as extracted by tracking the phase
of the peak in $\Psi_4$ at an extraction surface placed at $r=50M$.
The long-dash (red) line labeled $\omega_{\rm D2}$ displays the
dominant frequency in $_{-2}C_{22}$ obtained using
Eq.~(\ref{eq:omega_Dm}).  The dash-dash-dot (blue) line labeled
$\omega_{\rm NQC}$ displays the orbital angular velocity obtained from
$_{-\!2}C_{22}(t)$ using Eq.~(\ref{eq:omegafromPsi4}).
Finally, the vertical dotted (blue) line marks the approximate time
that a common AH forms.}
\end{center}
\end{figure}

This result can be used in two ways.  First, if we assume that
Eq.~(\ref{eq:circ_orb_22}) provides a good approximation to the
waveform, then we can extract $\omega(t)$ from the waveform via
\begin{equation}\label{eq:omegafromPsi4}
  M\,\omega(t) = \left(\frac1{32\nu}\sqrt{\frac5{\pi}}
  |{}_{-\!2}C_{2\pm\!2}(t)|\right)^{3/8}\,.
\end{equation}
In Fig.~\ref{Fig:OmegaNQC22d16-19} we have also plotted $\omega_{\rm
NQC}$ obtained by the Newtonian quadrupole circular orbit
approximation of Eq.~(\ref{eq:omegafromPsi4}). As noted previously,
near the beginning of each evolution, artifacts from the initial data
dominate the waveform and this leads to large inaccuracies in
$\omega_\lambda$ and $\omega_{\rm D2}$.  Similar inaccuracies at early
time are also seen in $\omega_{\rm NQC}$.  Near the end of the
inspiral, when the orbital motion is no longer close to circular, we
should not expect Eq.~(\ref{eq:omegafromPsi4}) to yield an accurate
value for $M\omega$ and we see that the NQC method is systematically
underestimating the value of $M\omega$.

The inconsistency of these methods for determining $\omega$ near the
end of the inspiral should remind us that the very notion of ``orbital
angular frequency'' becomes poorly defined after the formation of a
common horizon.  $\omega_c$ terminates near the end of the inspiral as
a common horizon forms.  The various methods agree quite well until
about a quarter of an orbit before the formation of a common horizon
(see Fig.~\ref{Fig:InspNQC22d16-19} below).  At this point, which we
refer to as the ``decoupling point'', $\omega_c$ separates from
$\omega_\lambda$ and $\omega_{D2}$ and begins to rise more rapidly.
Determining the precise point of decoupling is difficult due to
numerical noise in the frequencies, but it seems to occur close to a
value of $M\omega_{\rm dec}=0.15 \pm 0.01$. Incidentally, this moment
of decoupling also seems to coincide with the time the centers of the individual
AH's cross what we estimate to be the late time co-rotating light ring
of the final BH (see Sec.\ref{secmerger}). Finally, in
Secs.~\ref{sec:adiabatic-PN} and \ref{secEOB}, we will again examine the orbital angular
frequency of the numerical models and find that $\omega$ from 3PN-adiabatic 
and EOB circular orbits agrees well with $\omega_c$.

The second way that the NQC approximation can be used is to
estimate $_{-\!2}C_{2\pm\!2}(t)$ by using $\omega$ extracted from
the evolution.  Figure~\ref{Fig:InspNQC22d16-19} compares the real part
of $_{-\!2}C_{22}(t)$ with the waveform estimated using the
Newtonian quadrupole circular orbit approximation of
Eq.~(\ref{eq:circ_orb_22}) for both the $d=16$ and $19$ cases.  In the
upper half of each plot, we use $\omega_c$ for the orbital angular
frequency.  In the lower half of each plot, $\omega_\lambda$ is used.
We do not consider reconstructing the waveform from $\omega_{\rm D2}$
or $\omega_{\rm NQC}$ because these where themselves derived from
$_{-\!2}C_{22}(t)$.

A benefit of examining these plots is that they give a clear
indication of how much of the initial waveform is contaminated by
artifacts from the initial data.  This can be most clearly seen in
Fig.~\ref{Fig:InspNQC22d16-19} in the comparison with $\omega_c$ where
we find that the {\em estimated} waveform begins later than the {\em
extracted} waveform.  The reason is that $\omega_c$ is a function of
``coordinate time'' while the extracted waveform is a function of
``retarded time'' at the extraction radius.  
So, the beginning of the {\em estimated} waveform
marks the earliest time that a waveform produced by the numerically
evolved inspiral motion could begin.  The numerical
signal preceding this is due entirely to the unphysical initial radiative
content of the initial data.  This signal precedes the inspiral
waveform because it originates from spatial locations in the domain
that are closer than the binary to the extraction sphere.  An initial
segment of the true inspiral signal is also contaminated because
of initial-data artifacts propagating to the extraction sphere from
{\em beyond} the binary.  
If we make the reasonable assumption that the most significant
contributions to the initial-data artifacts originate within the
extraction sphere located at $r=50M$, then we should expect a total of
around $2\times50M$ of the signal to be contaminated as measured in
the retarded time of the extraction sphere.
This number cannot be exact since we expect
the initial-data artifacts to be strongest close to the center
of the extraction sphere, and also because of variations in the
coordinate speed of light in the strong field region.  

Both halves of the plots in Fig.~\ref{Fig:InspNQC22d16-19} show a
clear mismatch at early times.  Because $\omega_\lambda$ is
constructed from information at the extraction sphere, it shows an
initial pulse of radiation that is clearly an artifact of the initial
data.  However, the level of contamination of the wavform decays quickly
following this initial pulse and appears to have become insignificant
by a time of $30M$ to $50M$ following this ``initial-data pulse'' 
\footnote{We also note that if we were computing $_{-\!2}C_{22}$ without 
assuming the Keplerian relation $\omega^2\,r^3/M=1$, the agreement 
would be better at earlier times because the Keplerian relation 
has its largest error there (see Fig.~\ref{d19_coord_e}).}.

A striking feature of these figures is the excellent agreement between
the estimated and extracted waveforms following the initial
contamination and up to a short time before the formation of a common
AH.  During this phase of the inspiral, Fig.~\ref{orbits} clearly
shows that the motion of the binary is not circular.  Nor is it the
smooth adiabatic inspiral that we would expect from an astrophysical binary that
has evolved from much larger separation.  In fact, the observed motion
exhibits a small radial oscillation about this ``desired'' motion.
This effect is most easily seen in the longer $d=19$ evolution, and also
in the plots of $\omega$ in
Fig.~\ref{Fig:OmegaNQC22d16-19} (see also Fig.~\ref{d19_coord_e},
 and Fig.~\ref{omega1} which
includes a fit to the PN form for $\omega$ expected for adiabatic
inspiral from large separation).  The point we want to emphasize is
that Eq.~(\ref{eq:circ_orb_22}) {\em gives an excellent approximation
for the waveform, even when the motion is clearly non-circular, so
long as the phase $\phi(t)$ and orbital angular velocity $\omega(t)$
accurately incorporate the non-circular aspects of the orbital
motion}.  As mentioned above, the NQC approximation appears to work
quite well up until about $1/4$ of an orbit (half a full wave cycle) before
the appearance of a common horizon.

\begin{figure}
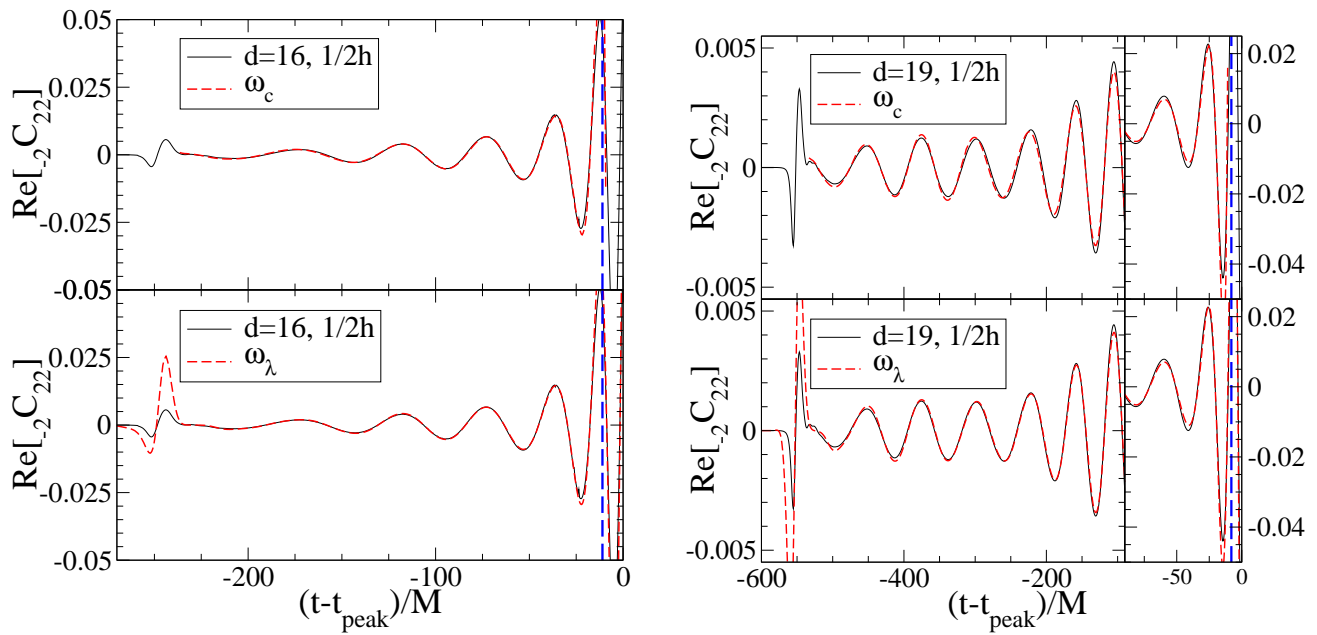

\begin{center}
\includegraphics[width=3.25in,clip]{InspiralNQC22d16}
\hspace{0.5cm}
\includegraphics[width=3.25in,clip]{InspiralNQC22d19_a}
\caption{\label{Fig:InspNQC22d16-19} Comparison of numerical and NQC
inspiral waveforms for the $d=16$ and $19$ cases.  In all plots, the
solid (black) line displays $_{-\!2}C_{22}(t)$ from the numerical
waveform and the vertical long-dashed (blue) line marks the
approximate time that a common AH forms.  In the upper plots, the
dashed (red) line displays $_{-\!2}C_{22}(t)$ as computed from
Eq.~(\ref{eq:circ_orb_22}) using $\omega_c$ and $\phi(t)$ is obtained
from Eq.~(\ref{eq:phifromomega}).  In the lower plots, the dashed
(red) line displays $_{-\!2}C_{22}(t)$ as computed from
Eq.~(\ref{eq:circ_orb_22}) using $\omega_\lambda$. The longer
evolution of the $d=19$ run leads to significant changes in scale for
$_{-\!2}C_{22}(t)$.  Therefore, we change scales for the final $90M$
of the evolution.}
\end{center}
\end{figure}

\begin{figure}
\begin{center}
\includegraphics[width=3.25in,clip]{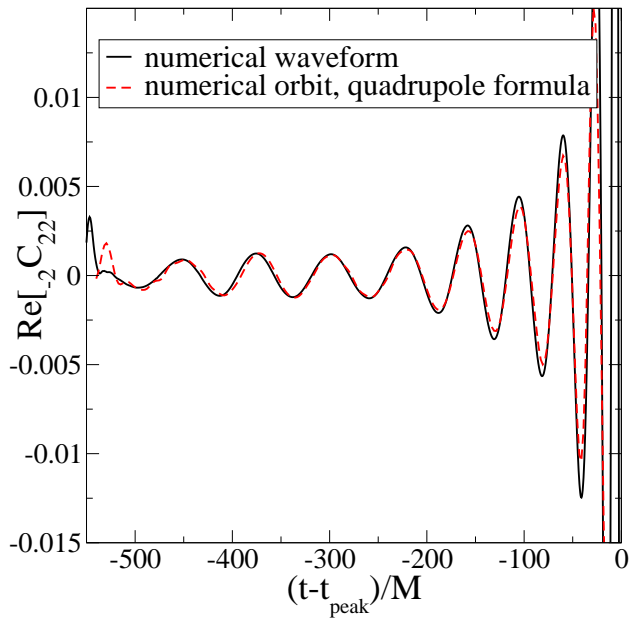}
\caption{\label{Fig:d19_c22_quad_comp} 
Comparison of the real part of the $_{-2}C_{22}$ component of the waveform 
from the $d=19$ ($h/2$)
numerical simulation (solid curve) versus the same quantity calculated by 
using the quadrupole formula (dashed curve) for two point particles,
each of mass $M/2$ and following trajectories
given by the {\em coordinate} motion of the AH's in the simulation (see Fig.~\ref{orbits}).
The waveform from the orbit ends once a common AH forms.}
\end{center}
\end{figure}

Another rather intriguing example demonstrating the adequacy of the quadrupole
formula, and how well adapted the numerical coordinate system 
is in describing the binary motion is
shown in Fig.~\ref{Fig:d19_c22_quad_comp}. For brevity we focus on
a single example here, comparing the real part of the
$_{-2}C_{22}$ component of the $d=19$ waveform
to the same component of a waveform calculated using the
quadrupole formula~\footnote{Here we mean taking directly four time derivatives 
of the coordinate motion of the centers of the individual AHs.} for two point sources of mass $M/2$ following
trajectories given by the coordinate locations of the AH's from the
simulation. The latter curve ends when a common AH forms, and 
has again been shifted in time by a constant amount to account for the
propagation time for the wave to reach the extraction surface. The difference
between this comparison and the preceding NQC comparison
is we have not assumed circular orbits, using instead the
detailed orbit motion obtained from the simulation. Not
only does the good agreement testify to the well-suited nature
of the coordinates, it shows that the quadrupole formula does a
remarkably good job of capturing the dominant physics of GW
emission during the entire merger regime prior to common AH
formation.

\subsection{Adiabatic post-Newtonian model}
\label{sec:adiabatic-PN}

The post-Newtonian approximation to the two-body dynamics of compact 
objects provides the most accurate predictions for the motion and 
the GW emission during the inspiral phase, when the weak-field and 
slow-motion assumptions hold. 

In a more rigorous analysis we would compare the numerical and analytical 
dynamical quantities expressed in the same coordinate system and gauge. 
Here and in Appendix \ref{appendix_comp}, we limit the comparison to a 
few gauge-invariant dynamical quantities, such as 
the orbital frequency $\omega$, the orbital phase $\phi$, the energy flux 
$F_E$ and the angular-momentum flux $F_J$, expressed in terms 
of the instantaneous orbital frequency and/or the time of a stationary 
observer at infinity. This is also motivated by the fact 
that previous studies~\cite{DIS98,BD2} have shown that PN-approximants to 
dynamical quantities are more {\it robust} (under change of PN order) 
if expressed in terms of gauge-invariant quantities, notably the 
instantaneous orbital-frequency $M \omega$. 
At the present time, PN calculations provide the orbital frequency through 
3.5PN order~\cite{35PNnospin} if spins are neglected, and through 2.5PN 
order~\cite{25PNspin} if spins are included. As mentioned in Sec.~\ref{sec1}
and shown in Table~\ref{tab_idparams}, the numerical initial data 
describe BHs which carry a small spin aligned with 
the direction of the orbital angular-momentum $\ell$. For this 
reason we include spin effects in the PN approximants. 

In this section we limit the analysis to the so-called 
PN-adiabatic model. In Sec.~\ref{secEOB} we shall investigate the 
comparisons with the EOB model which goes beyond the adiabatic approximation.
In the PN-adiabatic model the waveforms are computed 
assuming that the motion proceeds along an adiabatic sequence of quasi-circular orbits. 
More specifically, one assumes $\dot{r}=0$ and evaluates 
the variation in time of the orbital frequency $\omega$ from 
the energy-balance equation $d E/dt = - F_E$, where $E$ is the two-body energy 
and $F_E$ is the GW energy flux. In particular, $E$ and $F_E$ are 
first computed for circular orbits and written as a power expansion in $M\,\omega$, 
then $\dot{\omega}(t) = - F_E(\omega)/(dE(\omega)/d\omega)$. The adiabatic sequence of 
circular orbits ends at the {\it conservative} Innermost Circular Orbit (ICO), 
i.e., the ICO evaluated from the conservative dynamics by imposing ${(d E/d \omega)}
{_{\rm ICO}}=0$~\cite{LB}. The study of Ref.~\cite{BD2} (see in particular Figs. 4 and 5 
and discussion around them) and Ref.~\cite{MM}, showed that waveforms computed in the adiabatic 
approximation (which are very accurate at large separations) 
can have a non-negligible phase difference with respect to waveforms 
computed in the non-adiabatic approximation, even before reaching the last 
stable orbit. The accuracy of our numerical simulations and the nature of
the initial data will not allow us to explore these phase
differences.  Nevertheless, we have found it useful to use the adiabatic PN 
model as a diagnostic of the last few cycles of the numerical evolution. 

As discussed in Sec.~\ref{sec:Newt_Quadrupole}, it takes a certain time for the evolution to 
settle to a quasi-circular orbit. Moreover, the numerical results contain a 
non-negligible amount of eccentricity. For these reasons we shall evaluate the PN-adiabatic approximant 
which best {\it averages} the numerical orbital frequency until either the dynamical ISCO, the decoupling 
frequency, or the CAH.  Again, these issues will be overcome when numerical simulations starting at larger 
separation, and from initial conditions that more accurately model an adiabatic inspiral, become available.
We notice that in principle there could be non-negligible differences 
between the instantaneous orbital frequency as defined in PN theory and in the numerical simulation. 
In the latter, the orbital angular frequency is calculated 
from the coordinate locations of the centers of the AHs of each individual BH.  
However, since in Sec.~\ref{sec:adiabatic-PN} we have found that the numerical orbital-frequency agrees 
quite well with the numerical GW frequency extracted at larger
radii, we expect that the differences are small.

\begin{figure}[t]
\begin{center}
\includegraphics[width=3.25in,clip]{omegaHd16PN}
\hspace{0.5cm}
\includegraphics[width=3.25in,clip]{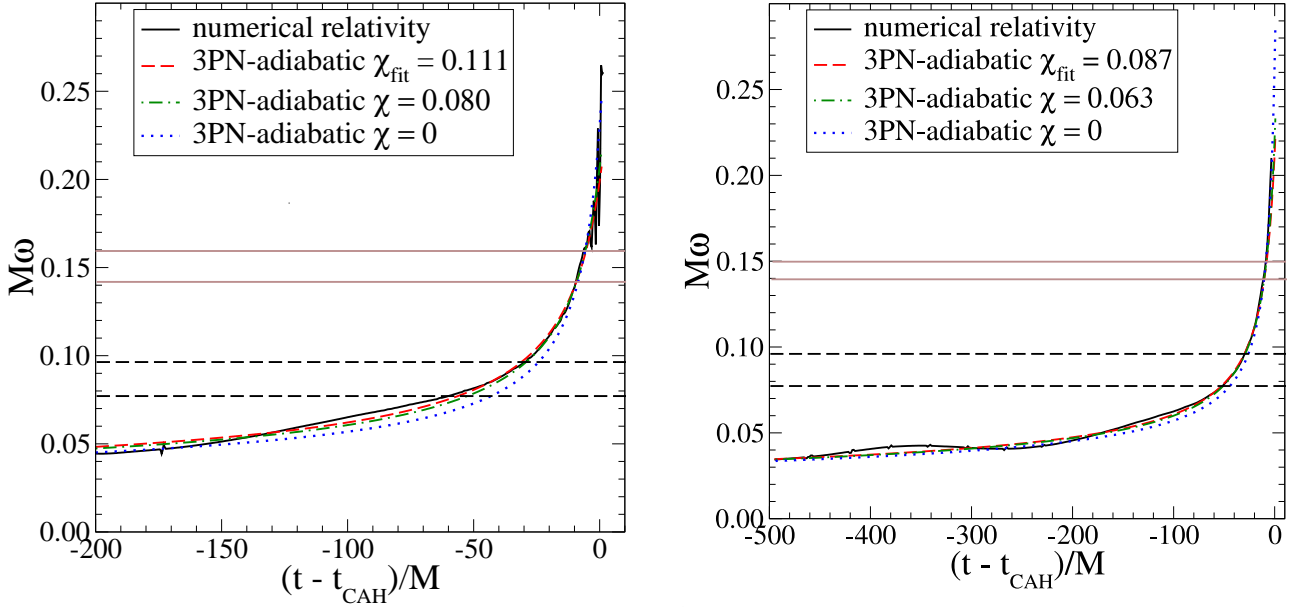}
\caption{We compare the NR and three analytical orbital frequencies 
obtained by fitting (i) both $t_c$ and $\chi$ (dashed line), (ii) only $t_c$ and 
using the nominal $\chi$ value from Table \ref{tab_idparams} (dot-dashed line) 
and (iii) only $t_c$ and assuming $\chi=0$ (dot line). 
The left panel refers to the run $d=16$, the right panel to $d=19$. From top to bottom, 
the continuous horizontal lines mark the ICO frequencies as predicted 
by the 3PN-adiabatic energy and the decoupling frequency. The dashed horizontal 
line in the right panel span the frequency range when a dynamical ISCO could be present. 
\label{omega1}}
\end{center}
\end{figure}

Defining $\tau = \nu\,(t_c-t)/5M$, where $t_c$ is the time 
at which the orbital-frequency diverges (time of coalescence, not to be
confused with the decoupling time) we have 
\begin{eqnarray}
\label{omega}
M\,\omega  &=& {1\over 8}\tau^{-3/8}\biggl\{ 1 + \left( \frac{743}{2688}
 +\frac{11}{32}\nu\right)\tau^{-1/4} -
 \frac{3}{10}\pi\tau^{-3/8}\nonumber\\ 
&+& \left ( \frac{47}{40}\, \frac{S_\ell}{M^2}+\frac{15}{32}
\frac{\delta M}{M}\,\frac{\Sigma_\ell}{M^2} 
\right )\,\tau^{-3/8} \nonumber \\
&+& \left( \frac{1855099}{14450688}
 + \frac{56975}{258048} \nu + \frac{371}{2048} \nu^2 \right) \tau^{-1/2}
 +\left(-\frac{7729}{21504} +\frac{13}{256}\nu\right) \pi
 \tau^{-5/8}\nonumber\\
&+& \left [\left ( \frac{101653}{32256}+\frac{733}{896}\nu \right ) \,\frac{S_\ell}{M^2}+ 
\left (\frac{7453}{7168}+\frac{347}{896}\nu \right ) \,\frac{\delta M}{M}\,\frac{\Sigma_\ell}{M^2} 
\right ]\,\tau^{-5/8} \nonumber \\
 &+&\left(-\frac{720817631400877}{288412611379200}+\frac{53}{200}\pi^2
 +\frac{107}{280}C
 -\frac{107}{2240}\ln\left(\frac{\tau}{256}\right)\right.\nonumber\\
 &+&\left.\left[\frac{25302017977}{4161798144}-\frac{451}{2048}\pi^2 \right]\nu
 -\frac{30913}{1835008}\nu^2+\frac{235925}{1769472}\nu^3\right)
 \tau^{-3/4}\nonumber\\ &+&
 \left(-\frac{188516689}{433520640}-\frac{97765}{258048}\nu
 +\frac{141769}{1290240}\nu^2\right)\pi \tau^{-7/8}\biggr\}\;,
\end{eqnarray}
where $\delta M = m_1-m_2$, the spin variables are   
\begin{eqnarray}
\label{spins}
&& \mathbf{S} \equiv \mathbf{S}_1 + \mathbf{S}_2\,,\\ 
&& \mathbf{\Sigma} \equiv M\left[\frac{\mathbf{S}_2}{m_2} -
\frac{\mathbf{S}_1}{m_1}\right]\,,
\end{eqnarray}
with $\mathbf{S}_i = \chi_i\,m_i^2\,\hat{\mathbf{S}}_i$. In Eq.~(\ref{omega}) we have denoted with 
$S_\ell$ and $\Sigma_\ell$ the spin components along the direction of the orbital angular-momentum 
$\ell$~\cite{25PNspin}. The orbital phase through 3.5PN order reads 
\begin{eqnarray}
\label{phase}
\phi &=& \phi_0 - {1\over \nu} \biggl\{ \tau^{5/8} + \left( \frac{3715}{8064}
  + \frac{55}{96} \nu \right) \tau^{3/8} - \frac{3}{4}\pi\tau^{1/4}
  \nonumber\\ 
 &+& \left ( \frac{47}{16}\, \frac{S_\ell}{M^2}+\frac{75}{64}
\frac{\delta M}{M}\,\frac{\Sigma_\ell}{M^2} 
\right )\,\tau^{1/4}
\nonumber \\ 
&+& \left( \frac{9275495}{14450688} +
  \frac{284875}{258048} \nu + \frac{1855}{2048} \nu^2 \right)
  \tau^{1/8} + \left( -\frac{38645}{172032} + \frac{65}{2048} \nu
  \right) \pi \ln \left({\tau}\right)\nonumber\\
 &+& \left [\left ( \frac{508265}{258048}+\frac{3665}{7168}\nu \right ) \,\frac{S_\ell}{M^2}+ 
\left (\frac{37265}{57344}+\frac{1735}{7168}\nu \right ) \,\frac{\delta M}{M}\,\frac{\Sigma_\ell}{M^2} 
\right ]\,\ln \left({\tau}\right)\nonumber \\
&+&\left(\frac{831032450749357}{57682522275840}-\frac{53}{40}\pi^2
  -\frac{107}{56}C+\frac{107}{448}\ln\left(\frac{\tau}{256}\right)
  \right.\nonumber\\
  &+&\left.\left[-\frac{126510089885}{4161798144}+\frac{2255}{2048}\pi^2 \right]\nu
  +\frac{154565}{1835008}\nu^2-\frac{1179625}{1769472}\nu^3\right)
  \tau^{-1/8}\nonumber\\ &+&
  \left(\frac{188516689}{173408256}+\frac{488825}{516096}\nu
  -\frac{141769}{516096}\nu^2\right)\pi \tau^{-1/4}\biggr\}\;,
\end{eqnarray}
where $C=0.577 \cdots$ is the Euler constant and $\phi_0$ is an arbitrary constant. 
The non-spin terms in Eqs.~(\ref{omega}) and (\ref{phase}) are given by Eqs.~(12) and (13) 
of Refs.~\cite{BFIJ,35PNnospin}, while we evaluated the spin terms through 2.5PN order 
using the recent results of Ref.~\cite{25PNspin} (we use the constant spin variables 
as defined in Ref.~\cite{25PNspin}), and we neglected spin-spin contributions. 

\begin{figure}[t]
\begin{center}
\includegraphics[width=3.25in,clip]{ReC22Hd16PN}
\hspace{0.5cm}
\includegraphics[width=3.25in,clip]{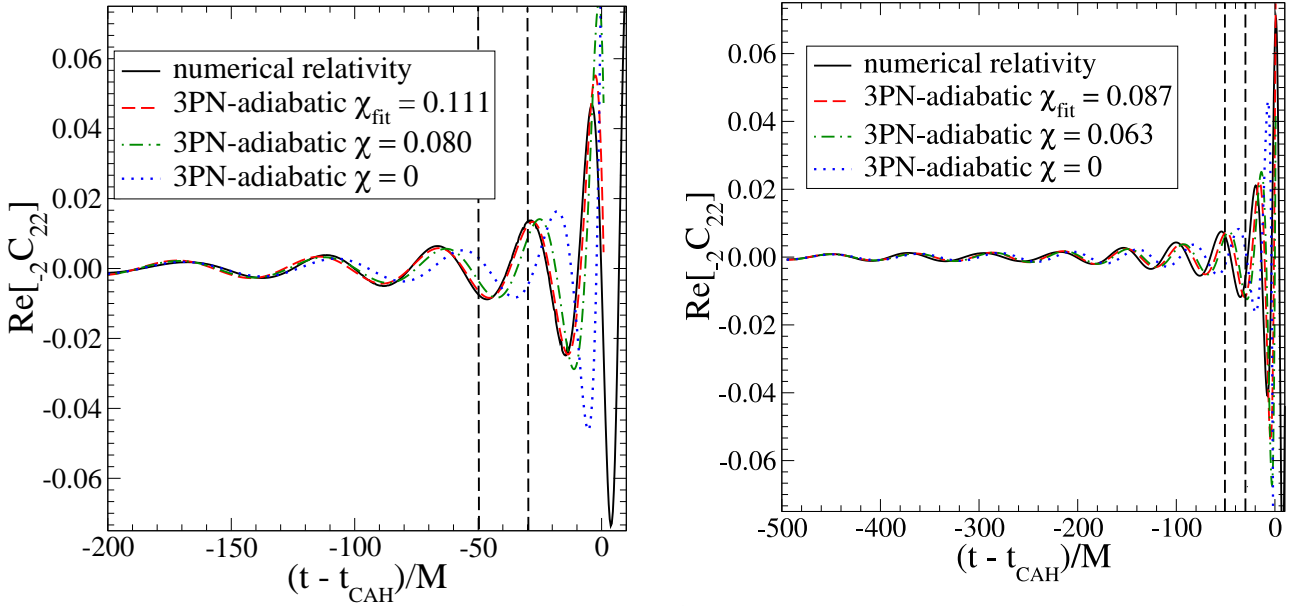}
\caption{We compare the NR and analytical ${\rm Re}[{}_{-2}C_{2 2}]$ 
obtained by fitting the orbital frequencies. The fit is done for  (i) 
both $t_c$ and $\chi$ (dashed line), (ii) only for $t_c$ and 
using the nominal value for $\chi$ from Table \ref{tab_idparams} (dot-dashed line) 
and (iii) only for $t_c$ and assuming $\chi=0$ (dot line). 
The left panel refers to the run $d=16$, the right panel to $d=19$. 
The dashed vertical lines span the region during which a dynamical ISCO 
could be present. The waveform cycle beyond this point could be associated 
with the plunge cycle.   
\label{ReC221}}
\end{center}
\end{figure}

For each of the three runs, we determine the time of coalescence $t_c$ and the spin-magnitude 
$\chi=\chi_1=\chi_2$ by fitting the PN orbital frequency $(\ref{omega})$ to the numerical orbital 
frequency $\omega_c$ using a non-linear least-squares method.
Figure~\ref{omega1} shows the results for $d=16$ and $d=19$. 
Due to the initial burst of radiation related to the initial conditions, 
we remove from the numerical data the first $\approx 22M$  
and $\approx 34M$, for the $d=16$ and $d=19$ runs, respectively. 
In Sec.~\ref{sec2.2} we discussed the possible presence of a blurred dynamical 
ISCO~\cite{BD2} occurring $30\mbox{--}50 M$ before the CAH forms, at $M \,\omega_{\rm dyn\,ISCO} = 
0.078\mbox{--}0.097$. The orbital frequency at the conservative ICO evaluated at 3PN order with 
$\chi = 0.080$ is $M\,\omega_{\rm ICO} = 0.143$ and with 
$\chi = 0.063$ is $M\,\omega_{\rm ICO} = 0.140$.  
We mark all these frequencies in Fig.~\ref{omega1} along 
with the decoupling frequency. In principle, the {\it adiabatic} PN waveform
should be used {\it only} until the last stable orbit, since it was derived from the 
balance equation for circular orbits which ends at this last stable orbit. 
However, since the two-body motion predicted by the numerical simulations is rather 
adiabatic and quasi-circular until the CAH forms~\cite{BD2}, we extend the PN waveforms through 
the plunge until almost that point.  
More specifically, we fit until the time at which the numerical orbital frequency 
{\it decouples} from the GW frequency, $M\omega_{\rm dec} \approx 0.15\pm0.01$.   
Notice that by fitting the time of coalescence $t_c$ and $\chi$ we are fitting the 
initial value of the orbital frequency. We find that the 3PN-approximant best fits the data 
with $\chi^{\rm d=16}=0.111$ and $t_c^{\rm d=16}= 220.1\, M$ [$M\,\omega(t=0)=0.04789$], 
$\chi^{\rm d = 19}=0.0874$  and $t_c^{\rm d=19}= 509.1\, M$ [$M\,\omega(t=0)=0.03477$].
If we fit until the CAH time, we find $\chi^{\rm d=16}=0.0812$ and $t_c^{\rm d=16}= 217.6\, M$ 
[$M\,\omega(t=0)=0.04715$], $\chi^{\rm d = 19}=0.0626$  and $t_c^{\rm d=19}= 506.2\, m$ 
[$M\,\omega(t=0)=0.03447$]. Those values are closer to the nominal $\chi$ values of 
Table \ref{tab_idparams}. However, we think this is accidental. 
Finally, notice that if we fit until the dynamcal ISCO $M\omega_{\rm dyn\,ISCO} 
= 0.096$ we find $\chi^{\rm d=19}=0.113$ and $t_c^{\rm d=19}= 514.3\, M$ 
[$M\,\omega(t=0)=0.03502$].

In Fig.~\ref{omega1} we also show curves evaluated using the nominal $\chi$ 
value of Table \ref{tab_idparams}. To understand how spins affect 
the PN-adiabatic orbital frequency, we show in Fig.~\ref{omega1} also the case in which we fix 
$\chi = 0$ and fit only $t_c$. The latter values produce a difference of $\approx 0.3$ 
GW cycles at the CAH time with respect to the case where we fit both $t_c$ and $\chi$. 
Although we can use the numerical results to discriminate between several 
PN-adiabatic models with spin, it is not clear which 
role the spin variable is playing in fitting the data. In fact, the spin values obtained 
from the fit are also affected by the eccentricity present in the numerical data but 
absent in the analytical model. 

Using the time of coalescence and spin values obtained from the fits, we plot in 
Fig.~\ref{ReC221} the waveforms ${\rm Re}[{}_{-2}C_{2 2}]$. Notice that the 
GW phase differences between the fitted models is smaller than the maximum 
GW phase error estimated in Table \ref{tab_simnums} using lower resolution 
runs. This gives a concrete example of how cumulative phase error in a 
numerical simulation translates to uncertainties associating PN model
parameters with the numerical waveform, despite the deceptively small
phase error after maquillage.
PN-adiabatic models can also fit the data of lower resolution 
runs and give initial orbital frequencies larger than those found for
the high resolution runs.

\begin{figure}
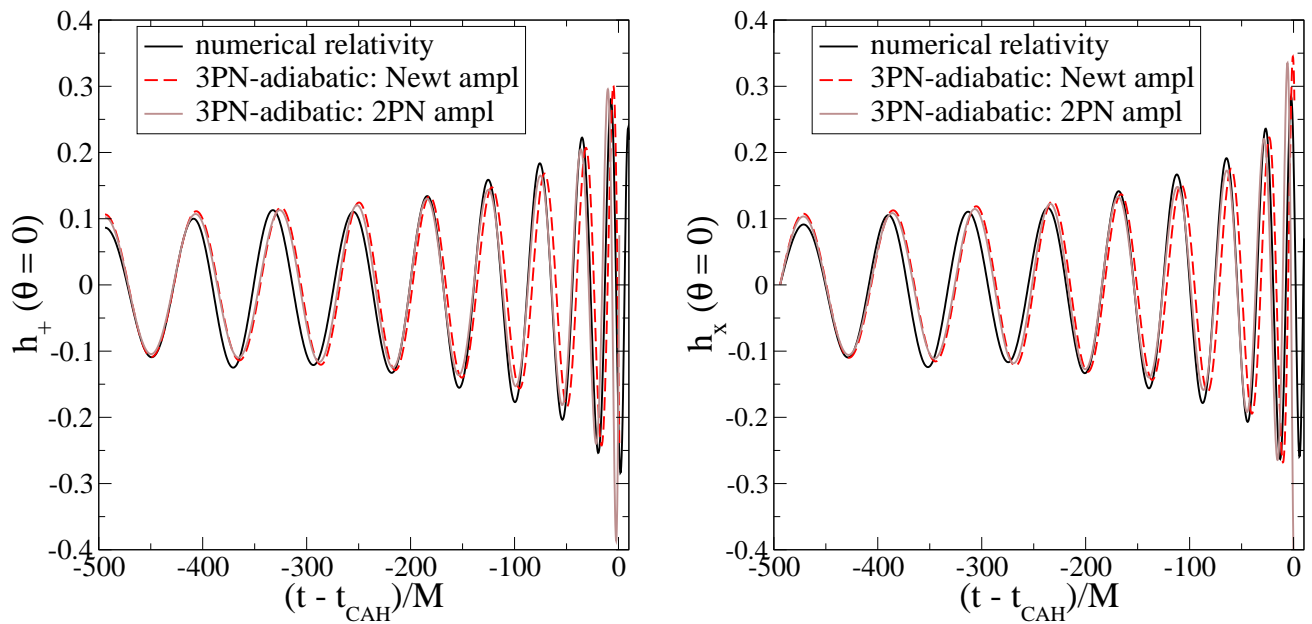

\begin{center}
\includegraphics[width=3.25in,clip]{hpd19}
\hspace{0.5cm}
\includegraphics[width=3.25in,clip]{hcd19}
\caption{We plot the numerical $h_+$ (left panel) and $h_\times$
  (right panel) extracted along the direction perpendicular to the
  orbital plane (dark continuous line) and compare them with
  PN-adiabatic predictions with the phase evaluated at 3PN order and
  the amplitude evaluated either at Newtonian order (dashed line) or
  through 2PN order (light continuous line). These results refer to
  the $d = 19$ run. \label{hphc19}}
\end{center}
\end{figure}

So far, when comparing with numerical waveforms, we have neglected higher-order PN corrections to the 
GW amplitude and have restricted the comparison to ${\rm Re}[{}_{-2}C_{2 2}]$, 
i.e., we used waveforms in the so-called {\it restricted approximation}. 
In Ref.~\cite{BIWW} the authors evaluated the ready-to-use PN waveforms $h_+$ and 
$h_\times$ through 2PN order in the amplitude (see Ref.~\cite{ABQI} where this computation 
has been pushed through 2.5PN order). In Figs.~\ref{hphc19} 
we compare the numerical $h_+$ and $h_\times$, which are obtained by integrating 
$\psi_4$ twice in time, with the analytical $h_+$ and $h_\times$ as 
given by Eqs.~(2)--(4) of Ref.~\cite{BIWW}. The phase is computed from Eq.~(\ref{phase}) at 3PN order 
with the values of $t_c$ and $\chi$ obtained from the best fit. The waves are extracted along the direction 
perpendicular to the orbital plane. Because the BH masses are equal, only 
the $2^{\rm nd}$ harmonic is present. We would conclude that higher-order PN amplitude corrections have 
a mild effect in the waveform emitted by equal-mass binaries. However, we notice an oscillating 
behavior in the PN approximation to $h_+$ and $h_\times$. For example the 1PN correction is 
rather large and opposite in sign to the Newtonian correction, resulting in 
a significant reduction of the amplitude of the signal. The 1.5PN and 2PN corrections 
undo this effect. This oscillating behavior seems to also affect the higher multipoles 
$C_{l m}$. In fact, we checked that $C_{44}$ is well approximated by the 3PN-adiabatic 
model for the phase, if computed at (leading) 1PN order in the amplitude, but 
the agreement become worse when 2PN corrections are added. 
We plan to investigate in more detail the effect of higher-order 
PN corrections and higher multipoles in the future. 

To obtain more robust comparisons between PN and numerical predictions, it would be preferable 
to start the numerical evolution where we are confident that the PN expansion can be safely applied.
First, we notice that if we were computing $\omega$ as a function of time instead of from Eq.~(\ref{omega}), then 
integrating numerically the following equation~\cite{35PNnospin,25PNspin}
\bea 
&&\frac{\dot{\omega}}{\omega^2}=\frac{96}{5}\,\nu\,(M\omega)^{5/3}\Bigg
\{1-\frac{743+924\,\nu}{336}\,(m\omega)^{2/3} +
\frac{(M \omega)}{M^2}\left[-\frac{47}{3}S_\ell
-\frac{25}{4}\frac{\delta M}{M}\Sigma_\ell\right]
+ 4\pi (M\omega) \nonumber \\ 
&& + \Bigg
(\frac{34\,103}{18\,144}+\frac{13\,661}{2\,016}\,\nu+\frac{59}{18}\,\nu^2
\Bigg )\,(M\omega)^{4/3} +\frac{(M\omega)^{5/3}}{M^2}\left[
\left(-\frac{31811}{1008}+\frac{5039}{84}\nu\right)S_\ell
+\left(-\frac{473}{84}+\frac{1231}{56}\nu\right)\frac{\delta
M}{M}\Sigma_\ell\right] \nonumber \\ 
&& -\frac{1}{672}\,(4\,159 +
15\,876\,\nu)\,\pi\,(M\omega)^{5/3} + \Bigg[
\left(\frac{16\,447\,322\,263}{139\,708\,800}-\frac{1\,712}{105}\gamma_E+\frac{16}{3}\pi^2\right)+
\left(-\frac{56\,198\,689}{217\,728}+\frac{451}{48}\pi^2 \right)\nu \nonumber \\
&&+\frac{541}{896}\nu^2-\frac{5\,605}{2\,592}\nu^3
-\frac{856}{105}\log\left[16(M\omega)^{2/3}\right] \Bigg] (M\omega)^2
+ \Bigg (
-\frac{4\,415}{4\,032}+\frac{358\,675}{6\,048}\,\nu+\frac{91\,495}{1\,512}\,\nu^2
\Bigg )\,\pi\,(M\omega)^{7/3} \Bigg \}\,,
\label{omegadot}
\eea
\begin{figure}
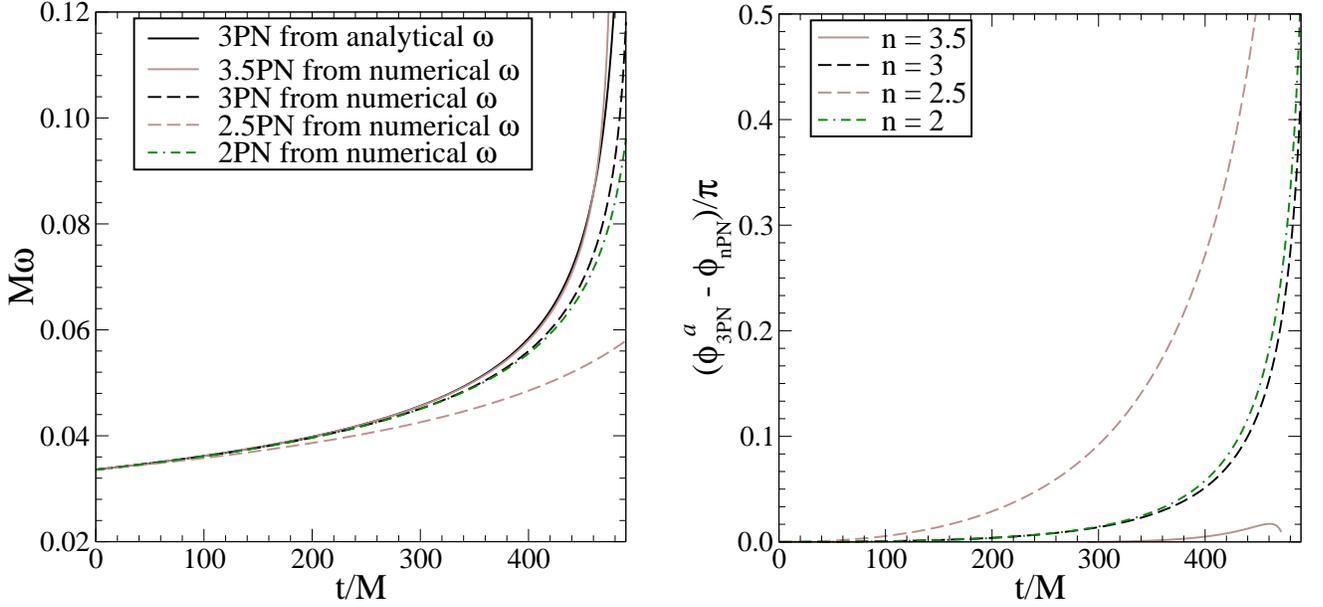

\begin{center}
\includegraphics[width=3.25in,clip]{diffomegad19}
\hspace{0.5cm}
\includegraphics[width=3.25in,clip]{diffphid19}
\caption{In the left panel we plot the orbital frequency for several PN approximants 
computed from Eqs.~(\ref{omega}) and (\ref{omegadot}) whose frequencies coincide at time $t=0$. 
In the right panel we plot the 
differences between the number of GW cycles at 3PN computed from the 
analytical Eq.~(\ref{omega}), and at nPN order computed from Eq.~(\ref{omegadot}). 
The initial frequency coincides with the one computed from the fit of 
the $d = 19$ run when $\chi = 0$. \label{diff}}
\end{center}
\end{figure}
which can be derived expanding $\dot{\omega} = - F(\omega)/(dE/d\omega)$ in powers 
of $(M\omega)$, we find some differences from Eq.~(\ref{omega}). In particular, at 3.5PN 
and 2.5PN orders, $\omega$ derived from Eq.~(\ref{omega}) reaches a maximum and then starts 
decreasing, becoming negative. By contrast, this behavior does not occur in the 
$\omega$ derived by numerically integrating Eq.~(\ref{omegadot}). 
In Fig.~\ref{diff} we show the differences in the orbital frequency and the number of GW cycles 
if the latter quantities were computed from Eq.~(\ref{omegadot}) 
at different PN orders but with the same initial frequency $\omega_0 = 0.03361$. 
We consider here a non-spinning binary. 
In the right panel of Fig.~\ref{diff} we compute the differences in the number of 
GW cycles between the 3PN $\omega$ from the analytical Eq.~(\ref{omega}) and 
several PN $\omega$ from Eq.~(\ref{omegadot}). Quite interestingly 
the 3.5PN order computed numerically is very close to the 3PN order 
computed analytically, whereas the 3.5PN order computed analytically has 
almost half a cycle of difference at the end of the inspiral.  
These differences are a consequence of the fact that at such (close) initial 
separation, $M \,\omega \sim 0.033$, the differences between 
PN-approximants are not negligible. This fact is better 
illustrated in Fig.~\ref{diffphiomega}.  
We start the evolutions at $M \,\omega = 0.004$ (left panel) 
and $M \,\omega = 0.02$ (right panel) and 
plot the differences between the number of GW cycles 
at 3.5PN and at nPN order versus the $\omega$ computed 
at 3.5PN order. All quantities are obtained by numerically integrating Eq.~(\ref{omega}) 
with spins set to zero. 2PN and 2.5PN approximants accumulate large differences 
from the 3.5PN-approximant when evolving from larger separations (left panel).
 
Thus, summarizing, the phase computed at 3PN order from Eq.~(\ref{omega}) 
or at 3.5PN order from Eq.~(\ref{omegadot}) best-fit the numerical results.    
If we wanted to investigate to which PN order, notably 3PN or 3.5PN, the NR orbital 
frequency is closest to, we would need to start the numerical evolution at frequencies smaller 
then the one used in the $d=19$ run, which is $\sim 0.033$. 
\begin{figure}
\begin{center}
\includegraphics[width=3.25in,clip]{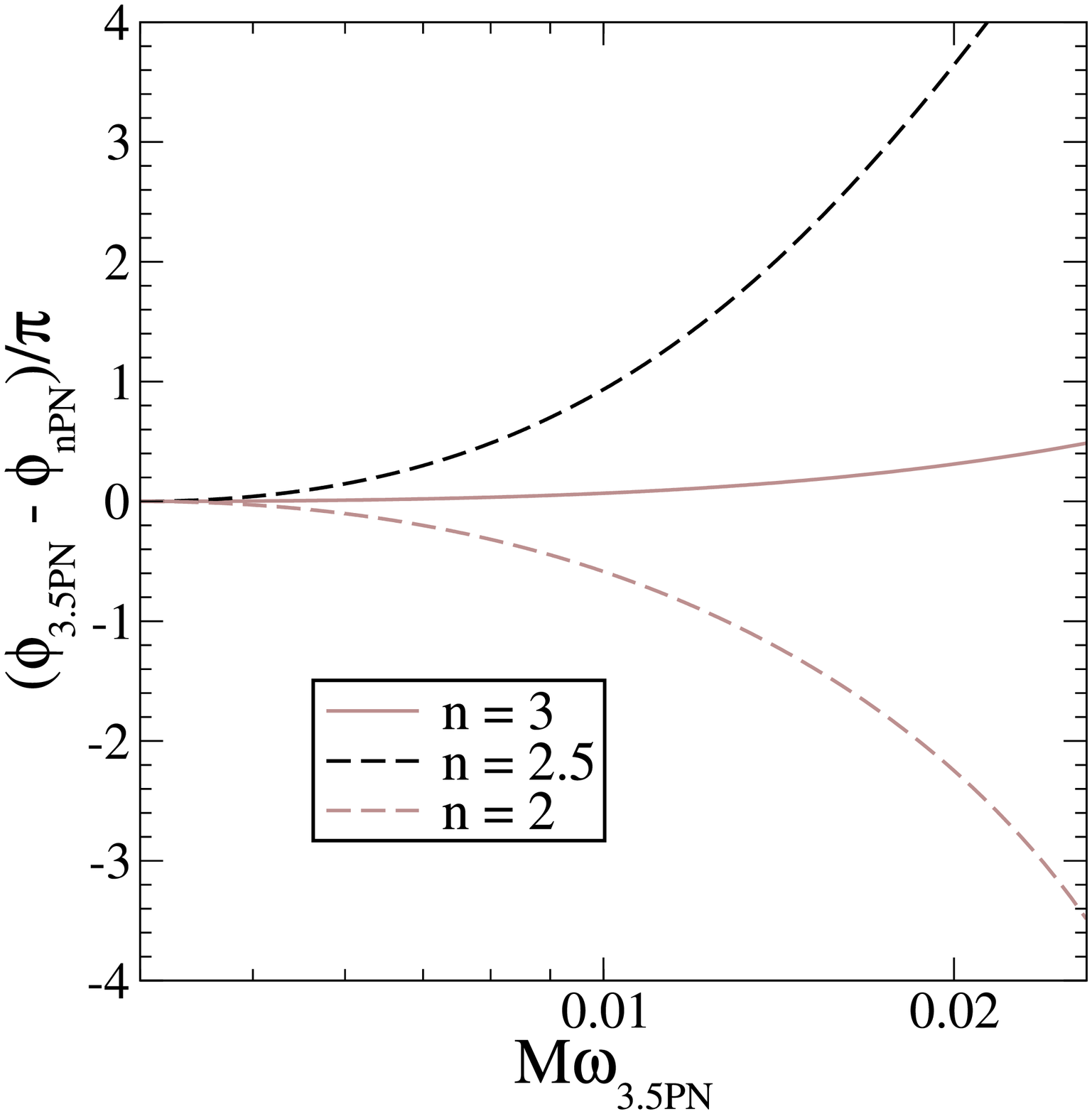}
\hspace{0.5cm}
\includegraphics[width=3.25in,clip]{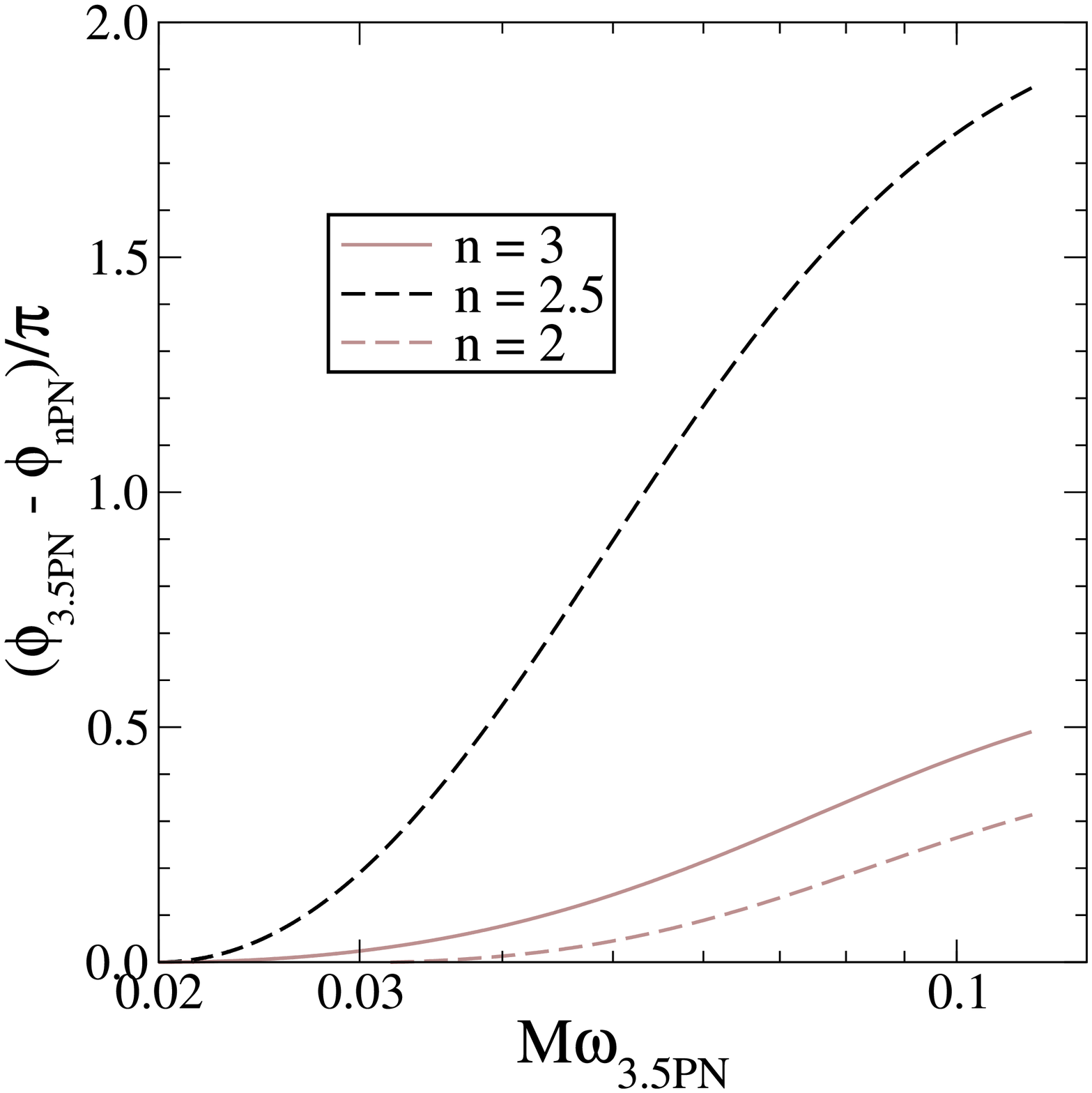}
\caption{We plot the differences between the number of GW cycles 
at 3.5PN and at nPN order versus the 3.5PN $\omega$. All 
quantities are computed integrating numerically Eq.~(\ref{omegadot}) 
with spins set to zero. 
The initial and final frequencies are $M \omega = 0.004$ and $M \omega = 0.026$
in the left panel, $M \omega = 0.02$ and $M \omega = 0.13$ in the right panel. 
They correspond to an equal-mass binary sweeping in the most sensitive 
frequency band of LIGO from $\sim 43$ Hz to $\sim 280$ Hz of mass 
$(3+3)M_\odot$ in the left panel and of mass $(15+15)M_\odot$ in the right  panel.
\label{diffphiomega}} 
\end{center}
\end{figure}

\section{The ring-down phase}
\label{sec4}

During the ring-down, the GW can be decomposed in
terms of the quasi-normal modes (QNM) of 
Kerr\cite{L85,E89,brandt_seidel}.  These modes are
distinguished by their longitudinal and azimuthal indices $\ell$ and
$m$, as well as by their overtone number $n$.  Each mode has a
particular frequency $\omega_{\ell mn}$ and decay constant $\tau_{\ell
mn}$ which are functions of the Kerr parameter $a$ and total mass
$M_f$ of the background BH that is being perturbed.  To
shorten the notation, we will introduce the complex frequency
$\hat\omega_{\ell mn}$ and use ${}^*$ to denote complex conjugation:
\begin{equation}
  \hat\omega_{\ell mn} \equiv \omega_{\ell mn} - i/\tau_{\ell mn}.
\end{equation}

Following Ref.~\cite{BCW}, the ring down can be
expressed, in terms of the Weyl scalar as
\begin{eqnarray}
\label{eq:ring-down_gen_spheroidal}
  r M \Psi_4 &=& \sum_{\ell mn}\Bigl\{
      {\cal C}_{\ell mn}e^{-i(\hat\omega_{\ell mn}t + \phi_{\ell mn})}
          S_{\ell mn}
    + {\cal C}^\prime_{\ell mn}e^{i(\hat\omega^*_{\ell mn}t +\phi^\prime_{\ell mn})}
          S^*_{\ell mn} 
\Bigr\},
\end{eqnarray}
where ${\cal C}_{\ell mn}$, ${\cal C}^\prime_{\ell mn}$, $\phi_{\ell
mn}$, and $\phi^\prime_{\ell mn}$ are real constants, and $S_{\ell mn}
= S_{\ell m}(a\hat\omega_{\ell mn})$ are the spin-weight -2 {\em
spheroidal} harmonics\footnote{We note that the $\theta$
depencence of spheroidal harmonics is connected to the separability of
the Kerr metric in terms of Boyer-Lindquist coordinates.  While our
spherical coordinate system is not Boyer-Lindquist, the differences
are not significant in the wave zone where the waveform is extracted.}
implicitly evaluated at the complex QNM
frequencies.  The primed terms are necessary because, for a given
$(\ell,m,n)$ and a fixed non-vanishing angular momentum, there are two
solutions of the eigenvalue problem.  To make the notation as clear as
possible, we will always take the real frequencies and decay constants
to be non-negative, so $\omega_{\ell mn}\ge0$ and $\tau_{\ell mn}
\ge0$.  Of the two solutions to the eigenvalue problem for fixed
$(\ell,m,n)$, one solution has positive frequency and one negative,
and the complex frequencies are related by $-\hat\omega_{\ell mn} =
\hat\omega^*_{\ell-\!mn}$ (see Ref.~\cite{BCW} for a full discussion).
Because of this relationship, it is only necessary to determine the
positive (or negative) frequency modes.  In Ref.~\cite{BCW} the
authors compute the positive frequency modes and choose the convention
that $\omega_{\ell m n} \ge \omega_{\ell-\!mn}$, with equality in the
case that $m=0$ or $a=0$.  Finally, we note that with these
conventions, it is necessary to introduce an overall sign change on
the real frequency in the equations of Ref.~\cite{BCW} in order for
the signs of the frequencies of the various modes to agree with
numerical simulations.

The decomposition of $\Psi_4$ in terms of spin-weight -2 {\em
spherical} harmonics is given by Eq.~(\ref{eq:psi4Ylmdef}).  In order to
relate the expansion coefficients in
Eqs~(\ref{eq:ring-down_gen_spheroidal}) and (\ref{eq:psi4Ylmdef}), we
need the expansion of the spheriodal harmonics in terms of the
spherical harmonics.  Following Press and Teukolsky~\cite{PT},
\begin{equation}
  S_{\ell mn} = \sum_{\ell^{\prime\prime}}
     {\cal A}_{\ell\ell^{\prime\prime} mn}{}_{-\!2}Y_{\ell^{\prime\prime} m}.
\end{equation}
Using the orthonormality of spin-weighted spherical harmonics, we find
that
\begin{eqnarray}\label{eq:fullQNM_Cexp}
_{-\!2}C_{\ell m} &=& \sum_{\ell^{\prime\prime} n}\Bigl\{
          {\cal C}_{\ell^{\prime\prime} mn}
	  {\cal A}_{\ell\ell^{\prime\prime} mn}
	  e^{-i(\hat\omega_{\ell^{\prime\prime} mn}t + 
	                           \phi_{\ell^{\prime\prime} mn})}
	  + {\cal C}^\prime_{\ell^{\prime\prime}-\!mn}
	  {\cal A}^*_{\ell\ell^{\prime\prime}-\!mn}
	  e^{i(\hat\omega^*_{\ell^{\prime\prime}-\!mn}t + 
	                           \phi^\prime_{\ell^{\prime\prime}-\!mn})}
	  \Bigr\},\\
\label{eq:RealQNM_Cexp}
&\equiv& \sum_{\ell^{\prime\prime} n}\Bigl\{
          {\cal C}_{\ell\ell^{\prime\prime} mn}
	  e^{-i(\hat\omega_{\ell^{\prime\prime} mn}t + 
	                           \phi_{\ell\ell^{\prime\prime} mn})}
	  + {\cal C}^\prime_{\ell\ell^{\prime\prime}mn}
	  e^{i(\hat\omega^*_{\ell^{\prime\prime}-\!mn}t + 
	                           \phi^\prime_{\ell\ell^{\prime\prime} mn})}
	  \Bigr\}.
\end{eqnarray}
So, in principle, a spherical harmonic mode amplitude
$_{-\!2}C_{\ell m}$ of the ring-down signal can contain a
contribution from any of the negative frequency modes with azimuthal
index $m$ {\em and} from any of the positive frequency modes with
azimuthal index $-m$.  Note that in the second version of this
expansion, the complex coefficients ${\cal A}_{\ell\ell^{\prime\prime}mn}$
have been absorbed into the new real expansion coefficients 
${\cal C}_{\ell\ell^{\prime\prime}mn}$, 
${\cal C}^\prime_{\ell\ell^{\prime\prime}mn}$,
$\phi_{\ell\ell^{\prime\prime} mn}$, and 
$\phi^\prime_{\ell\ell^{\prime\prime} mn}$, where each coefficient
now has four indices.

The expansion coefficients ${\cal A}_{\ell\ell^{\prime\prime} mn}$
depend on the product of the Kerr parameter and the complex QNM
frequency $a\hat\omega_{\ell mn}$ and for sufficiently small values
can be determined via perturbation theory (cf.\ Ref.~\cite{PT}).  For
example, using first order perturbation theory, we find
\begin{eqnarray}
_{-\!2}C_{22} &=& \sum_{n}\Bigl\{
     {\cal C}_{22n}
          e^{-i(\hat\omega_{22n}t + \phi_{22n})}
   + {\cal C}^\prime_{2-\!2n}
          e^{i(\hat\omega^*_{2-\!2n}t + \phi^\prime_{2-\!2n})}
\\ && \mbox{}\hspace{0.25in}
	  + \frac1{18}\sqrt{\frac57}a\hat\omega_{32n}(4+a\hat\omega_{32n})
	  {\cal C}_{32n} e^{-i(\hat\omega_{32n}t + \phi_{32n})}
\\ && \mbox{}\hspace{0.25in}
	  - \frac1{18}\sqrt{\frac57}a\hat\omega^*_{3-\!2n}(4+a\hat\omega^*_{3-\!2n})
	  {\cal C}^\prime_{3-\!2n} e^{i(\hat\omega^*_{3-\!2n}t + \phi^\prime_{3-\!2n})}
\\ && \mbox{}\hspace{0.25in}
	  + \frac{\sqrt{5}}{294}(a\hat\omega_{42n})^2{\cal C}_{42n}
	  e^{-i(\hat\omega_{42n}t + \phi_{42n})}
\\ && \mbox{}\hspace{0.25in}
	  + \frac{\sqrt{5}}{294}(a\hat\omega^*_{4-\!2n})^2{\cal C}^\prime_{4-\!2n}
	  e^{i(\hat\omega_{4-\!2n}t + \phi^\prime_{4-\!2n})}
\Bigr\}.
\end{eqnarray}

We have extracted the various QNM contributions to the
$_{-\!2}C_{22}(t)$ ring-down signal in the following way~\footnote{After
the work to fit the ring-down modes was completed, a similar approach was 
posted in the preprint archives~\cite{DBDST}.}.  At late
times, we expect the $\ell=2$, $m=2$, $n=0$ QNM to dominate.  We fit
the signal {\em after} time $t_r+t_{\rm peak}$ to this single mode
using non-linear regression and choose $t_r$ to minimize the error in
the fit.
There are four dimensionless parameters in this non-linear
fit: ${\cal C}_{220}$, $\phi_{220}$, $m\omega_{220}$, and
$\tau_{220}/m$.  However, instead of fitting directly for these four
parameters, we treat $m\omega_{\ell mn}$ and $\tau_{\ell mn}/m$ as
functions of $a/M_f$ and $M_f/m$ which can be obtained via
interpolation from tabulated values (cf. Tables II--IV of Ref.~\cite{BCW})
or via approximating functions (cf. Tables VIII--X of Ref.~\cite{BCW}).
The advantage of using $(a/M_f,M_f/m,{\cal C}_{220},\phi_{220})$ for
the set of fitting parameters comes when we fit to additional modes.

If we knew $a/M_f$ and $M_f/M$ precisely from the fit to the dominant
mode, then we could directly
compute the values for $M\omega_{\ell mn}$ and $\tau_{\ell mn}/M$ for
all additional contributing modes.
All that would be necessary for determining the contribution of each
additional mode would be to fit for its amplitude and phase.  However,
we will not know $a/M_f$ and $M_f/M$ precisely from the fit to the
dominant mode.  Therefore, we treat the frequency and damping constant
for each mode as functions of $a/M_f$ and $M_f/M$ so that they are
determined consistently when we fit multiple modes to a given 
signal.

To make our procedure more explicit, we note that for the
$_{-\!2}C_{22}(t)$ ring-down signal, the contributions of modes with
$\ell>2$ seem to be near the level of numerical precision, as are the
positive frequency modes with $\ell=2$ and $m=-2$.  Therefore, we take
as our fitting function:
\begin{equation}\label{eq:C22Nfitfunc}
_{-\!2}C_{22} = \sum_{n=0}^N
     {\cal C}_{222n}
          e^{-i(\hat\omega_{22n}(a/M_f,M_f/m)(t - t_{\rm peak}) 
	    + \phi_{222n})}.
\end{equation}
We fit separately the real ${\rm Re}[{}_{-\!2}C_{22}]$ and imaginary ${\rm Im}[{}_{-\!2}C_{22}]$
parts of the $_{-\!2}C_{22}(t)$ ring-down signal without any
phase shifting of the numerical waveform.  Separate fits were
performed because simple non-linear least-squares fitting was used.
During each fit, numerical data for the waveforms
for times prior to $t_r+t_{\rm peak}$ are removed from the signal.
Starting with $N=0$, we fit the data for a sequence of values for
$t_r$ and choose as our final $t_r$ the value that produces the
smallest error estimate for $a/M_f$ and $M_f/M$.  We include
additional overtones ($N>0$) successively, using results from 
$N=0$ fits as seeds for the $N=1$ fits, and so forth.  
For each value of $N$, we
refit the entire function, so for $N=0$ there are 4 parameters in the
fit, for $N=1$ there are 6, for $N=2$ there are 8, and so forth 
(see
Tables~\ref{tab:C22Nd13}, \ref{tab:C22Nd16}, and \ref{tab:C22Nd19}
for and explicit list of the parameters being fit for each $N$). 
At
each set, we determine new values for $a/M_f$ and $M_f/M$ that are
used consistently for all of the modes.  Also, each time we include a
new mode in the fit, we also fit the data for a sequence of values for
$t_r$ and again choose our final $t_r$ for that set of modes by the
value that produces the smallest error estimates for $a/M_f$ and
$M_f/M$.

Tables~\ref{tab:C22Nd13}, \ref{tab:C22Nd16}, and \ref{tab:C22Nd19} in
Appendix~\ref{appendix_tables} display the fit parameters for the
$_{-\!2}C_{22}(t)$ waveform obtained from initial data with separation
parameter $d=13$, $d=16$ and $d=19$.  These tables show the fits for
only the high resolution ($\frac12h$) runs and give result to 3
significant figures.  In most cases, the errors in the fit suggest
that only two significant figures can be trusted, but we display the
additional digit in order to clearly illustrate the level of
consistency in the fits.  However, even if the accuracy of the
individual fits were higher, we must still take into account the
discretization error when estimating the value of parameters from the
fits.  Using Richardson techniques, we can for example estimate the value
and error of the angular momentum and mass of the BH at the
end of the ring-down phase by using fits to the medium ($\frac34h$)
and high ($\frac12h$) resolution runs.  Table~\ref{tab:C22RDdata}
shows the results of this analysis for the angular momentum ratio
$a/M_f$ and final mass ratio $M_f/m$ and includes results for the
$d=13$, $d=16$, and $d=19$ cases.  There is considerable consistency
in the value of the final mass ratio, with $M_f/M\approx0.95$ for all 
separations.  However, there is a
discernible decrease in $a/M_f$ as the separation increases.  In fact
each case differs by about $0.01$ in value from its neighboring
separation.  This variation in the final spin of the coalesced BH 
is, in fact, completely consistent with the change in the spins
of the initial {\em corotating} BHs.  If we compute the total
angular momentum contained in the spin of the individual BHs 
$S$, then we find respectively for the $d=(13,16,19)$ cases
$S/M_f^2 = (0.06,0.04,0.03)$.

\begin{table}
\caption{\label{tab:C22RDdata} Richardson extrapolated values for the
angular momentum ratio $a/M_f$ and final mass ratio $M_f/M$ from the
ring-down fits to ${\rm Re}[{}_{-\!2}C_{22}]$ for the $d=13$, $d=16$,
and $d=19$ cases.  For each separation we provide estimates when $N$
overtones of the $\ell=2,m=2$ QNMs are used.}
\begin{ruledtabular}
\begin{tabular}{c|cc|cc|cc}
 & \multicolumn{2}{c|}{$d=13$} & \multicolumn{2}{c|}{$d=16$} 
 & \multicolumn{2}{c}{$d=19$} \\
N & $a/M_f$ & $M_f/M$ & $a/M_f$ & $M_f/M$ & $a/M_f$ & $M_f/M$ \\
\hline
0 & $0.724\pm0.002$ & $0.948\pm0.006$ 
  & $0.72\pm0.01$ & $0.945\pm0.003$ 
  & $0.702\pm0.007$ & $0.946\pm0.002$ \\
1 & $0.723\pm0.002$ & $0.945\pm0.008$ 
  & $0.72\pm0.01$ & $0.942\pm0.001$ 
  & $0.706\pm0.004$ & $0.947\pm0.001$ \\
2 & $0.735\pm0.002$ & $0.955\pm0.006$ 
  & $0.725\pm0.007$ & $0.946\pm0.002$ 
  & $0.711\pm0.002$ & $0.949\pm0.002$ \\
3 & $0.732\pm0.006$ & $0.951\pm0.003$ 
  & $0.725\pm0.007$ & $0.946\pm0.002$ 
  & $0.709\pm0.003$ & $0.947\pm0.001$ \\
\end{tabular}
\end{ruledtabular}
\end{table}

\begin{figure}
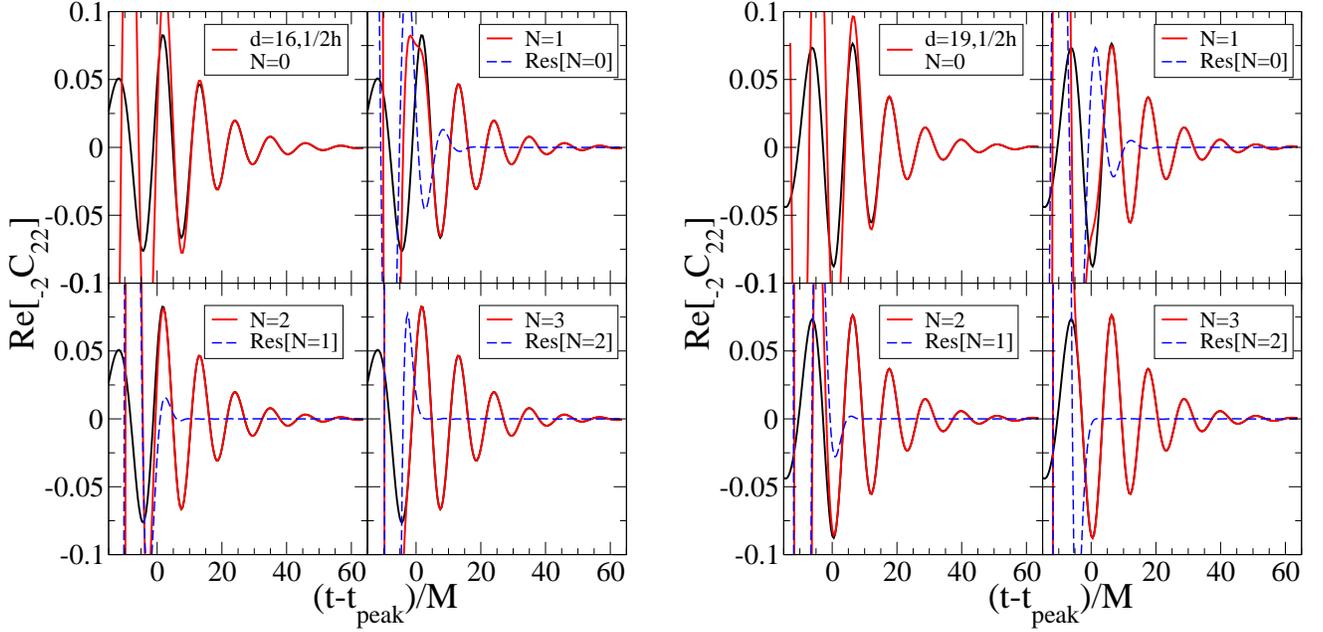

\begin{center}
\includegraphics[width=3.25in,clip]{RDNC22rd16}
\hspace{0.5cm}
\includegraphics[width=3.25in,clip]{RDNC22rd19}
\caption{\label{Fig:RDNC22rd16-19} Comparison of numerical and QNM
$_{-\!2}C_{22}$ ring-down waveforms for $d=16$ and $19$.  All plots
show the numerical ring-down waveform as a thin solid (black) line.
The thick solid (red) line displays the fit of the ring-down signal
using the first $N$ $\ell=2,m=2$ overtones beyond the fundamental.
For plots containing $N>0$ overtones, we also include the fit residual
from the previous value of $N$.  This is displayed as a dashed (blue)
line.  The coefficients for the displayed fits are found in
Tables~\ref{tab:C22Nd16} and \ref{tab:C22Nd19}.  }
\end{center}
\end{figure}

Figure~\ref{Fig:RDNC22rd16-19} shows the quality of the fit to ${\rm
Re}[{}_{-\!2}C_{22}(t)]$ for the cases $N=0,1,2,3$ and for the
separations $d=16$ and $19$.  We note that by including modes through
the $n=3$ overtone, we can fit the ring-down quite well to times {\em
preceding} the point where $|{}_{-\!2}C_{22}|$ reaches its peak.  For
each case beyond the fit to the fundamental $n=0$ mode, we include the
residual of the previous fit.  To be explicit, the residual displayed
for $N=1$ is defined as the difference between the numerical signal
and the fit obtained using the fundamental mode.  The residual
displayed for $N=2$ is the difference between the
numerical signal and the $n=0,1$ modes used in the $N=1$ fit.
This residual gives an estimate of the remaining signal that is being
fit.  However, it is important to remember that for each value of $N$,
the entire signal is actually being fit, including a redetermination
of $a/M_f$ and $M_f/M$ for all the modes.  The most important point to
notice from the residuals is that for each value of $N$ there is a
clear signal that is being fit.

A close examination of Tables~\ref{tab:C22Nd13}--\ref{tab:C22Nd19}
reveals a significant level of consistency to the fits.  For each
separation $d$, the spin and mass ratios remain very consistent and
the ${\cal C}_{22n}$ and $\phi_{22n}$ coefficients remain quite
consistent, as we increase the number of overtones included in the
fits.  This is true individually within the separate fits of the real
and imaginary parts of $_{-\!2}C_{22}$, and consistency is also seen
between the fits of the real and imaginary parts.  While the $\ell=2$,
$m=2$ QNMs seem to dominate the ring down signal in $_{-\!2}C_{22}$,
the $\ell=2$, $m=-2$ modes and the modes with $\ell>2$ should be
present.  However, the remaining residual after the $N=3$ fit (not
shown in any figure) has very low amplitude at times after the peak in
$|{}_{-\!2}C_{22}|$.  While there are some hints to structure, there
is insufficient signal and the simple approach we have used for
fitting does not yield consistent fits when additional modes are
included.

However, if we fix the values for $a/M_f$ and $M_f/M$ to the values
obtained from the $\ell=2$, $m=2$ fits, we can fit for the ${\cal
C}_{\ell\pm\!2n}$ and $\phi_{\ell\pm\!2n}$ coefficients for a range of
modes.  Doing so, we find that the fundamental QNM with $\ell=3$,
$m=-2$ has the most significant contribution, followed by the
$\ell=4$, $m=-2$ and $\ell=3$, $m=2$ fundamental modes at roughly
comparable levels.  Unlike the case of fitting only the $\ell=2$,
$m=2$ modes, adding in higher overtones when an increased spectrum of
modes was considered did not lead to consistent fits.  Part of the
difficulty in finding consistent fits to the subdominant modes is
likely due to the fact that the signal associated with these modes is
close to the level of numerical precision in the waveform.  However,
it is also likely that more sophisticated fitting methods are needed.
In particular, it would be useful to fit the real and imaginary parts
of the waveform simultaneously.  It may also be helpful to fit several
$_{-\!2}C_{\ell m}$ modes simultaneously.

While fitting multiple modes is problematic in some cases, it is
essential in others.  For the case of $_{-\!2}C_{32}(t)$, the
dominant QNMs include {\em both} $\ell=2$ and $\ell=3$, both with
$m=2$.  In fact, it was not possible to fit the ring-down signal of
$_{-\!2}C_{32}(t)$ without fitting simultaneously for these two
modes.  To be explicit, we take as our fitting function:
\begin{eqnarray}\label{eq:C32Nfitfunc}
_{-\!2}C_{32} &=& \sum_{n=0}^N\bigl\{
     {\cal C}_{322n}
          e^{-i(\hat\omega_{22n}(a/M_f,M_f/M)(t - t_{\rm peak}) 
	    + \phi_{322n})}
\\ && \mbox{} \hspace{0.25in}
     + {\cal C}_{332n}
          e^{-i(\hat\omega_{32n}(a/M_f,M_f/M)(t - t_{\rm peak}) 
	    + \phi_{332n})}\bigr\}. \nonumber
\end{eqnarray}
Fitting proceeds as with $_{-\!2}C_{22}$, starting with $N=0$ and
then adding successive overtones which allow us to fit to successively
earlier times in the ring down.

\begin{figure}
\begin{center}
\includegraphics[width=3.25in,clip]{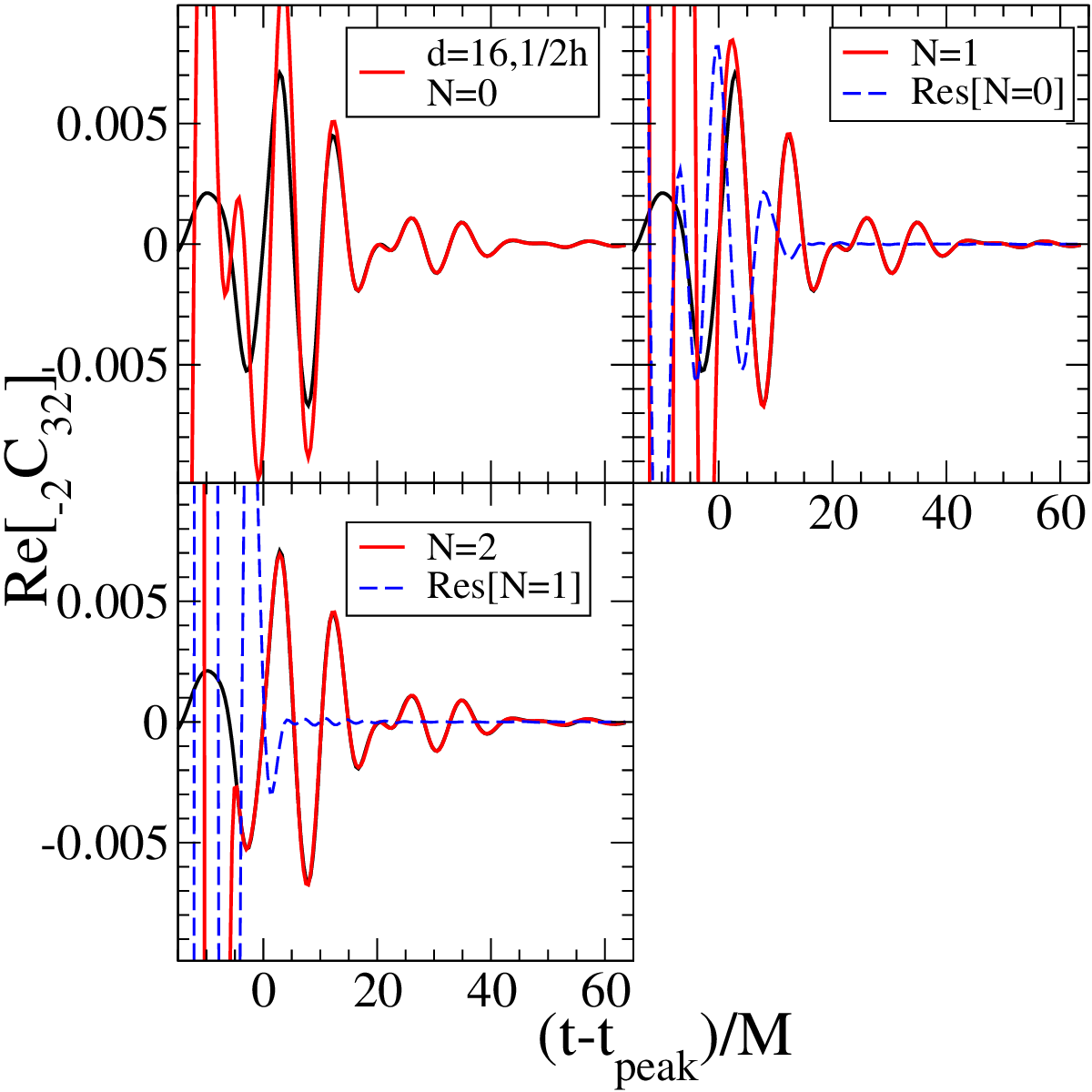}
\hspace{0.5cm}
\includegraphics[width=3.25in,clip]{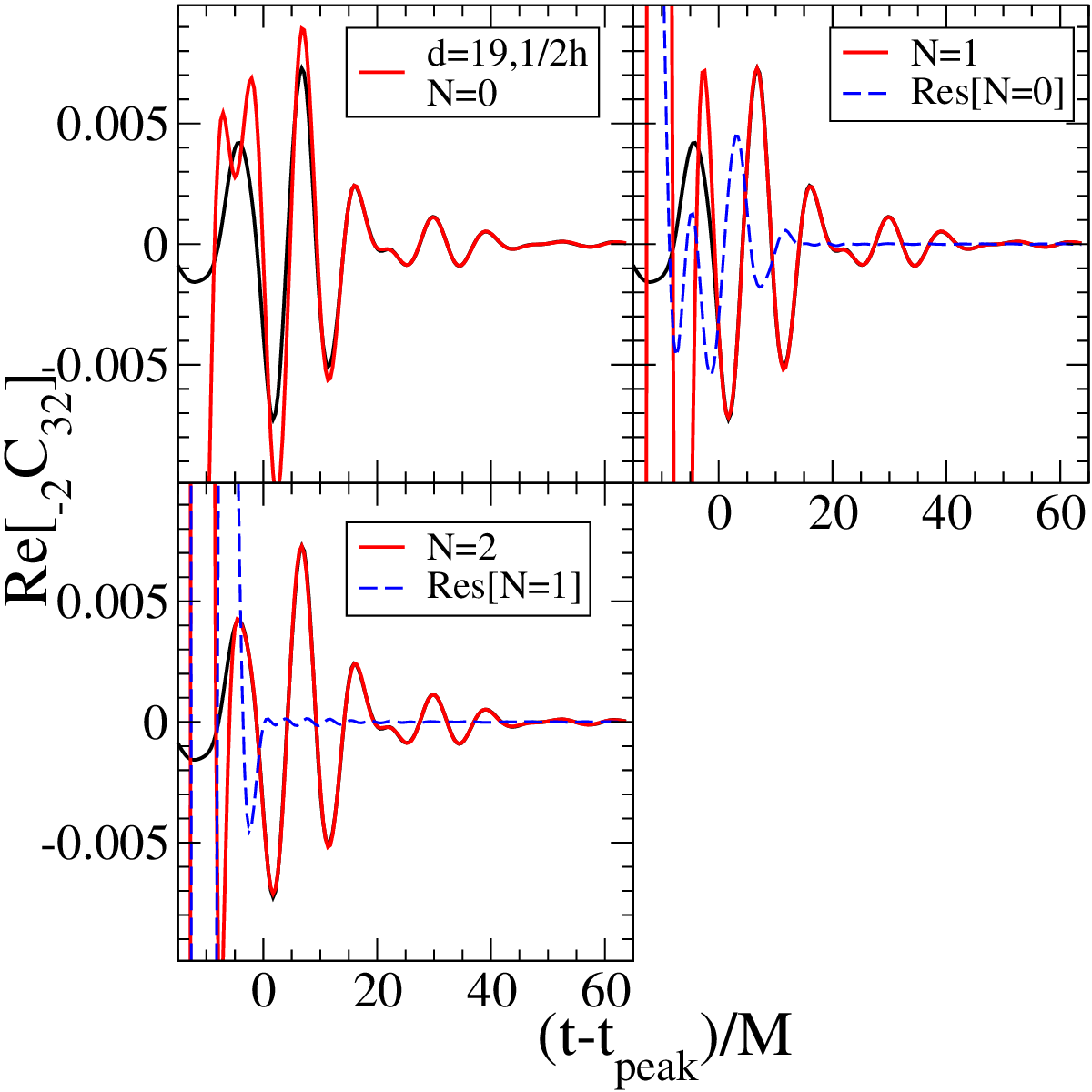}
\caption{\label{Fig:RDNC32rd16-19} Comparison of numerical and QNM
$_{-\!2}C_{32}$ ring-down waveforms for $d=16$ and $19$.  All plots
show the numerical ring-down waveform as a thin solid (black) line.
The thick solid (red) line displays the fit of the ring-down signal
using the first $N$ $\ell=2,m=2$ and $\ell=3,m=2$ overtones beyond the
fundamental.  For plots containing $N>0$ overtones, we also include
the fit residual from the previoius value of $N$.  This is displayed
as a dashed (blue) line.  The coefficients for the displayed fits are
found in Tables~\ref{tab:C32Nd16} and \ref{tab:C32Nd19}}
\end{center}
\end{figure}

Tables~\ref{tab:C32Nd13}, \ref{tab:C32Nd16}, and \ref{tab:C32Nd19} in
Appendix~\ref{appendix_tables} display the fit parameters for the
$_{-\!2}C_{32}(t)$ waveform obtained from initial data with separation
parameter $d=13$, $d=16$ and $d=19$.  Figure~\ref{Fig:RDNC32rd16-19}
shows the quality of the fit to ${\rm Re}[{}_{-\!2}C_{32}(t)]$ for the
cases $N=0,1,2$ and for the separations $d=16$ and $19$.  We note that
by including modes through the $n=2$ overtone, we can again fit the
ring down quite well to times {\em preceding} the point where
$|{}_{-\!2}C_{22}|$ reaches its peak.  As for ${\rm
Re}[{}_{-\!2}C_{22}(t)]$, we also include the residuals of the
previous fit.  We note that the level of consistency of the
fits, though significant, is not as high for $_{-\!2}C_{32}$ as seen
for $_{-\!2}C_{22}$.

During the ring-down phase, it is possible for a few percent of the
final mass $M_f$ and angular momentum $a M_f$ to be radiated away from
the system.  The Kerr QNM frequencies and decay constants are computed
assuming that the mass and angular momentum they carry away
constitute a negligible perturbation on the system.  This raises the
question as to whether or not the radiated energy and angular momentum
are affecting the QNM fits.  This issue will, of course, become more
significant as the fits are pushed to earlier times.  As we have seen
for the cases of $_{-\!2}C_{22}$ and $_{-\!2}C_{32}$, fitting to
earlier times in the ring down requires the use of higher overtones
($n>0$) with shorter decay times.  Because these higher overtones
dominate the waveform {\em only} at earlier times in the ring down,
we should expect some increase in the level of uncertainty in the
fits as we incorporate these overtones.

\begin{figure}
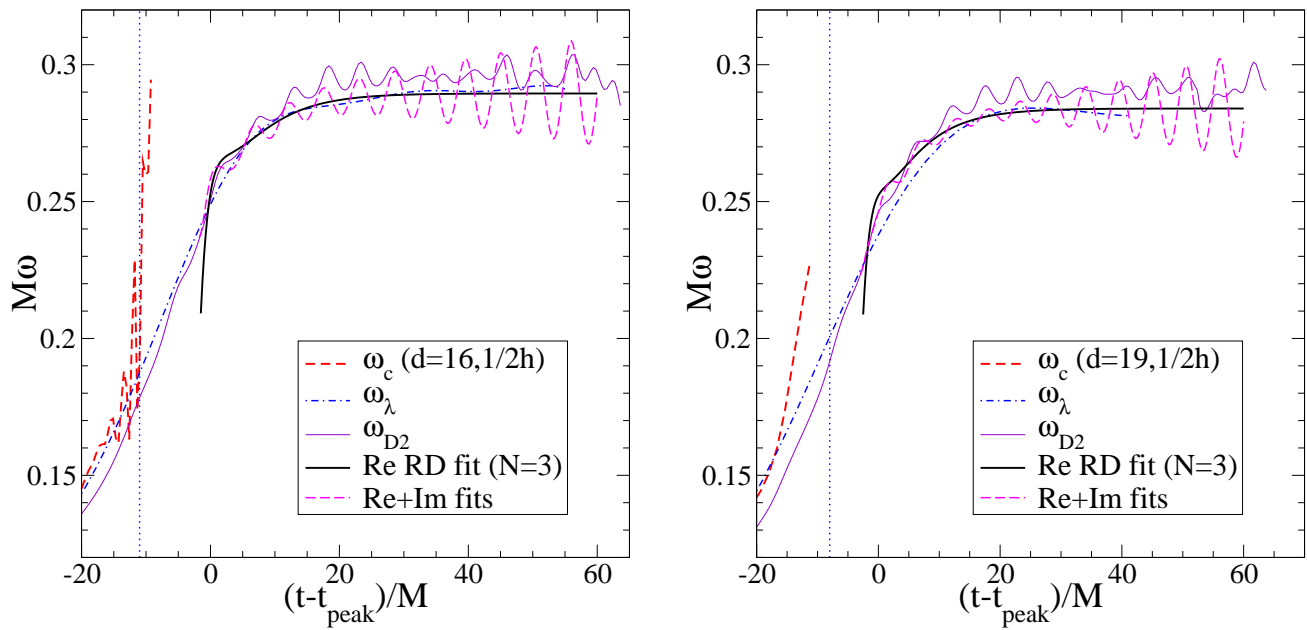

\begin{center}
\includegraphics[width=3.25in,clip]{mOmegad16D}
\hspace{0.5cm}
\includegraphics[width=3.25in,clip]{mOmegad19D}
\caption{\label{Fig:RDmOC22rd16-19} Dominant frequencies during the
ring down for the $d=16$ and $19$ cases evaluated using several
methods.  The short dashed (red) line is $\omega_c$.  The dot-dashed
(blue) line is $\omega_\lambda$.  And the thin solid (purple) line is
$\omega_{\rm D2}$.  These three lines were previously displayed in
Fig.~\ref{Fig:OmegaNQC22d16-19}.  The thick solid (black) line shows
the dominant frequency of Eq.~(\ref{eq:C22Nfitfunc}) as fit to the
real part of the $_{-\!2}C_{22}$ ring-down signal.  The long-dashed
(magenta) line shows the dominant frequency of a similar fit, but
consistently using the both the real and imaginary fits to
$_{-\!2}C_{22}$.  The vertical dotted (blue) line marks the 
approximate time that a common AH forms.}
\end{center}
\end{figure}

Finally we want to revisit the plots of the orbital angular frequency
displayed in Fig.~\ref{Fig:OmegaNQC22d16-19}.  The $\omega_\lambda$
and $\omega_{D2}$ frequencies continue beyond the inspiral phase and through the
ring down.  Beyond the inspiral phase, this frequency clearly cannot
be associated with the orbital angular frequency.  Rather they are
half of the dominant GW frequencies seen in $_{-\!2}C_{22}$.  In
Fig.~\ref{Fig:RDmOC22rd16-19}, we plot this dominant frequency
from a time about $10M$ before the formation of a common AH and
through the ring down.  In this range of times, $\omega_c$ clearly
decouples from $\omega_\lambda$ and $\omega_{\rm D2}$.  As the
dynamics transitions from the inspiral phase, the dominant
frequency rises very rapidly, finally reaching a plateau associated
with the dominant QNM ring-down frequency.
Both $\omega_\lambda$ and $\omega_{\rm D2}$ agree quite well
through both the transition and ring down, but we note that
$\omega_{D2}$ shows an unusual ``beating'' of the frequency 
during the ring down.

We also plot in Fig.~\ref{Fig:RDmOC22rd16-19} the dominant frequency
as measured by the fits to the ring down.  Using the ring-down fit
function in Eq.~(\ref{eq:C22Nfitfunc}) together with the fit value
given in Tables~\ref{tab:C22Nd13}--\ref{tab:C22Nd19} yields an
analytic expression for $_{-\!2}C_{22}$ through the ring-down phase.
Because we independently fit Eq.~(\ref{eq:C22Nfitfunc}) to the real
and imaginary parts of $_{-\!2}C_{22}(t)$ we have three different ways
that we can construct $_{-\!2}C_{22}$.  We can take the coefficients
for the fit from either ${\rm Re}[{}_{-\!2}C_{22}]$ or ${\rm
Im}[{}_{-\!2}C_{22}]$ and use that set of coefficients exclusively in
Eq.~(\ref{eq:C22Nfitfunc}).  Using the analytic representation of
$_{-\!2}C_{22}(t)$ we can compute the dominant frequency using
Eq.~(\ref{eq:omega_Dm}) with $m=2$.  A plot of this frequency using
the coefficient obtained form the fit of ${\rm Re}[{}_{-\!2}C_{22}]$
is shown if Fig.~\ref{Fig:RDmOC22rd16-19} with the label ``{\tt Re RD
fit (N=3)}.''  Notice that the frequencies agree well during the
ring-down phase and show a period following the peak in
$|{}_{-\!2}C_{22}|$ where the frequency increases before reaching its
plateau.  Also, we see no evidence of the beating seen in $\omega_{\rm
D2}$.

However, if we instead construct an analytic representation of
$_{-\!2}C_{22}$ using the coefficients from the fits to {\rm both}
${\rm Re}[{}_{-\!2}C_{22}]$ and ${\rm Im}[{}_{-\!2}C_{22}]$, we
recover the beating of the frequency.  To be clear, using
Eq.~(\ref{eq:omega_Dm}) to construct the dominant frequency
incorporates both the real and imaginary parts of $_{-\!2}C_{22}$.  If
we consistently use the coefficients from the fit to ${\rm
Re}[{}_{-\!2}C_{22}]$ when constructing an analytic representation for
${\rm Re}[{}_{-\!2}C_{22}]$ and use the coefficients from ${\rm
Im}[{}_{-\!2}C_{22}]$ for its representation, then we obtain the line
labeled ``{\tt Re+Im fits}'' in Fig.~\ref{Fig:RDmOC22rd16-19}.  The
plot of this frequency clearly shows a beating of the frequency and
this is due to a small mismatch between the real and imaginary fit
coefficients.  It seems clear that the beating we observe in
$\omega_{\rm D2}$ is caused by a similar effect.  Essentially, 
numerical error is leading to a non-physical mode that is
not circularly polarized, leading to the mismatch seen
between the real
and imaginary parts of $\Psi_4$.

\section{The (plunge and) merger}
\label{secmerger}

In Sec.~\ref{sec2.2} we discussed the possible presence of a rather blurred 
dynamical ISCO which marks the beginning of the plunge phase. The latter ends 
when the CAH forms. The plunge has a duration of $30\mbox{--} 50M$, corresponding to 
$1\mbox{--} 1.5$ GW cycles. The plunge cycle has a slightly different shape 
than the inspiralling cycles when viewed in ${\rm Re}[{}_{-\!2}C_{22}]$ (see Fig.~\ref{ReC221}), 
but it can barely be distinguished from the inspiralling cycles when 
viewed in $h_+$ and $h_\times$ (see Fig. \ref{hphc19}). Quite interestingly we notice that the 
onset of the plunge phase seems to happen soon after the ``knee'' in 
the frequency curve (see Fig.~\ref{Fig:OmegaNQC22d16-19}), and when the first 
change in the slope of the GW energy flux occurs (see Fig.~\ref{flux19}, especially 
the right panel). 
A second change of slope in the frequency and GW energy flux seems to happen roughly 
around the CAH, the third change occurs at the peak of the radiation.  
 
\begin{figure}
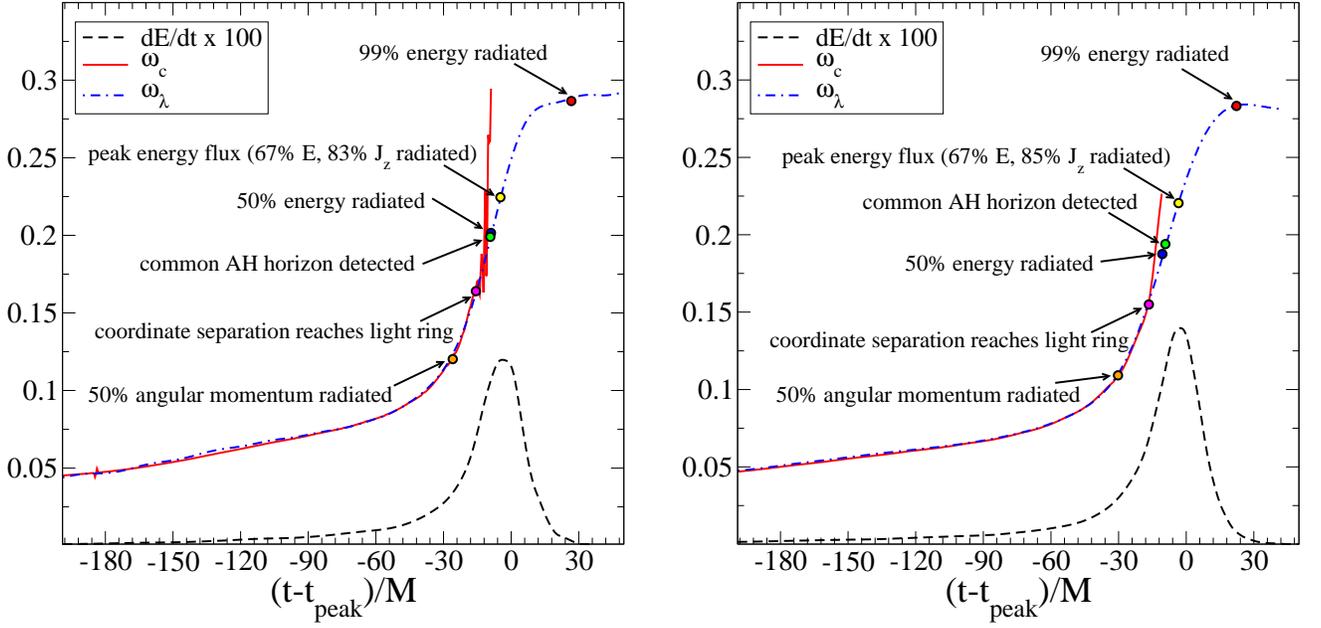

\begin{center}
\includegraphics[width=3.25in,clip]{d16_phases}
\hspace{0.5cm}
\includegraphics[width=3.25in,clip]{d19_phases}
\caption{Features of the merger phase. We plot the frequency evaluated from the orbit and the wave, 
and the GW energy flux. We mark with circles the time when the common AH of the final
BH first appears, when the binary separation reaches the light-ring 
of the final BH, the peak of the radiation flux (which occurs around $3-4M$ before the
peak in the amplitude of the waveform), when $50\%$ of the energy
and angular momentum have been radiated, and the time when $99\%$ of the
energy has been radiated ($99\%$ of the angular momentum appears to be
radiated around $5-10M$ before this, though due to the oscillations
in $dJ_z/dt$ we are much less certain exactly when this occurs---see Fig.\ref{jdot_edot_comp}). 
The left panel refers to the $d=16$ run and 
the right panel to the $d=19$ run. 
\label{merger16-19}}
\end{center}
\end{figure}

In Fig.~\ref{merger16-19} we illustrate some other interesting features of 
the inspiral to ring-down transition, i.e., the binary BH merger. We plot the frequencies 
$\omega_c$ and $\omega_\lambda$, and the GW energy flux (multiplied by 100).
Circles mark the position (time and frequency) at which the CAH forms and 
show when the coordinate separation between the BHs 
become less that the estimated co-rotating light ring of the final BH.
The latter are coordinate dependent quantities. The light ring is an unstable circular
null geodesic of the Kerr geometry in the equatorial plane of the BH,
and we estimate the position of the light ring by noting that for a Kerr BH 
with $a=0.70$, in Boyer-Lindquist coordinates the co-rotating light ring is 
a radial distance of $\approx 1.17$ times that of the outer horizon~\cite{BM}. For an
estimate of the light-ring location in the generalized harmonic coordinates of the
simulation we took the late-time coordinate radius of the final AH multiplied by $1.17$. 
Of course the different coordinates used to arrive at this value make it a rather
rough estimate, though given that the notion of a light
ring is not well-defined near coalescence it would not add much if we found
the exact location.
When the equal-mass binary reaches the CAH, {\it only} half of the total 
energy has been released. A little while before this point we observe that $\omega_\lambda$  
{\it decouples} from $\omega_c$. Note also that at the peak of the radiation, $23\%$ of 
the total energy has yet to be released. Here and in the following, the total 
energy refers to the energy radiated from the beginning of the simulation until the end. Thus, it doesn't include 
the energy radiated during the long inspiral preceding the initial time 
of the simulation.

The results obtained in Sec.~\ref{sec4}, 
in particular Figs.~\ref{Fig:RDmOC22rd16-19} and discussion around them, suggests that  
the GW emission soon after the peak of radiation is caused by the excitation of the QNMs of the final 
Kerr BH. The frequency of the least damped QNM $l=2,m=2, n=0$ is responsible for
the plateau, and  the higher overtones ($l=2,m=2, n> 0$), which have smaller frequencies 
and smaller decay times, should be responsible for {\it raising} the frequency from 
the peak of the radiation to the plateau. It is still not completely clear to us 
whether the higher overtones and/or other QNMs, e.g., $l=2, m \neq 2, n \geq 0$, are able to smoothly connect the 
decoupling frequency with the plateau. In fact, soon after the decoupling, there could be a very short 
{\it non-linear} phase, perhaps with strong mode mixing, that would preclude
a description in terms of QNMs.  To clarify those issues, it would be 
interesting to compute the binary BH metric around the decoupling point or the CAH, 
decompose it as a single Kerr metric plus perturbations, e.g., as done in the  
close limit approximation~\cite{PP94,AC,GNPP,Pullin99}, and determine more
precisely when 
the perturbative regime starts. In the next section, following a more phenomenological 
approach aimed at providing templates for GW detection, we will see how the 
ring-down phase could be matched to the inspiral phase in the EOB model, 
{\it assuming} that the QNMs are responsible for raising the frequency from the decoupling 
to the plateau. 

Finally, if we denote by {\it merger} the phase from roughly the decoupling point when 
$M\,\omega_{\rm dec} \sim 0.14 \mbox{--} 0.16$ to the peak of 
${\rm Re}[{}_{-\!2}C_{22}]$ when $M\,\omega_{\rm peak} \sim 0.2$ or the peak 
of radiation, the merger occurs in a very short time $\approx 10\mbox{--}15 M$, corresponding to 
$\approx 0.5\mbox{--}0.75$ GW cycles. During this phase the frequency increases by 
$\approx 45\%$, causing the GW spectrum to spread over a large frequency 
range (see Figs.~\ref{FFligo} and \ref{FFlisa}). We shall discuss how this 
will affect the detectability of GWs from equal-mass binaries in Sec.~\ref{sec5}. 

\section{Effective-one-body approach to inspiral--(plunge)--merger--ring-down}
\label{secEOB}

The Taylor-expanded Hamiltonian for a two-body system was computed at 3PN order in 
Refs.~\cite{DJSdr,JS}.  It took several years to compute the GW energy 
flux at 3.5PN order~\cite{35PNnospin}. Before the 3.5PN dynamics was completed  
the Taylor-expanded PN predictions for the GW energy flux and the phasing of {\it equal-mass} 
binaries were not accurate enough to obtain robust predictions of the GW signal during the {\it last stages} 
of inspiral and plunge. For example, through  2.5PN order, 
the PN-approximants of some of the crucial ingredients entering the 
GW signal, such as the GW energy flux, differ significantly when 
evaluated at subsequent PN orders in the typical frequency 
band of ground-based detectors~\cite{DIS98}. On the other hand, the signal-to-noise 
ratio of ground-based interferometers (especially the LIGOs) reaches its maximum 
around comparable-mass binaries. Thus, likely, 
the first detection may come from a coalescence 
of stellar--comparable-mass BHs merging in the most 
sensitive region of the detector's frequency band. 

\begin{figure}
\begin{center}
\includegraphics[width=3.25in,clip]{omegaHd16EOB}
\hspace{0.5cm}
\includegraphics[width=3.25in,clip]{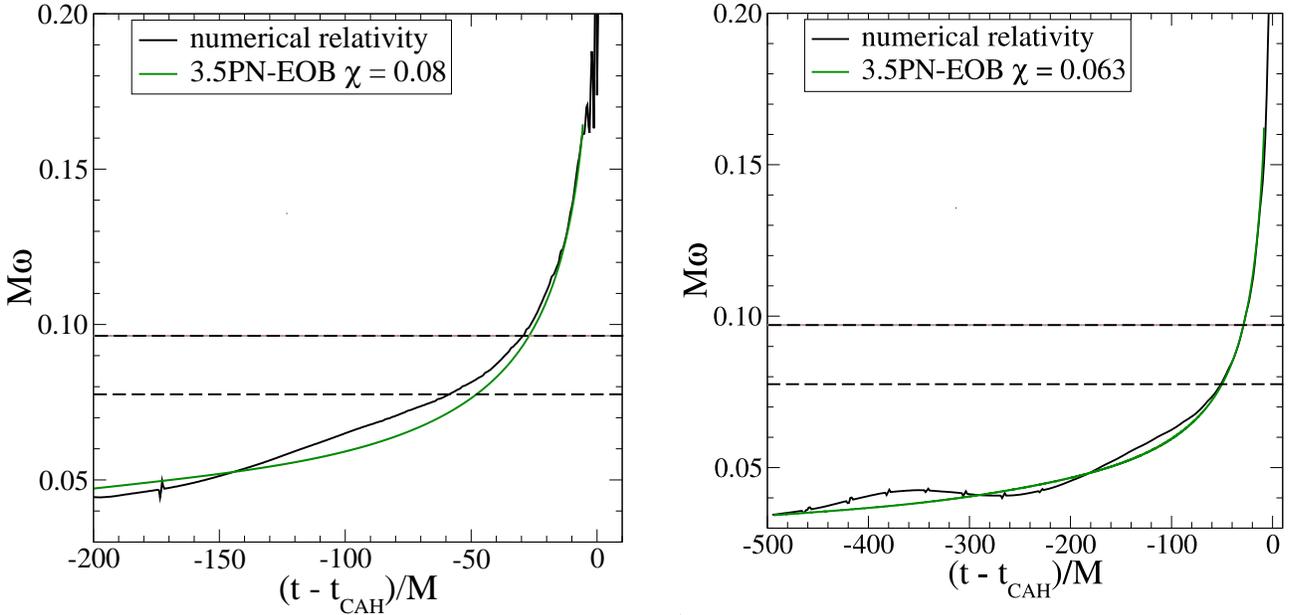}
\caption{We compare the NR orbital frequency and the EOB orbital frequency  
obtained using the nominal $\chi$ value 
from Table \ref{tab_idparams}. The left panel refers to the run $d=16$, the right panel to $d=19$. 
The horizontal light continuous line marks the ISCO frequency, as predicted by the conservative 
3PN-EOB Hamiltonian. The two  dashed lines span the region in which a dynamical ISCO might be present. 
\label{eobomega}}
\end{center}
\end{figure}

At the beginning of 2000, in absence of NR results and under the 
urgency of providing templates to search for comparable-mass BHs, 
some resummation techniques of the PN series were proposed. 
In Ref.~\cite{DIS98}, the authors proposed the Pad\'e 
resummation of the two-body energy and the GW energy flux, and 
in Refs.~\cite{BD1} a specific resummation of the PN Hamiltonian 
was proposed, the so-called effective-one-body (EOB) Hamiltonian. 
Later on, in Ref.~\cite{BD2} the last stages of inspiral and plunge were
modeled by combining the EOB Hamiltonian with the Pad\'e 
resummation of radiation-reaction effects, providing the GW signal 
which includes effects {\it beyond the adiabatic approximation}. 
The EOB Hamiltonian was then extended at 3PN order without spin effects 
in Ref.~\cite{DJS} and with spin effects in Ref.~\cite{TD}. More recently, 
the transition from inspiral to plunge including spin couplings has 
been modeled in Ref.~\cite{BCD}. These analytical 
studies predicted that (i) the two-body motion would be quasi-circular 
throughout the last stages of inspiral and plunge, until the light-ring 
(see Fig.~1 in Ref.~\cite{BD2}), (ii) the ISCO for an equal-mass binary 
is a rather blurred concept, 
taking place roughly during half of a GW cycle (see Fig.~12 in Ref.~\cite{BD2}) 
and that (iii) the adiabatic plunge lasts {\it only} for almost 
one GW cycle (see Fig.~12 in Ref.~\cite{BD2}). 

In Refs.~\cite{BD2,BCD}, the authors also provided an {\it example} of the full 
waveform by modeling the merger as a very short (instantaneous) phase 
and by matching the natural end of the EOB plunge (around the light-ring) 
with the ring-down phase (see Ref.~\cite{lazarus} where similar 
ideas subsequently developed also in NR). The matching was done using {\it only} the least 
damped QNM whose mass and spin were determined by the binary BH 
energy and angular momentum at the end of the EOB plunge. 
The choice of the light-ring at $\approx 3M$ for shifting the description 
between a (quasi-circular) binary motion and a deformed Kerr BH, 
was inspired by two considerations~\cite{BD2}. First, in the test-mass limit, 
$\nu \ll 1$, Refs.~\cite{Davis, Press} (see also Ref.~\cite{BM}) 
realized a long time ago that the basic physical reason underlying 
the presence of a universal merger signal was that 
when a test particle falls below $ 3 M$ (which 
is also the unstable light storage ring of Schwarzschild), the GW 
it generates is strongly filtered by the effective potential 
barrier centered around it. Secondly, for the equal-mass case 
$\nu = 1/4$, the close limit approximation~\cite{PP94,AC,GNPP,Pullin99,AHSSS}
suggests a matching between the two-body and the perturbed-black-hole 
descriptions when the distance modulus $\mu_0 \simeq 2$, which would correspond 
to a Schwarzschild-like radial distance $\simeq 2.6 M$.

For non-spinning, equal-mass binaries, the EOB approach predicts at 
3PN order~\cite{BCD}: $a_{\rm end}/M_{\rm end} \simeq 0.77$, $M_{\rm end} \simeq 0.974\,M$,  
an energy released of $ \simeq 1.3\% M$ until $t_{\rm end}$ and $ \approx 1\%$ 
from the least-damped QNM phase~\cite{BD3,BCD}. Here and henceforth we denote 
by the subscript ``${\rm end}$'' the time at the end of the EOB plunge. Depending 
on the PN order the ending time can occur slightly before the conservative light-ring. 
$t_{\rm end}$ is the time at which the quasi-circular assumption used in building the EOB 
equations of motion breaks down~\cite{BCD}.

\begin{figure}
\begin{center}
\includegraphics[width=3.25in,clip]{ReC22Hd16EOB}
\hspace{0.5cm}
\includegraphics[width=3.25in,clip]{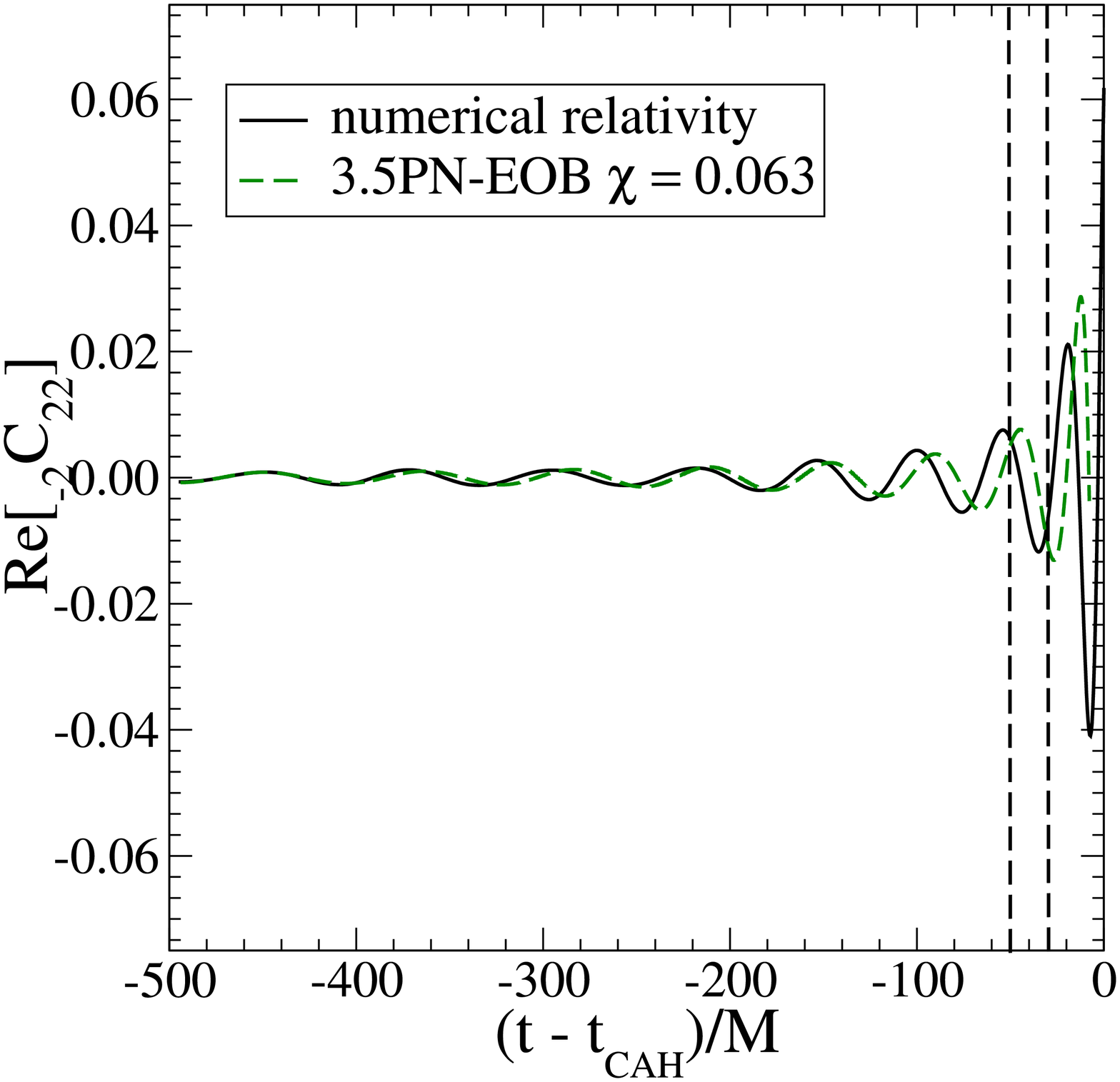}
\caption{We compare the NR ${\rm Re}[{_{-2}C_{22}}]$ and the EOB ${\rm Re}[{_{-2}C_{22}}]$ waveforms
obtained using the nominal $\chi$ value from Table \ref{tab_idparams}. 
The left panel refers to the run $d=16$, the right panel to $d=19$. The two vertical 
dashed lines span the region in which a dynamical ISCO might be present. 
\label{eobC22}}
\end{center}
\end{figure}
We show now some first-order comparisons between the numerical waveforms and the EOB waveforms, 
evaluated with the EOB conservative dynamics at 3PN order with spin-orbit and spin-spin 
effects included through 2PN order, and Pad\'e radiation-reaction 
energy flux at 3.5PN order~\cite{BCD}. 
In contrast to the PN adiabatic model discussed in Sec.~\ref{sec:adiabatic-PN}, 
which should be used with confidence only until the last stable orbit, the EOB model extends beyond it 
through the plunge. 

In Sec.~\ref{sec:adiabatic-PN} a comparison of the PN orbital frequency with the numerical results was obtained by 
{\em fitting} to the orbital frequency $\omega_c$. 
In the EOB model the orbital frequency is obtained by solving the EOB 
equations of motion, thus it is more complicated to implement a least-square 
fit. Here, we simply determine {\it by hand} which initial EOB orbital frequency best 
matches, on average, the numerical orbital frequency $\omega_c$ from the initial time to the 
light-ring or CAH (see Sec.~\ref{secmerger}), and compute the corresponding 
wave. In Fig.~\ref{eobomega} we show the results for the $d=16$ and $d=19$ runs 
assuming the nominal $\chi$ value from Table \ref{tab_idparams}. 
At the initial time $t=0$, we find $M\,\omega_0 = 0.047$, $E_0/M = 0.986$, $J_0/M^2 =0.896$ for $d=16$ and 
$M\,\omega_0 =0.034 $, $E_0/M = 0.988$, $J_0/M^2 = 0.945$ for $d=19$. 
The evolution ends at $r_{\rm end} \simeq 2.5\, M$, where $M\,\omega_{\rm end} \simeq 0.16$, 
$E_{\rm end}/M \simeq 0.971$, $J_{\rm end}/M^2 \simeq 0.741$ for the $d=16$ run, and 
at $r_{\rm end} \simeq 2.5\,M$ where $M\,\omega_{\rm end} \simeq 0.16 $, $E_{\rm end}/M 
\simeq 0.971$, $J_{\rm end}/M^2 \simeq 0.737$ for the $d=19$ run. We notice that the EOB 
conservative ISCO for $\chi = 0.063$ is at $M\,\omega_{\rm ISCO} 
\sim 0.096$~\cite{TD,BCD}, rather close to the frequency range 
$M\omega_{\rm dyn\,ISCO} = 0.078\mbox{--} 0.097$ of the dynamical 
ISCO discussed in Sec.~\ref{sec2.2}.  
In Fig.~\ref{eobC22} we compare the NR and EOB ${\rm Re}[{_{-2}C_{22}}]$ waveforms.  
The two vertical dashed lines in Fig.~\ref{eobC22} mark the region during 
which a dynamical ISCO may be present. 
We compute the EOB waveform using Eq.~(\ref{eq:circ_orb_22}), which is valid 
in the adiabatic circular-orbit case. We checked that by relaxing 
this assumption and computing ${\rm Re}[{_{-2}C_{22}}]$ by taking derivatives 
of the binary quadrupole moment, the wave does not change much, except at the very end.

By assuming the merger is a very short phase, 
the authors of Ref.~\cite{BD2} simply joined the GW signal at the end of the inspiral 
to the least-damped QNM. As said above this modeling was inspired by the idea 
that once beyond the light-ring (i.e., inside the potential barrier), 
the GW emission is quickly dominated by the excitation of the QNM of 
the newly-formed BH. The choice of matching only one QNM inevitably creates 
a sudden jump of the GW frequency at the matching point. However, 
a smoother transition can be obtained by including higher overtones. 
As discussed in Sec.~\ref{secmerger}, the analysis done in Sec.~\ref{sec4} 
would suggest that the QNM production starts a bit later, around the peak 
of the radiation. At this stage we do not know whether a linear superposition 
of QNM can be responsible for raising the frequency from around the light ring 
(see Fig.~\ref{merger16-19}) to the peak of radiation. In the spirit of an 
effective approach aimed at modeling the GW signal for detection,  
we push this idea further by including higher overtones when matching to 
the ring-down phase and discuss the consequences. The inclusion of higher 
overtones when matching to the ring-downn phase has also recently been
adopted in Ref.~\cite{DNT}, where the authors computed the transition  
inspiral--plunge--ring-down of a test particle in Schwarzschild. 
By including higher overtones, the authors could successfully match  
the {\rm exact numerical} rise of frequency from the light ring 
to the least damped QNM as obtained from the Zerilli equation. 

\begin{figure}
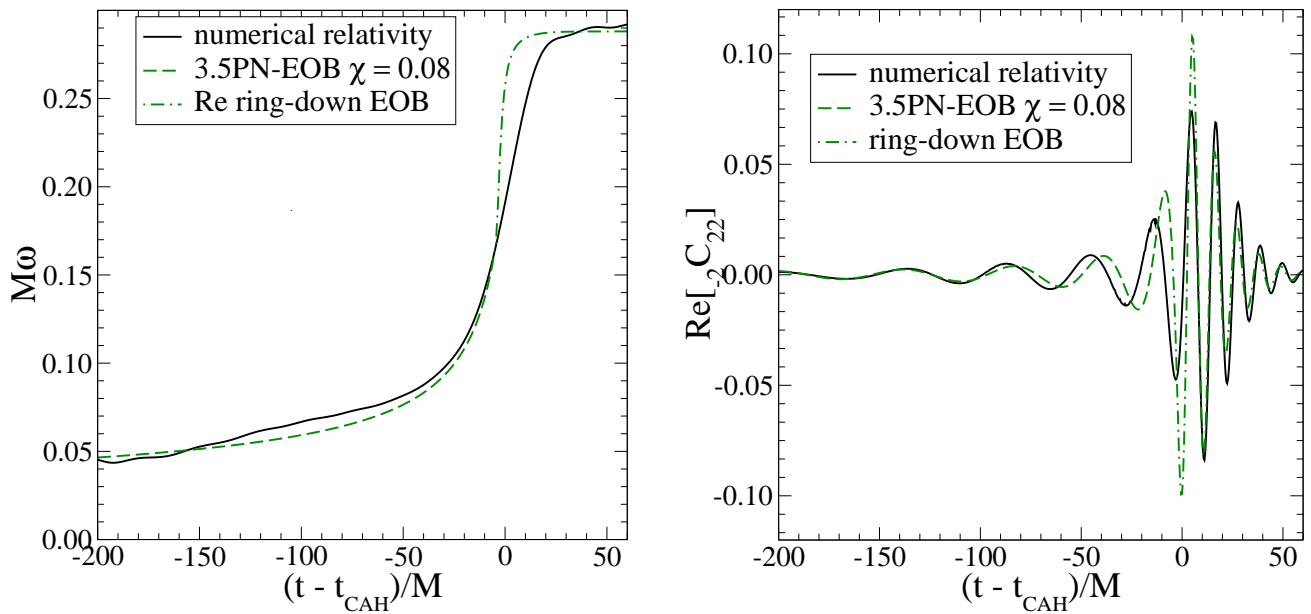

\begin{center}
\includegraphics[width=3.25in,clip]{omegaHd16EOBMatch}
\hspace{0.5cm}
\includegraphics[width=3.25in,clip]{ReC22Hd16EOBMatch}
\caption{We compare the NR and EOB frequency and  ${\rm Re}[{_{-2}C_{22}}]$ waveforms throughout the 
entire inspiral--merger--ring-down evolution. The data refers to the $d=16$ run. 
\label{Fulleob16}}
\end{center}
\end{figure}

First, we evaluate the BH mass and angular-momentum at the end of the 
EOB plunge, finding $M_{\rm end} \equiv E_{\rm end} \simeq 0.971\,M$ and 
$a_{\rm end}/M_{\rm end} \equiv J_{\rm end}/E^2_{\rm end} \simeq 0.785$. Then, we notice that 
those values {\it are not} the final BH mass and angular momentum because 
the binary has yet to emit energy and angular-momentum 
from the light-ring or CAH  to the least-damped mode (see discussions 
around Fig.~\ref{merger16-19}).  
Future NR simulations will provide predictions for different mass ratios and 
spins. Here, guided by the results of Sec.~\ref{secmerger} and Fig.~\ref{merger16-19},   
we assume that $(M_{\rm end} - M_{f})/M_{\rm end}=1.5\% $ and 
$(a_{\rm end}-a_{f})/a_{\rm end} = 6\%$. Thus, we obtain $M_{f} = 0.956\,M$ 
and $a_{f}/M_f = 0.738$. Using Ref.~\cite{BCW}, 
we determine the frequency and the decay time of the fundamental 
mode and the first two overtones, 
finding $M\,\omega_{220}= 0.576$, $M\,\omega_{221} = 0.565$, $M\,\omega_{222} = 0.545$, 
$\tau_{220}/M = 0.0828$, $\tau_{221}/M =0.250 $ and $\tau_{222}/M =0.422$. 
We then determine the three unknown amplitudes and three unknown 
phases of the three QNMs by imposing the continuity of 
the frequency $\omega_{\rm D2}$, Eq.~(\ref{eq:omega_Dm}), 
the wave and its first five derivatives at $t=t_{\rm end}$. 
We note that this matching procedure is rather sensitive to the time of 
matching because the frequency is increasing very quickly around $t_{\rm end}$. 
In Fig.~\ref{Fulleob16} we compare the frequency and 
the inspiral--(plunge)--merger--ring-down wave ${\rm Re}[{_{-2}C_{22}}]$ 
of the EOB model with the NR results for the case $d=16$. 
The EOB ring-down frequency is computed from Eq.~(\ref{eq:omega_Dm}) where 
we used in ${\rm Im}[{_{-2}C_{22}}]$ the same three amplitudes and three phases 
of  ${\rm Re}[{_{-2}C_{22}}]$. Were we to use in ${\rm Im}[{_{-2}C_{22}}]$ 
the three amplitudes and phases obtained by matching it at $t_{\rm end}$, 
we would not obtain a good result. This is due to numerical errors introduced by  
matching separately ${\rm Re}[{_{-2}C_{22}}]$ and ${\rm Im}[{_{-2}C_{22}}]$ 
at $t_{\rm end}$.  

As seen in Fig.~\ref{Fulleob16}, by matching the fundamental QNM and 
the first two overtones, the frequency transition becomes smoother,  
but nevertheless it differs from the NR frequency $\omega_\lambda$. 
As we shall see in the next section, this effective way of including a short in time, 
but spread in frequency, merger phase, can mimic the frequency 
spread of the power spectrum of the NR waves, though with a slightly different power law.

\section{Detectability of the signal}
\label{sec5}

In this section we compute the Fourier transform of the numerical waveforms, 
compare the results with the analytical predictions and give an estimate of 
the optimal signal-to-noise ratio (SNR) for ground-based and space-based detectors. 

Frequency domain PN templates for the inspiral phase are generally computed in 
the so-called Stationary Phase Approximation (SPA). They read
\beq
h_{\rm SPA}(f) = {\cal A}\,f^{-7/6}\,e^{i \Psi_{\rm SPA} (f)}\,, 
\quad \quad {\cal A} = \frac{1}{\sqrt{30}\pi^{2/3}} \frac{{\cal M}_c^{5/6}}{D_{\rm L}}\, 
\label{spa}
\eeq
where $f$ is the frequency of the GWs, ${\cal M}_c = \nu^{3/5} M$ is the chirp mass 
and $D_{\rm L}$ is the luminosity distance to the source. In Eq.~(\ref{spa}) 
we have adopted the  standard ``restricted PN approximation'', in which 
the amplitude is expressed to the leading order in a 
PN expansion while the phasing $\Psi_{\rm SPA} (f)$, is expressed to the highest
PN order available. The phase is currently known 
through 3.5PN order. Here, we are only interested in computing the amplitude of the
Fourier transform of the signal and investigating how and when it starts 
deviating from the Newtonian prediction $f^{-7/6}$. We shall analyze the comparisons 
between the Fourier transform phases in the future.

\begin{figure}
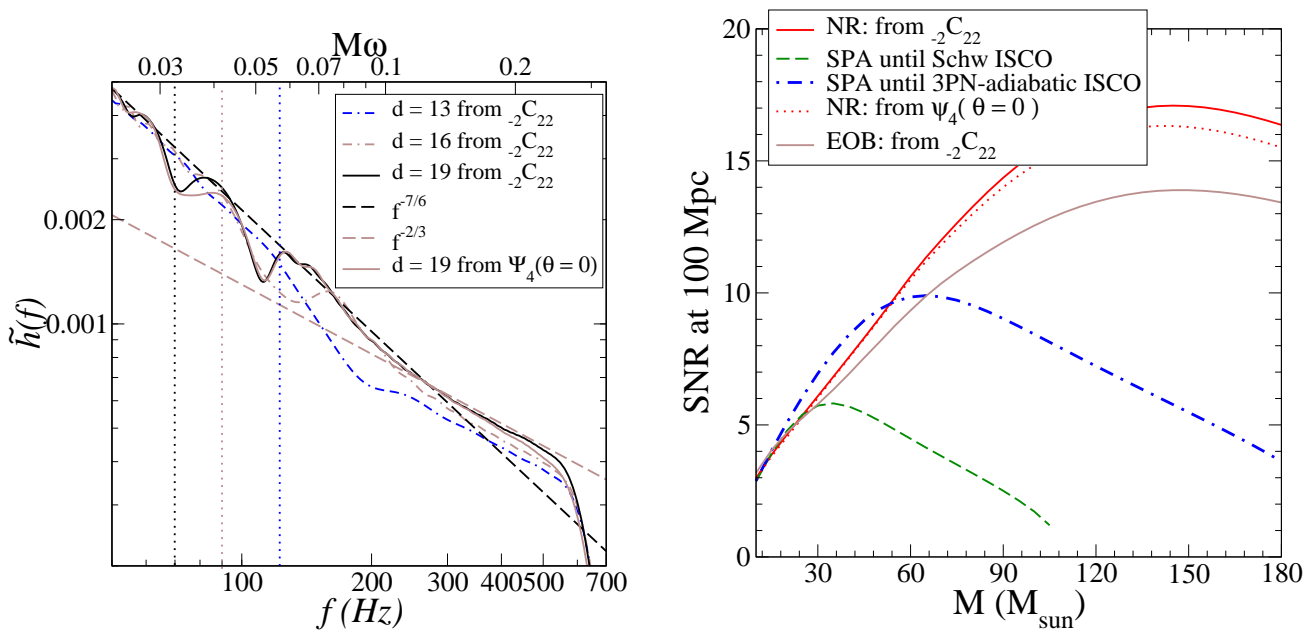

\begin{center}
\includegraphics[width=3.25in,clip]{FFligo}
\hspace{0.5cm}
\includegraphics[width=3.25in,clip]{SNRligo}
\caption{In the left panel we plot the amplitude of the Fourier transform of the numerical waveform 
for a binary with (redshift) mass $(15+15) M_\odot$. 
The lower horizontal axis marks the {\em gravitational wave} frequency in Hz, 
while the upper axis marks the dimensionless {\em orbital angular velocity}
that coincides with the instantaneous GW frequency.
The vertical dotted lines mark the frequencies 
at which the runs start. In the right panel, 
we show the average SNR for one detector versus the (redshift) total mass for an equal-mass binary 
at 100 Mpc. \label{FFligo}}
\end{center}
\end{figure}

In the left panel of Fig.~\ref{FFligo} we plot the Fourier transform amplitudes 
of the numerical waveform for the three runs, 
for a $(15+15) M_\odot$ binary which is a typical source for LIGO/VIRGO/GEO/TAMA.  
Using Table I we find that 
for this binary mass the initial GW frequency in the three runs is $121$ Hz, $89$ Hz 
and $71$ Hz (vertical dot lines in the left panel of Fig.~\ref{FFligo}) . 
To compute the Fourier transform we extrapolate the numerical waveforms 
at earlier time, for almost $6 \times 10^4$ m, by {\it attaching} to it the 3PN-adiabatic model which 
best-fits it. We compute the Fourier transform in three different ways, from ${\rm Re}[\Psi_4]$ and 
$h_+$ extracted along the direction perpendicular to the orbital plane, and from ${_{-2}C_{22}}$. 
Besides a normalization factor, the amplitudes computed from ${\rm Re}[\Psi_4]$ and 
${_{-2}C_{22}}$ must agree in the inspiral phase, where they satisfy more and more 
the restricted PN approximation, but they can differ in the last part of inspiral,  
merger and ring-down, when non-linear effects  and higher harmonics can become 
important. 

From Fig.~\ref{FFligo}, we see that at low-frequency, during the last 
stages of inspiral, the amplitude can be approximated by the Newtonian amplitude $f^{-7/6}$, 
with small bumps maybe due to the presence of eccentricity.   
At higher frequency, during the merger and ring-down the slope changes to $f^{-n}$.
At this stage we cannot uniquely determine the 
slope index or say if there is more than one change in the slope. 
We estimate $n \approx 0.6\mbox{--} 0.8$~\footnote{Similar results were also 
obtained independently in Ref.~\cite{Bakeretal}} and notice that the change in the 
slope can occur as early as the beginning of the plunge $M\omega \sim 0.1$ 
or as late as the decoupling time $M\,\omega \sim 0.16$.  
As an example in Fig.~\ref{FFligo} we show the case $n=2/3$.
Finally, the signal drops at higher frequencies, around the frequency of the 
fundamental QNM, i.e., $ 620$ Hz for the $(15+15) M_\odot$. 
Even if the merger occurs in a short time, corresponding to $\approx 0.5 \mbox{--} 0.75$ 
GW cycle, the frequency increases very quickly during this phase and then approaches  
the frequency of the fundamental QNM. As a consequence, the Fourier transform signal 
spreads over a large frequency band $300\mbox{--}600$ Hz. 
We also computed the Fourier transform amplitude from $\Psi_4$ extracted along 
directions $\theta \neq 0$, and found differences in the merger--ring-down amplitude 
slope. We shall discuss those interesting features in a future publication. 

In the right panel of Fig.~\ref{FFligo}, we show the sky averaged SNRs versus total mass, for an equal 
mass binary at 100 Mpc, and for initial LIGO. The dashed-dot and dashed curves 
are computed assuming the SPA inspiral signal (\ref{spa}) until the 
Schwarzschild ISCO $f_{\rm ISCO} = 4400/(M/M_\odot)$ and the ICO predicted 
by adiabatic PN theory at 3PN order~\cite{ICO}, respectively. 
The average SNR for one detector for an SPA signal is computed using
\beq
\sqrt{<{\rm SNR}^2>} = \frac{1}{D_L}\,\frac{1}{\pi^{2/3}}\,\sqrt{\frac{2}{15}}\,{\cal M}_c^{5/6}\,
\left [ \int_{f_{\rm low}}^{f_{\rm high}} 
\frac{f^{-7/3}}{S_n(f)} \right ]^{1/2}\,.
\eeq
The continuous light curve in Fig.~\ref{FFligo} is computed from 
the EOB inspiral--(plunge)--merger--ring-down wave shown 
in Fig.~\ref{Fulleob16}. 
The change in the slope that we observe in the left panel of Fig.~\ref{FFligo}, 
together with the inclusion of the signal beyond the ISCO, causes 
an increase of the SNR for large masses~\cite{BD2,TD,BCD,FH}. 
For total masses lower than $(15+15) M_\odot$, the end of 
the inspiral occurs around the most sensitive LIGO frequency, while the merger--ring-down  
is pushed to higher frequency where the sensitivity is much lower. As a consequence, 
the average SNR which includes merger and ring-down phases does not differ much from the one 
that includes only the inspiral phase. Astrophysical observations and theoretical 
predictions suggest that stellar mass BHs have a total mass ranging between 
$6 \mbox{--} 30 M_\odot$. If binary BHs of larger total mass exist,
they could be detected by initial LIGO with very high SNR.  The EOB 
model with the instantaneous matching to three QNMs predicts a SNR close 
to the NR result. It is smaller because the effective matching of the 
inspiral to the three QNMs (see Fig.~\ref{Fulleob16}) gives a Fourier transform amplitude 
which extends until the QNM frequency, but with a slope index slightly larger than 
$\approx -0.6\mbox{--}0.8$.  

\begin{figure}
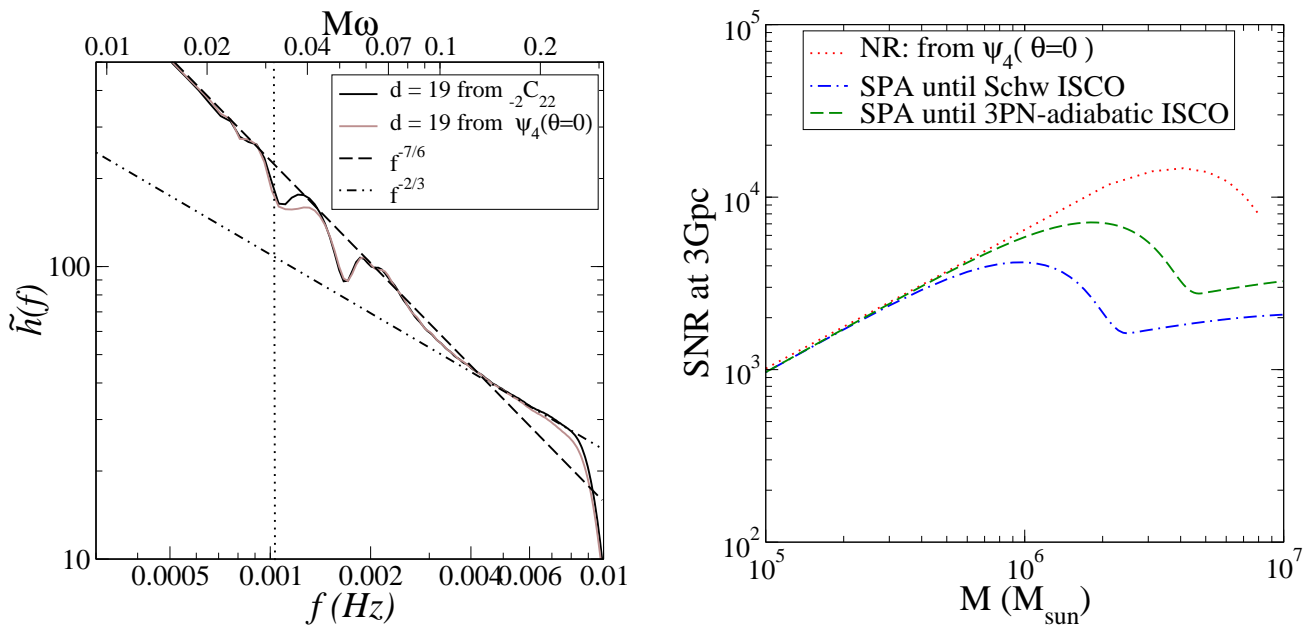

\begin{center}
\includegraphics[width=3.25in,clip]{FFlisa}
\hspace{0.5cm}
\includegraphics[width=3.25in,clip]{SNRlisa}
\caption{In the left panel we plot the amplitude of the Fourier transform of the numerical waveform 
for a binary with (redshift) mass $(10^6+10^6) M_\odot$. 
The lower horizontal axis marks the {\em gravitational wave} frequency in Hz, 
while the upper axis marks the dimensionless {\em orbital angular velocity}
that coincides with the instantaneous GW frequency.
The vertical dotted line marks the frequency at which the $d=19$ run starts. In the right panel 
we show the average SNR versus the (redshift) total mass for an equal-mass binary 
at 3 Gpc ($z = 0.54$). \label{FFlisa}}
\end{center}
\end{figure}

In the left panel of Fig.~\ref{FFlisa} we plot the amplitude of the Fourier transform 
for a signal typical of LISA, a $(10^6 + 10^6)M_\odot$ supermassive BH binary.  
In the right panel we show the average SNR for one Michelson LISA configuration versus 
the total (redshift) mass for an equal-mass binary at 3 Gpc. The dip in the plot is due to 
the WD-WD confusion noise~\cite{BBW}. Due to the inclusion of merger and ring-down phases, 
the SNR increases considerably for total masses larger than $2 \times 10^6 M_\odot$. 
We notice that Fig.~\ref{FFlisa} is consistent with Fig. 7 of Ref.~\cite{BCW} 
where the authors computed the SNR due to the ring-down phase, assuming $\sim 3\%m$ 
of energy released during the merger.  

\section{Conclusions}
\label{sec:discussion}

In this paper we have analyzed the data of several numerical 
simulations of the inspiral--merger--ring-down of an equal-mass binary 
carrying a very small spin aligned with the orbital angular momentum. 

The combination of several effects including (i) limited resolution, 
(ii) relatively close initial configurations and (iii) 
lack of diagnostics to measure and compensate 
for possible coordinate artifacts, make it impossible to claim 
very high accuracy in the comparisons with analytical models 
and the analysis of the merger waveform. 
Nevertheless, the resolution studies performed in Sec.~\ref{sec2} 
suggest that we {\em are} in the convergent regime, and so
meaningful conclusions can be drawn from the data. Furthermore, 
the consistency with which several quantities have been 
measured by independent means suggest that adverse gauge effects are minor
and will not affect many of the conclusions reached. 
In particular, as Tables~\ref{tab_simnums} and \ref{tab:C22RDdata} show, 
the final mass and angular momentum extracted from the ring-down are very close to their values
measured through AH properties. Also,
the orbital frequency measured via AH motion is close to the one extracted from the GW, 
as Fig.~\ref{Fig:OmegaNQC22d16-19} shows. 
  
We found that one of the dominant forms of numerical error is a slow drift in the phase
of the waveform. With respect to detectability in a GW burst-search
this error does not seem to be very significant, as the cumulated
phase error up to the peak in the radiation can be factored out
by a constant phase shift and the subsequent coalescence/ring-down
waveform does not seem to be significantly affected by prior
phase error. However, for matching to PN models at earlier time and 
parameter estimation this error is significant, and directly 
translates into uncertainties in matching parameters. 

As Figs.~\ref{d13-16_wave} and \ref{d19_wave} show, the numerical evolution 
is characterized by a strong initial pulse of radiation. The timing of the pulse 
suggests that it is associated with the assumption of conformal flatness.
In fact it has very similar characteristics to the initial pulse
seen in scalar field collapse generated binaries~\cite{FP,FP3},
which also begin with a conformally flat spatial metric. 
Fortunately, the effects of this pulse of radiation seem to diminish rapidly.
The initial data does not put the binary on a {\it clean}  
circular inspiral path. As discussed in Sec.~\ref{sec2.2} (see Fig.~\ref{d19_coord_e}), 
the trajectory clearly oscillates 
about the desired trajectory. The effects of this oscillation can be interpreted as a small
eccentricity of the initial orbit. We estimated it to be $e \sim 0.02$ 
for the $d=19$ case.
While these oscillations could be due to an actual eccentricity in the initial data, they
could also be due to a lack of an appropriate initial radial
momentum.  The current simulations cannot determine which
effect, if either, is most significant. However, while there are 
oscillations in the inspiral trajectory, the resulting 
dynamics is still adiabatic (e.g., see Fig.~\ref{d19_vel}) and 
modeled well by circular orbits.

Concerning the inspiral phase, as described in Sec.~\ref{sec:Newt_Quadrupole}
it is remarkable how well the Newtonian quasi-circular approximation can match the numerical signal 
of an equal-mass binary. For this we mean that if the orbital phase is modeled well, the leading order
Newtonian term in the expansion of the waveform is able to model both the amplitude 
and phase of the GW quite accurately until close to the time of merger.
It is also striking how, despite possible coordinate artifacts,
a GW signal computed with the leading quadrupole formula using the {\em coordinate} motion
of the AH's matches the numerically
extracted wave to a reasonable degree (see Fig.~\ref{Fig:d19_c22_quad_comp}). 

As discussed in Sec.~\ref{sec:adiabatic-PN}, the PN adiabatic phase and 
frequency matches $\omega_c$ (frequency computed from the AH motion) well. 
In particular, if the analytical expressions for $\omega$ (\ref{omega}) and $\phi$ (\ref{phase}) 
are used, the 3PN-approximant best fits the data, whereas if the expression for $\dot\omega$ (\ref{omegadot}) is 
solved numerically, a 3.5PN-approximant also matchs the numerical data to a similar level. 
The accuracy of the PN-adiabatic model improves with increasing binary separation.  
We expect it deviates from non-adiabatic models when approaching the last stable 
orbit. Nevertheless, we found that if the PN-adiabatic model is extrapolated even 
to the formation of the CAH, it gives reasonable results. 
This is due to the fact that the numerical plunge cycle is very short 
and still quasi-circular for an equal-mass binary, as originally predicted in the 
EOB model~\cite{BD2}. In Sec.~\ref{secEOB} we compared the numerical results 
with the EOB model. Again we found that the 3.5PN-approximant best matches the 
data. The blurred ISCO phase of half of a GW cycle and subsequent 
plunge phase of almost one GW cycle predicted in the EOB model~\cite{BD2} 
seems to be consistent with the numerical results. 
By properly adjusting the mass and angular-momentum of the BH 
at the end of the EOB plunge, to account for the energy and angular-momentum 
released during the merger, and by including three QNMs, we could match 
the EOB inspiral-(plunge) wave to the ring-down wave (see Fig. \ref{Fulleob16}). 
This matching procedure is very sensitive to the time of matching and it 
is only partially effective (see the difference in the GW frequency in 
Fig.~\ref{Fulleob16}) since it does not capture the details of the merger, 
but could be improved in the future.  

As Fig.~\ref{Fig:OmegaNQC22d16-19} shows, $\omega_c$ decouples from the dominant 
GW frequency just before the point of formation of the common AH, 
around the light ring. We conjecture that this decoupling time marks 
the transition point between inspiral-(plunge) and merger. 
We found that the merger accounts for only a brief time 
($\sim 10\mbox{--}15M$) compared to the inspiral (arbitrarily long) or ring down
phases ($\sim 30 M$ as measured by adding $\tau_{220}$ to the time at
which higher-order modes and overtones become insignificant).
However, the dominant GW frequency rises very quickly and spans a
significant range of frequencies during the merger, as can be seen in 
the left panel of Fig.~\ref{FFligo}. 
With current data, the peak in $|{}_{-\!2}C_{22}|$ (which occurs a few $M$ after  
the peak of the radiation) seems to be a natural point to mark the 
transition between the merger and ring-down phases. 
It is possible that the higher-order ring-down
modes/overtones (with lower frequencies than $\omega_{220}$) are
excited by resonance with the dominant GW frequency as it rises
during the merger.  An open and important question remains
whether it would be possible to model the merger by using some 
kind of non-linear modification of the onset of each QNM.

Concerning the ring-down phase, we extracted the fundamental QNM and 
first few overtones. Results are shown in several Tables of 
Appendix \ref{appendix_tables} and Figs.~\ref{Fig:RDNC22rd16-19} and \ref{Fig:RDNC32rd16-19}. 
We fit to the individual $_{-\!2}C_{\ell m}$ modes instead of
to $\Psi_4$ directly because fitting to $_{-\!2}C_{\ell m}$ includes
information from all directions. 
The fit to each $_{-\!2}C_{\ell m}$ can, in principle, include
QNMs for all values of $\ell\ge|m|$ but only the negative frequency
modes with azimuthal index $m$ and positive frequency modes with
index $-m$. For $_{-\!2}C_{22}$, we found only the $\ell=2,m=2$ QNMs are
significant.  As seen in Fig.~\ref{Fig:RDmOC22rd16-19}, 
we could fit the ring-down signal to times slightly
before the peak in $|{}_{-\!2}C_{22}|$ by using $n=0\ldots3$.
For $_{-\!2}C_{32}$, we found that we must use both the
$\ell=2,m=2$ and $\ell=3,m=2$ QNMs, but other modes do not
contribute at a significant level.  Also in this case we can again fit the
ring-down signal to times slightly before the peak in
$|{}_{-\!2}C_{32}|$ by using $n=0\ldots2$.
More sophisticated ring-down fitting techniques will be
helpful in gaining a better understanding of the transition 
from merger to ring down.

Quite interestingly throughout the inspiral--merger-ring-down 
the balance equation $dE/dt = \omega\, dJ/dt$ holds quite well on 
average, as Fig.~\ref{jdot_edot_comp} 
in Appendix~\ref{appendix_comp} shows, with a {\it single} 
frequency always {\it dominating} the entire evolution 
(see Fig.~\ref{Fig:OmegaNQC22d16-19}). 

The analysis of the inspiral--merger--ring-down suggests that it should be 
possible to come up with good hybrid numerical/analytical 
waveforms, or even complete analytical waveforms where the
full numerics guides how we need to patch the
inspiral and ring-down waveforms together. To this end, it will be 
very important to devise a simple model of how the QNM's are excited
during the transition regime. Of course, all of this
will be moot if the relative simplicity of this merger
scenario breaks down for more interesting initial 
conditions ($m_1 \neq m_2$, $S_1 \neq S_2 \nsim 0$ and non-aligned with the 
orbital angular-momentum), so it will be important
to numerically evolve more varied classes of initial data.

Finally, during the relatively brief merger phase the dominant GW frequency 
rises quickly, generating a signal in the Fourier domain that is rather spread out in 
frequency. The left panels in Figs.~\ref{FFligo} and \ref{FFlisa} indicate 
a change of slope in the signal Fourier amplitude $\tilde{h}$. The slope during 
inspiral is $-7/6$. 
The slope during the transition merger--ring-down seems to 
be $\approx -0.6 \mbox{--} 0.8$. The inclusion of the merger--ring-down signal 
increases the SNR for large binary masses. If binary BHs 
of mass larger than $40 M_\odot$ exist, they could be detected by one 
single LIGO with SNR up to $\sim 15$ at 100 Mpc. LISA could detect supermassive 
BHs of masses $2 \times 10^6\mbox{--} 10^7$, with SNR up to $10^4$ at 1Gpc.

\acknowledgments 
We wish to thank Emanuele Berti, Yi Pan, Harald Pfeiffer, Bernard Whiting, 
Luis Lehner and Patrick Brady  for useful discussions. We are grateful 
to Emanuele Berti for providing us with some of the data described in Ref.~\cite{BCW}.
A.B.\ acknowledges support from NSF grant PHY-0603762 and from the 
Alfred Sloan Foundation. 
G.C.\ acknowledges support from the Z.\ Smith Reynolds Foundation.
F.P.\ acknowledges research support from the CIAR,
NSERC, and Alberta Ingenuity.
The simulations described here were performed on
the University of British Columbia's vnp4
cluster (supported by CFI and BCKDF), WestGrid machines
(supported by CFI, ASRI and BCKDF), and the Dell Lonestar cluster
at the University of Texas in Austin.
Some of the comparisons with PN and EOB models were obtained building on 
Mathematica codes developed in Refs.~\cite{BD1,BD2,BCV1,BCV2,BCD}. 

\appendix
\section{Additional multipole moments}
\label{appendix_mult}

\begin{figure}
\begin{center}
\includegraphics[height=3.2in,clip]{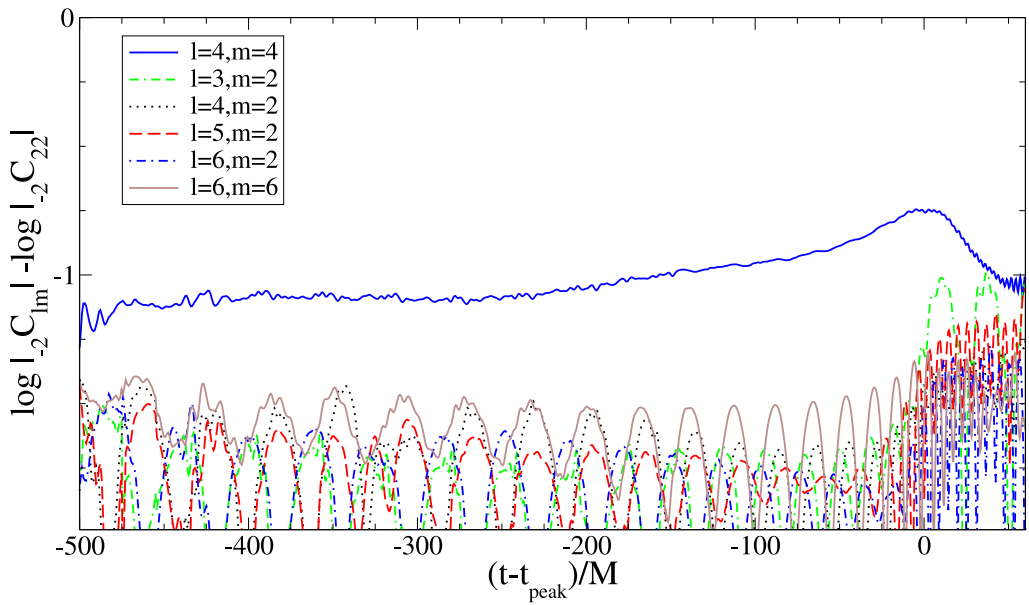}
\caption{\label{log_mag_d19_B} 
Magnitudes of
several sub-dominate components of the $d=19$ waveform
relative to the dominant $\ell=2,|m|=2$ component. 
Modes up to $\ell=6,|m|=6$ were extracted---those
not present in the plot were several times (at least)
smaller than any of the modes shown.
}
\end{center}
\end{figure}

For the majority of comparisons in this paper we focused on the dominant
$\ell=|m|=2$ spherical harmonic components of a waveform, as for quasi-circular,
equal mass binaries that are initially co-rotating
one does not expect other multipoles to
be present to any significant degree. In Fig.
\ref{log_mag_d19_B} we plot the magnitudes of several
higher multiple moments of the waveform relative to the $_{-2}C_{2,2}$
multipole moment for the $d=19$ $h/2$ run (the $d=13$ and $d=16$ have
a similar spectrum).  The next-to-leading order component of the
waveform is the $_{-2}C_{4,4}$ moment, which is about an order
of magnitude smaller than $_{-2}C_{2,2}$. At up 5 times
smaller than $_{-2}C_{4,4}$ are the $_{-2}C_{\ell\ge3,2}$, 
$_{-2}C_{6,4}$ and $_{-2}C_{6,6}$ modes.
Interestingly, during the ring-down part of the waveform the 
relative magnitude of the $_{-2}C_{\ell\ge3,2}$ modes 
grow and become almost as significant as the $_{-2}C_{4,4}$ mode.
We only calculated
components of $\Psi_4$ up to $\ell=|m|=6$---all other modes
not shown (including the axisymmetric $m=0$ modes) are at least
of factor of 3 times smaller than any of the modes in the plot.
Note however that the sub-dominant components of the waveform
are more susceptible to gauge effects and numerical truncation
error in the solution as their relative amplitudes are so
small; hence one should be wary of drawing any significant
conclusions from Fig.  \ref{log_mag_d19_B}. 

Even though the higher multipole moments are small, it will still
be interesting to compare them to the predictions of perturbative
calculations. Furthermore, for unequal mass binaries and/or binaries
with significant initial spins---in particular with spin components that are not aligned with
the orbital angular momentum---the higher multiple moments will
play a much more important role in the description of the waveforms. 
We leave it to future studies to analyze these modes in more detail.

\section{Energy and Angular Momentum Flux}
\label{appendix_comp}

\begin{figure}[t]
\begin{center}
\includegraphics[width=3.25in,clip]{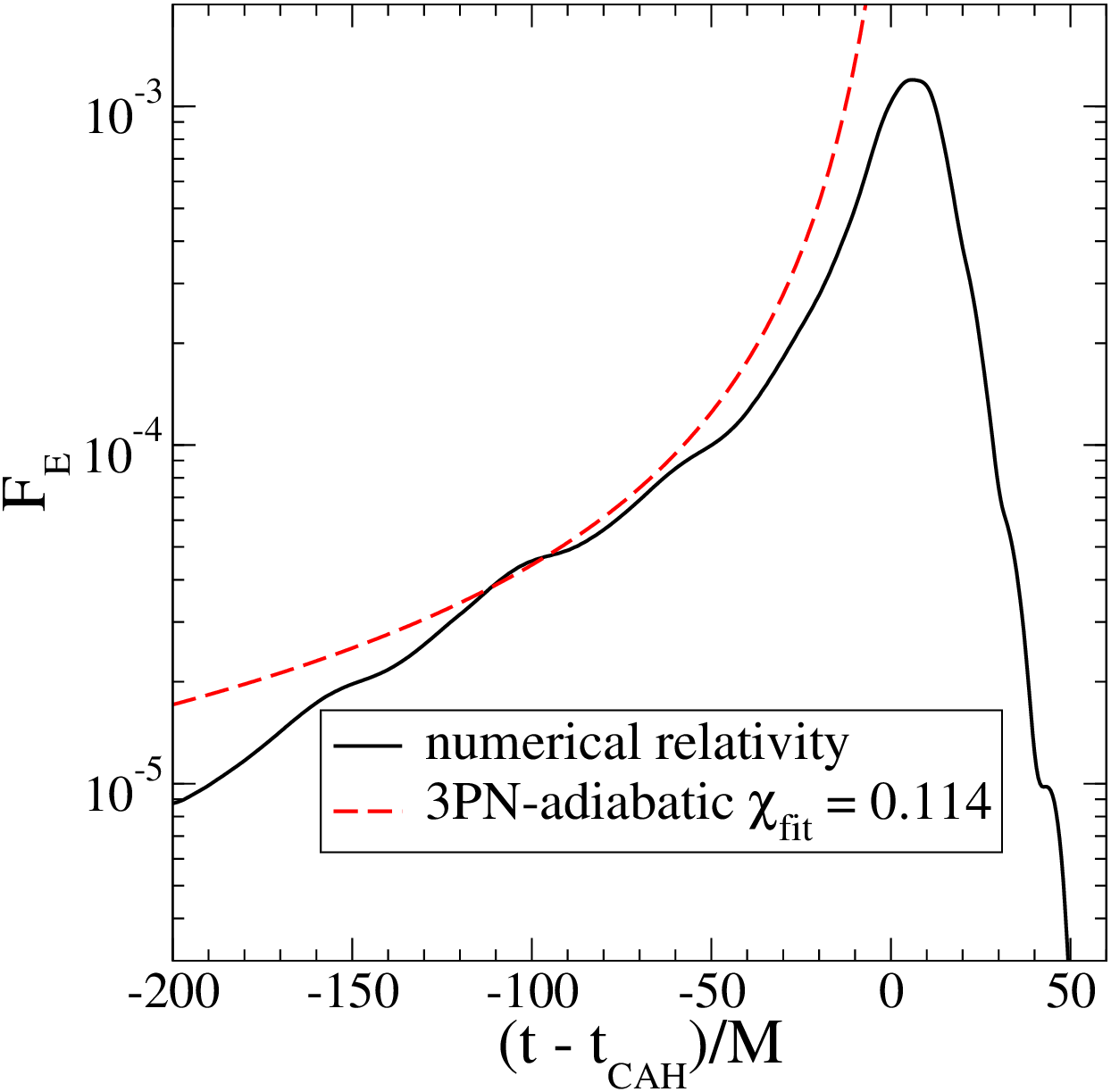}
\hspace{0.25cm}
\includegraphics[width=3.25in,clip]{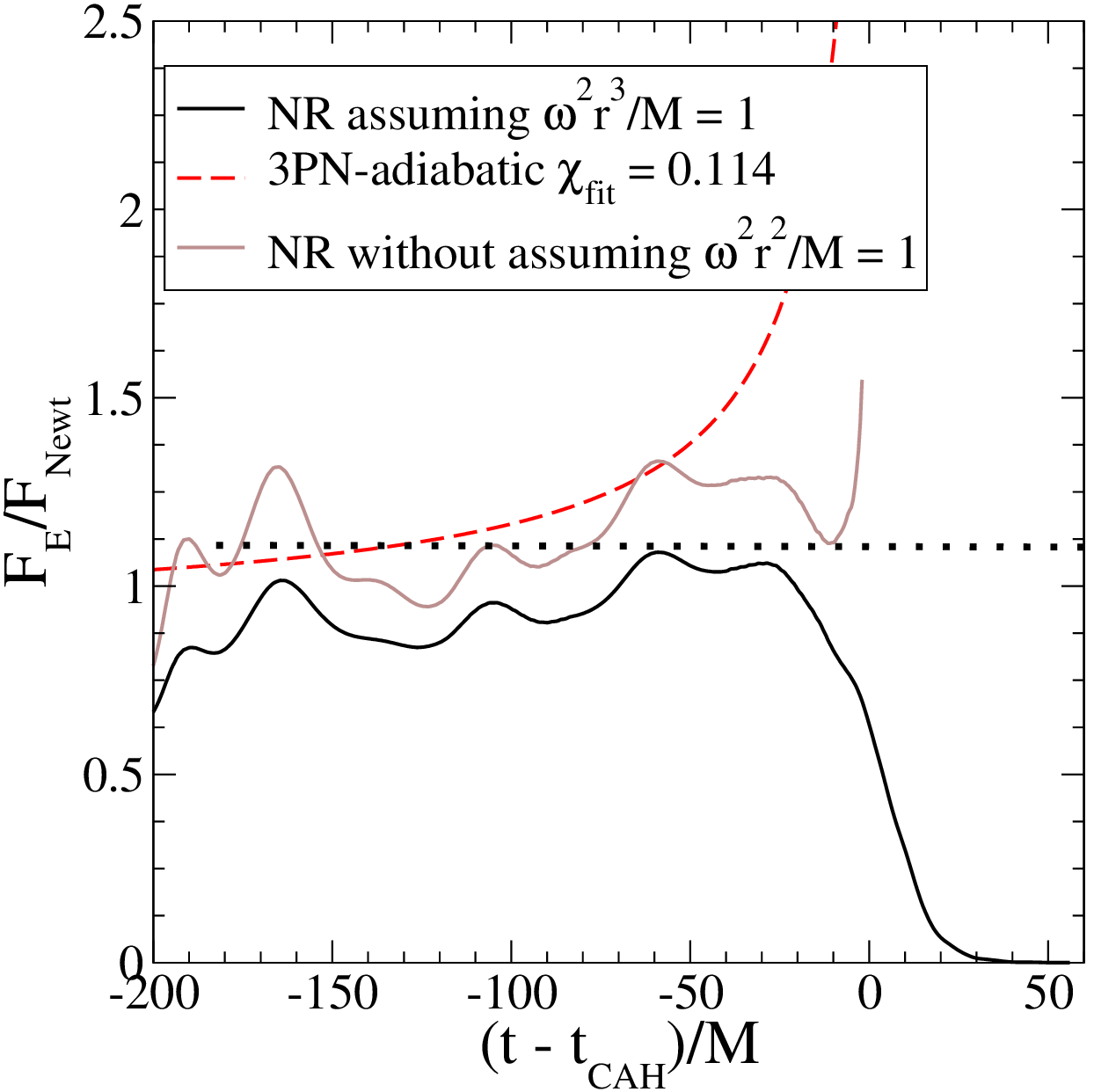}
\caption{In the left plot we compare the GW energy fluxes  
of the numerical simulation and PN-adiabatic model when the best-fit 
for the orbital frequency is used (see Sec.~\ref{sec:adiabatic-PN}). 
In the right plot, we show the same comparisons but we divide the fluxes 
by the Newtonian GW flux. For the numerical case we show the results 
assuming circular orbits, i.e., $F_{\rm Newt} = 32/5\,\nu^2(M\omega)^{10/3}$
and not assuming circular orbits, i.e., $F_{\rm Newt} = 32/5\,\nu^2(M\omega)^{6}\,(M/r)^{-4}$. 
The data refers to the $d=16$ run. \label{flux16}}
\end{center}
\end{figure}

The GW energy flux emitted by a binary moving along an adiabatic sequence of circular orbits 
is currently known through 3.5PN order~\cite{35PNnospin} for non-spinning BHs and through 2.5PN 
for spinning BHs~\cite{25PNspin}. It reads 
\begin{eqnarray}
\label{flux}
 F_{\rm E}  &=& {32\over 5}\nu^2 (M \omega)^{10/3} \biggl\{ 1 +
\left(-\frac{1247}{336}-\frac{35}{12}\nu \right) (M\,\omega)^{2/3} + 4\pi
(M\,\omega)\nonumber \\ &+&
\left(-\frac{44711}{9072}+\frac{9271}{504}\nu+\frac{65}{18}
\nu^2\right) (M\,\omega)^{4/3} +\left(-\frac{8191}{672}-\frac{583}{24}\nu\right)\pi
(M\,\omega)^{5/3}\nonumber \\
&+& \frac{(M\,\omega)}{G\,m^2}\left[-4S_\ell -\frac{5}{4}\frac{\delta
M}{M}\Sigma_\ell\right] \nonumber\\
&+& \frac{(M\,\omega)^{5/3}}{G\,m^2}\left[
\left(\frac{65}{14}+\frac{428}{63}\nu\right)S_\ell
+\left(\frac{51}{16}-\frac{67}{28}\nu\right)\frac{\delta M}{M}\Sigma_\ell\right]\nonumber \\
&+&\left(\frac{6643739519}{69854400}+\frac{16}{3}\pi^2-\frac{1712}{105}C
-\frac{856}{105}\ln (16~\!(M\,\omega)) \right.\nonumber\\
&+&\left.\left[-\frac{134543}{7776}+\frac{41}{48}\pi^2\right]\nu-\frac{94403}{3024}\nu^2
-\frac{775}{324}\nu^3\right) (M\,\omega)^2\nonumber\\ &+&
\left(-\frac{16285}{504}+\frac{214745}{1728}\nu
+\frac{193385}{3024}\nu^2\right)\pi (M\,\omega)^{7/3} \biggr\}\;.
\end{eqnarray}
The numerical energy flux is computed in terms of the mode amplitudes
$_{-\!2}C_{\ell m}(t)$ as
\begin{equation}
   \frac{dE}{dt} = \frac{1}{16\pi}\sum_{\ell m}{|D_{\ell m}(t)|^2},
\end{equation}
where $D_{\ell m}(t)$ is a dimensionless first time integral of
$_{-\!2}C_{\ell m}(t)$ defined by
\begin{equation}
   D_{\ell m}(t) \equiv \frac{1}{M}\int_0^t{dt^\prime
        {}_{-\!2}C_{\ell m}(t^\prime)}.
\end{equation}

\begin{figure}[t]
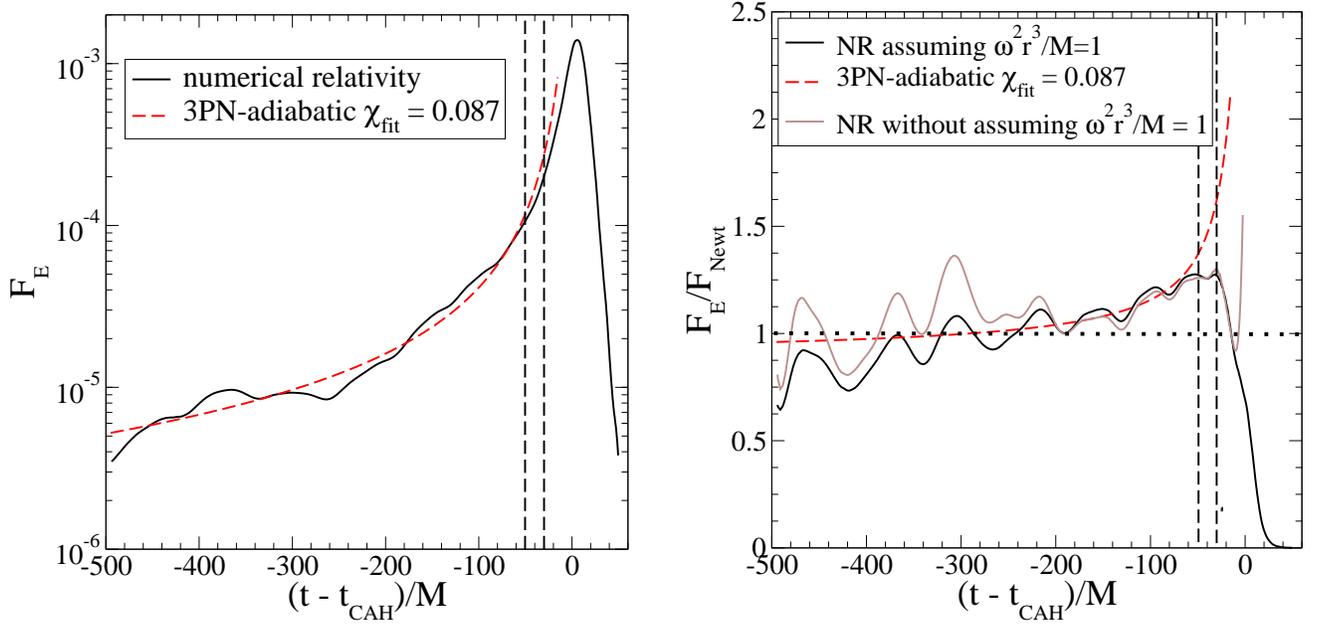

\begin{center}
\includegraphics[width=3.25in,clip]{FluxTPNd19}
\hspace{0.5cm}
\includegraphics[width=3.25in,clip]{FluxTnormPNd19}
\caption{In the left plot we compare the GW energy fluxes  
of the numerical simulation and PN-adiabatic model when the best-fit 
for the orbital frequency is used (see Sec.~\ref{sec:adiabatic-PN}). 
In the right plot, we show the same comparisons but we divide the fluxes 
by the Newtonian GW flux. For the numerical case we show the results 
assuming circular orbits, i.e., $F_{\rm Newt} = 32/5\,\nu^2(M\omega)^{10/3}$
and not assuming circular orbits, i.e., $F_{\rm Newt} = 32/5\,\nu^2(M\omega)^{6}\,(M/r)^{-4}$. 
The two vertical lines span the region in which a dynamical ISCO 
could exist. The data refers to the $d=19$ run. \label{flux19}}
\end{center}
\end{figure}

In the left panels of Figs.~\ref{flux16}, \ref{flux19}, we compare the PN-adiabatic and numerical fluxes for 
the runs $d=16$ and $d=19$. We compute the PN-adiabatic flux from Eq.~(\ref{flux}), 
using the best-fit $\omega$  derived in Sec.~\ref{sec:adiabatic-PN}. To better pinpoint 
the differences, we show in the right panels of Figs.~\ref{flux16}, \ref{flux19}, 
the GW energy fluxes normalized to the Newtonian GW flux, assuming 
the Keplerian relation $\omega^2 r^3/M=1$ valid for circular orbits. In addition, 
we show a curve where the flux has been calculated without assuming the Keplerian relation.
We notice that for the case $d=19$, except for the very first part of the evolution, 
the PN-adiabatic flux averages the numerical one until $\approx30\mbox{--}50M$ 
before the CAH forms. For the case $d=16$, the PN-adiabatic flux computed from the best-fit 
PN-adiabatic $\omega$ always overestimates the numerical one. 

Figures~\ref{flux16}, \ref{flux19} show that the frequency obtained by 
fitting the numerical frequency, as done in 
Sec.~\ref{sec:adiabatic-PN}, not only provides a GW signal that matches 
the numerical one quite well, but also provides an energy flux that is {\it consistent} with 
the numerical energy flux. However, this study does not give hints 
on which analytical model and/or PN order best fits the numerical flux. 
We leave for future work a more detail study of the comparison between numerical 
and analytical energy and angular-momentum fluxes, which includes fully relaxing the assumption 
of circular orbits and comparisons with the Pad\'e resummed fluxes.

In PN theory it is possible to show that the following relationship holds between the radiated
angular momentum flux $dJ_z/dt$ and energy flux $dE/dt$ for circular orbits:
\begin{equation}
\label{eqn:jdot_edot}
\frac{dE}{dt} = \omega\,\frac{dJ_z}{dt}\,.
\end{equation}
The above equation is used in building the EOB equations of motion~\cite{BD2,BCD}. 
The numerical angular momentum flux is computed in terms of the mode
amplitudes $_{-\!2}C_{\ell m}(t)$ as
\begin{equation}
   \frac{dJ_z}{dt} = -\frac{M}{16\pi}\sum_{\ell m}{m
        \left({\rm Im}[D_{\ell m}(t)E^*_{\ell m}(t)]\right)},
\end{equation}
where $E_{\ell m}(t)$ is a dimensionless second time integral of
$_{-\!2}C_{\ell m}(t)$ defined by
\begin{equation}
   E_{\ell m}(t) \equiv \frac{1}{M}\int_0^t{dt D_{\ell m}(t^\prime)}.
\end{equation}

\begin{figure}
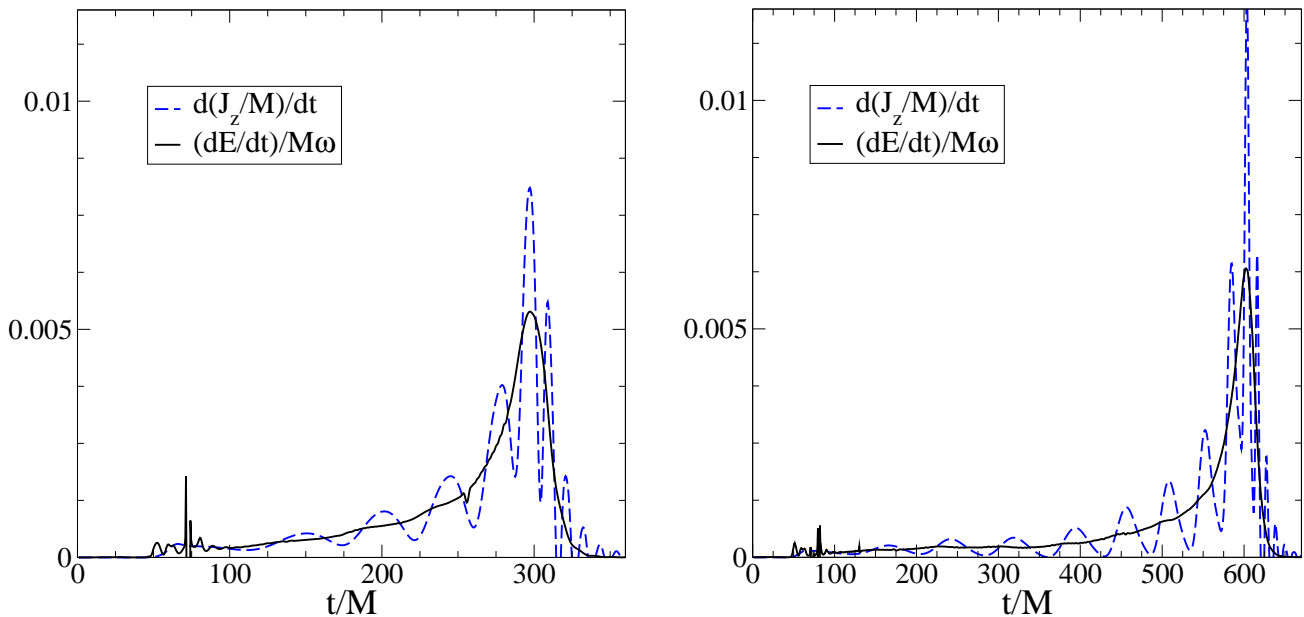

\begin{center}
\includegraphics[width=3.25in,clip]{dJ_dE_comp_d16}
\hspace{0.5cm}
\includegraphics[width=3.25in,clip]{dJ_dE_comp_d19}
\caption{Comparison of $dJ_z/dt$ and $(dE/dt)/(M\omega)$ from numerical data
for the $d=16$ and $d=19$ runs.
\label{jdot_edot_comp}}
\end{center}
\end{figure}

In Figure~\ref{jdot_edot_comp} we plot these two quantities calculated from
the angular momentum and energy fluxes of the numerical simulation,
using the frequency $\omega_{D2}$ (\ref{eq:omega_Dm}). The oscillations we
see in the numerical $dJ_z/dt$ are, we believe, due in part to improper
initial conditions in the double-time integral of $\Psi_4$ used to obtain
$dJ_z/dt$ (we used initial conditions that are valid at $t=-\infty$);
however, {\em on average} (\ref{eqn:jdot_edot}) appears to hold to a 
reasonable degree throughout the {\em entire} evolution.

\section{Effect of extraction radius on measured waveform}
\label{appendix_extract}

There are many coordinate related issues in extracting GWs from the
simulation, most of which have {\em not} been rigorously addressed for
this set of runs. The issues can be summarized by the question {\em
  how does the gauge used in the simulation affect the assumed
  relationship between $\Psi_4$ and the gravitational wave strain
  given in Eq. (\ref{eq:Psi4_ddh_defn})?}. Most certainly there are
differences that decay by some power of $r$, as only in the limit as
$r\rightarrow\infty$ is (\ref{eq:Psi4_ddh_defn}) strictly
satisfied. To test whether such differences are present and estimate
how significant they are, we can examine $r M \Psi_4(t,r,\theta,\phi)$,
or equivalently $_{-2}C_{\ell m}(t,r)$ as defined in
Eq. (\ref{eq:psi4Ylmdef}), as a function of extraction radius $r$.  In the
wave zone of a ``good'' coordinate system $_{-2}C_{\ell
  m}(t,r)= {}_{-2}C_{\ell m}(t-r)$ for large $r$, and so by comparing the
waveform at several extraction radii will give some hints as to the
adequacy of the coordinates.

\begin{figure}
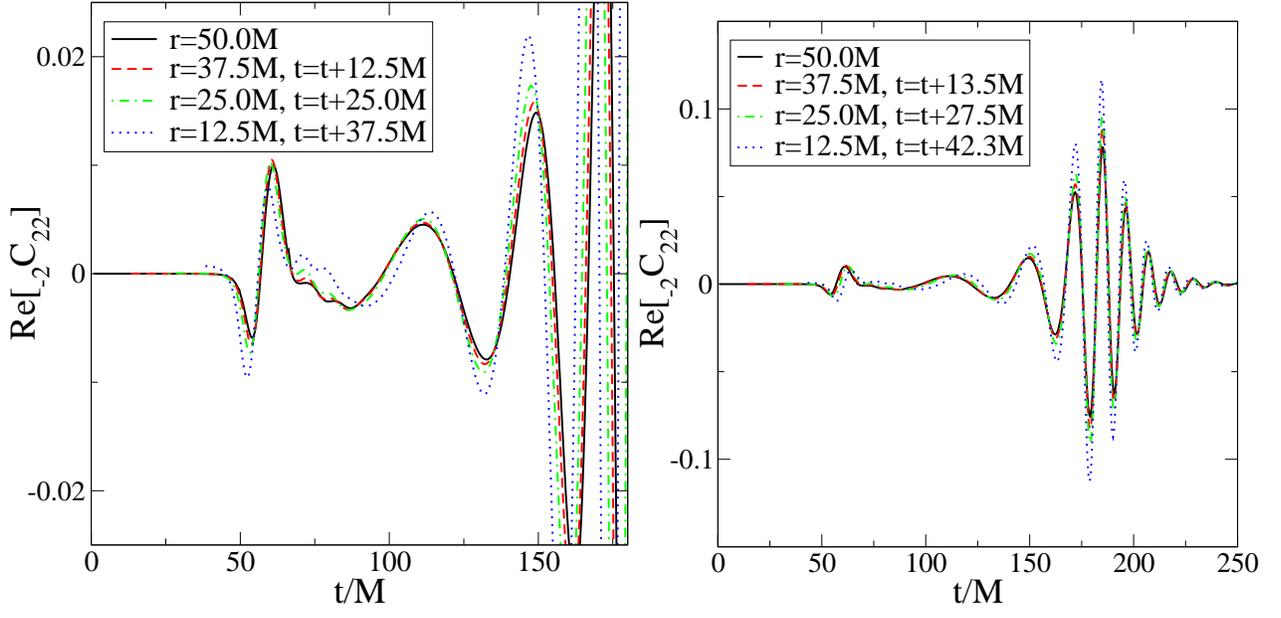

\begin{center}
\includegraphics[width=3.25in,clip]{d13_C22_vs_r}
\vspace{0.5cm}
\includegraphics[width=3.25in,clip]{d13_C22_vs_r_B}
\caption{\label{d13_C22_vs_r} A component of the waveform
of the d=13 (h/2) run measured at different extraction
radii, and shifted in time to account for wave
propagation speed. The figure to the left has time shifts
assuming unit coordinate speed, which for large extraction
radii seem to be a good assumption for the inspiral
part of the wave though produce some mismatch
around peak amplitude; the figure to the right depicts the time
shift necessary to produce the best overlap about 
peak amplitude. See Fig's \ref{d16_C22_vs_r} and
\ref{d19_C22_vs_r} for similar plots for d=16 and d=19
respectively.
}
\end{center}
\end{figure}

\begin{figure}
\begin{center}
\includegraphics[width=3.25in,clip]{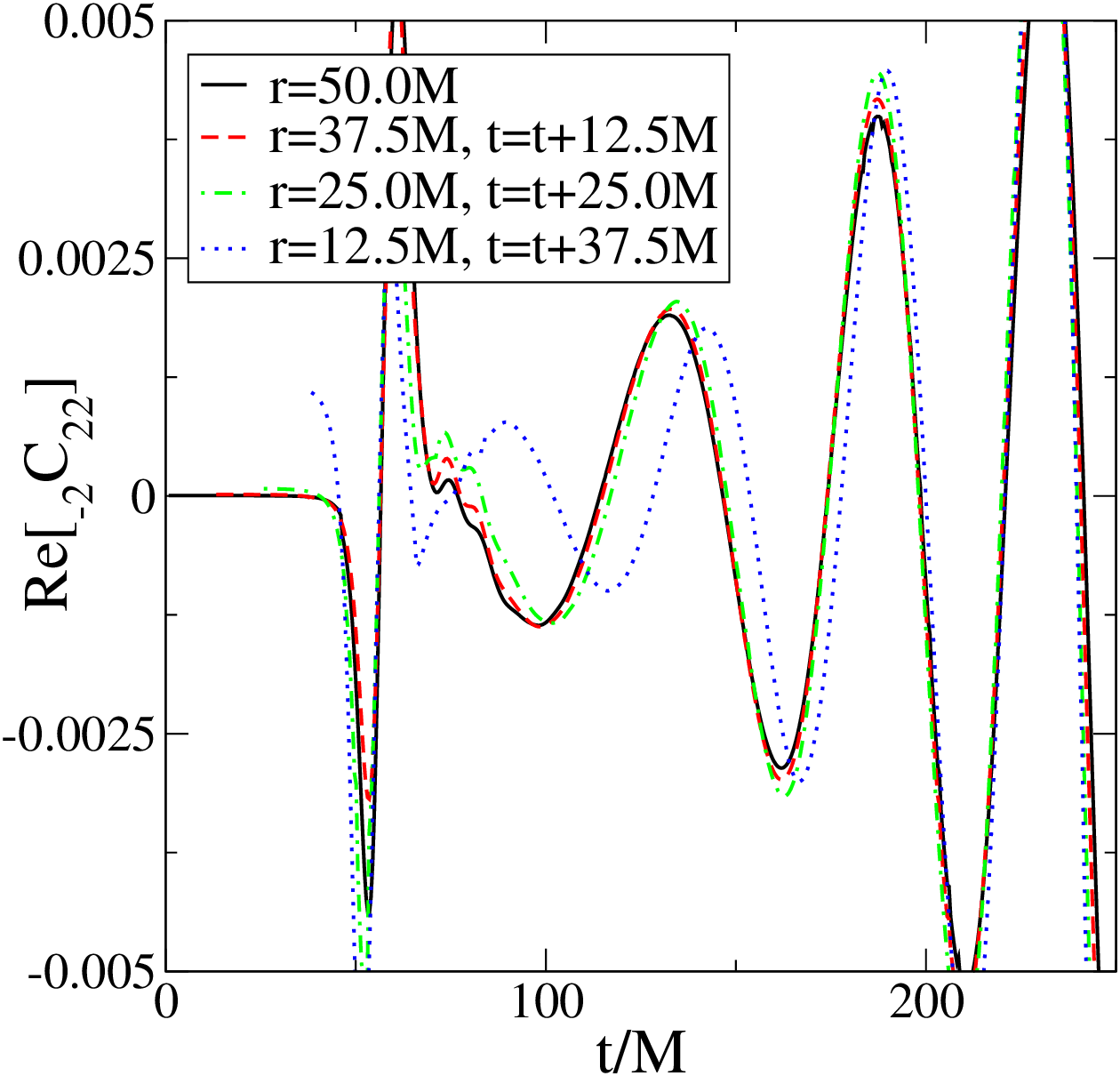}
\vspace{0.5cm}
\includegraphics[width=3.25in,clip]{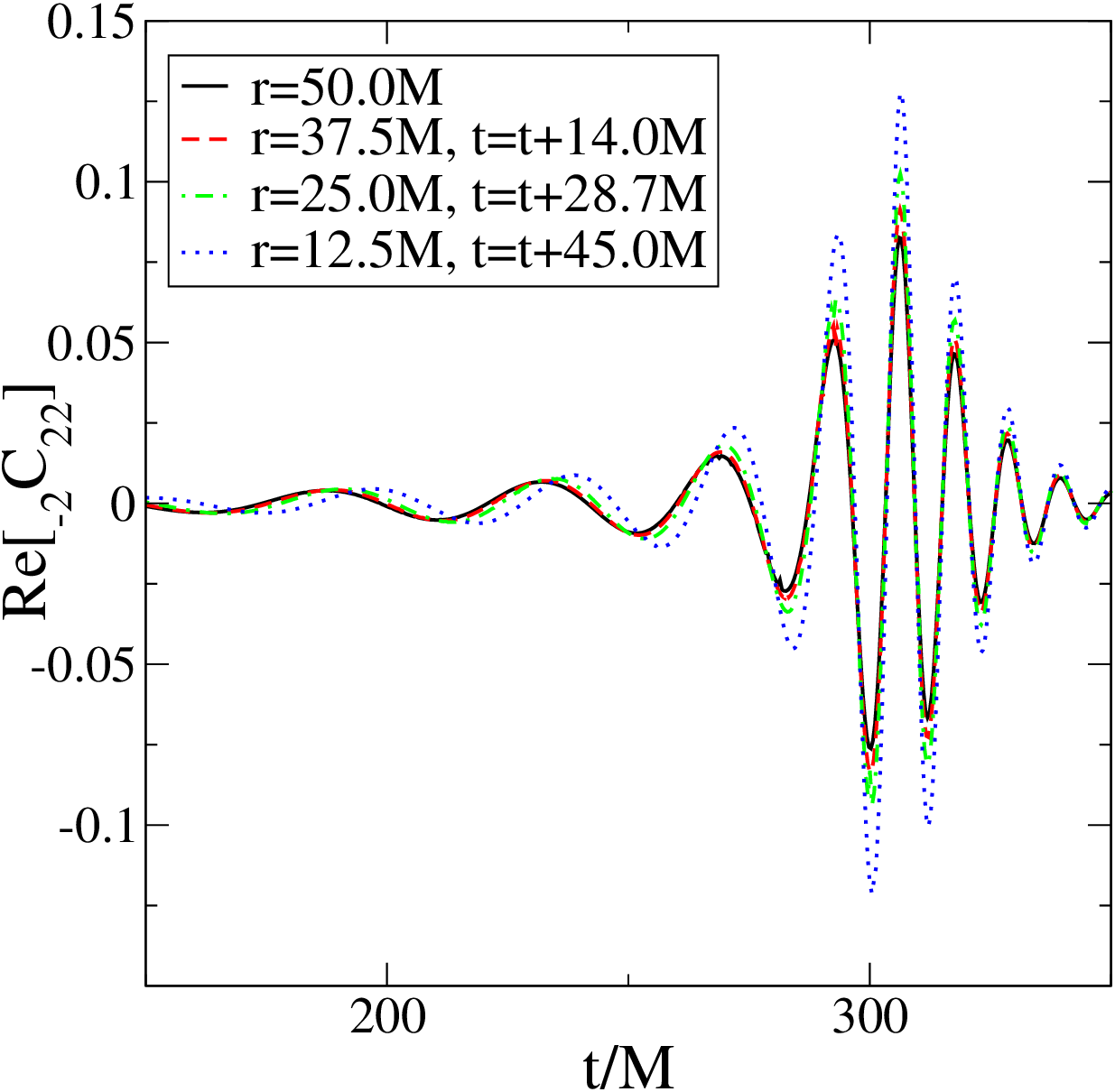}
\caption{\label{d16_C22_vs_r} Time-shifted waveforms versus
extraction radius for d=16 (see the caption describing the
the corresponding $d=13$ plot in Fig.\ref{d13_C22_vs_r})
}
\end{center}
\end{figure}

\begin{figure}
\begin{center}
\includegraphics[width=3.25in,clip]{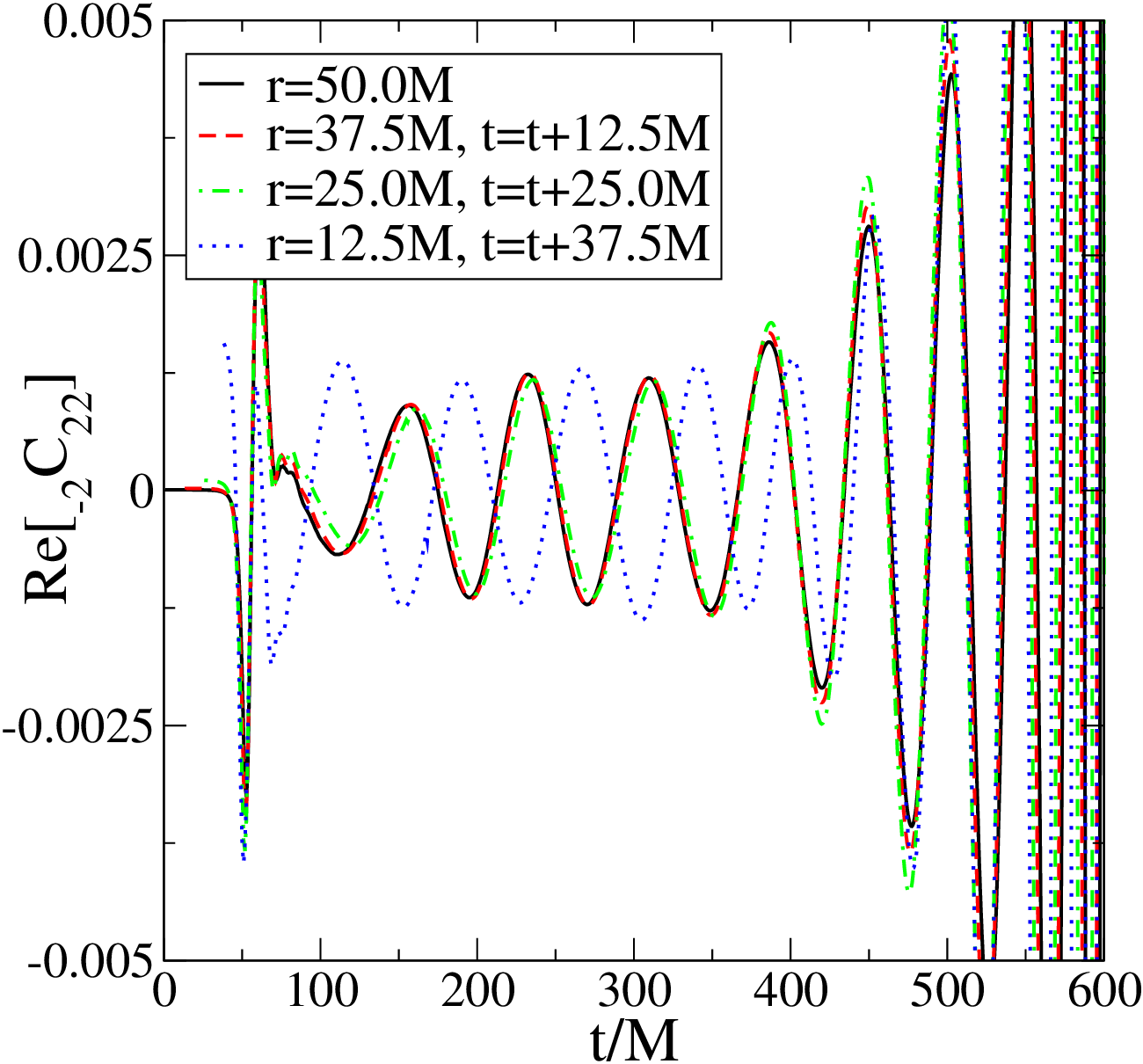}
\vspace{0.5cm}
\includegraphics[width=3.25in,clip]{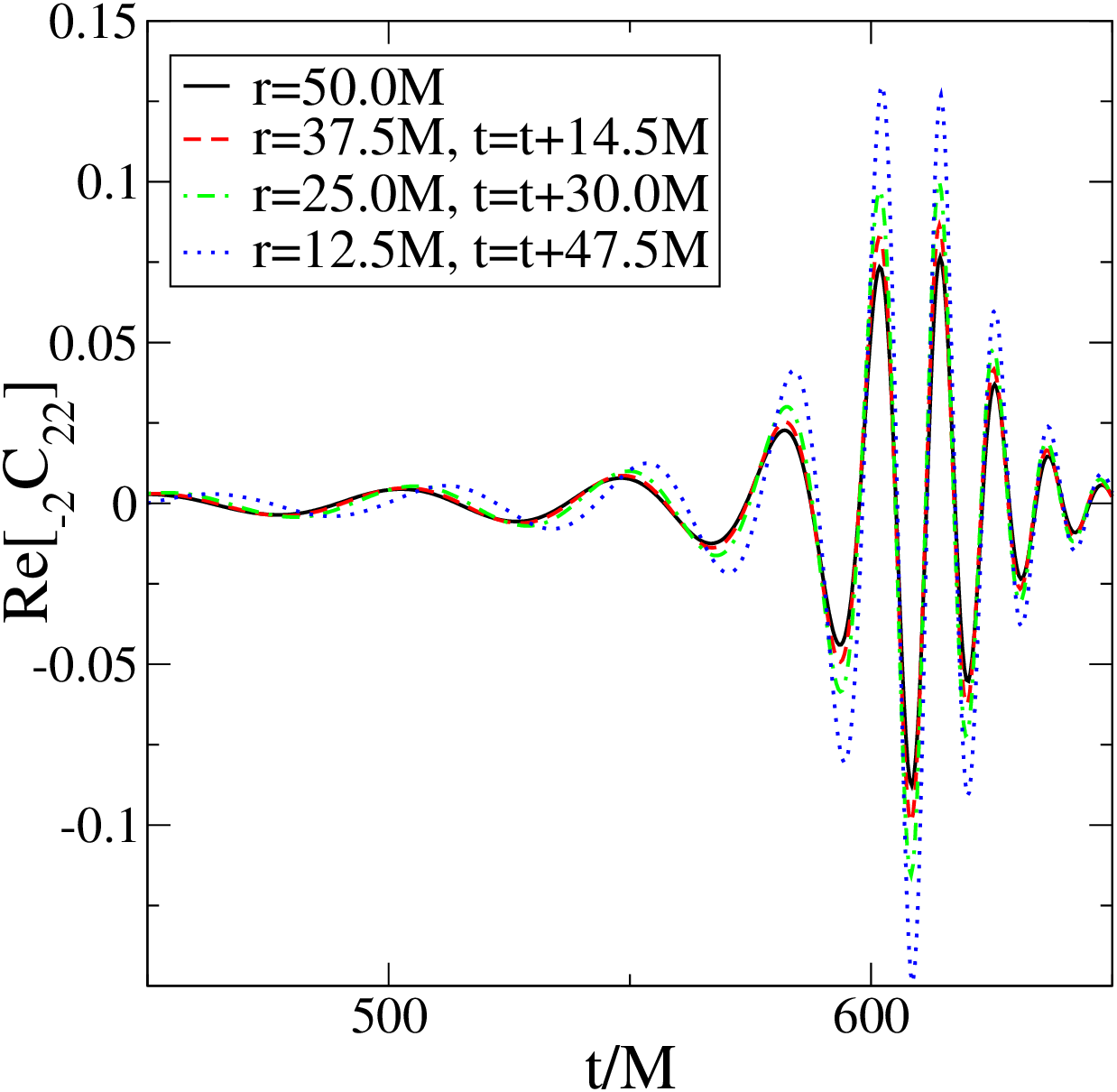}
\caption{\label{d19_C22_vs_r} Time-shifted waveforms versus
extraction radius for d=19 (see the caption describing the
the corresponding $d=13$ plot in Fig.\ref{d13_C22_vs_r})
}
\end{center}
\end{figure}

Figures~\ref{d13_C22_vs_r}--\ref{d19_C22_vs_r} show two
plots each of $_{-2}C_{2,2}(t,r)$ at four extraction radii, 
$r=12.5M,25M,37.5M$ and $50M$, for each of the three simulations 
(h/2 resolution). The plots on the left show the early part of
the waveform, shifted by an appropriate amount in time 
{\em assuming} that the wave is propagating with unit velocity.
As can be seen, going from an extraction radius of $37.5M$ to $50M$ the shifted
waveforms all overlap quite closely, in consistency with wave-like
propagation. However, these time-shifts do {\em not} give
such a good match in the part of the waveform associated with the
coalescence and ring-down of the binary. This is evident
from the plots on the right, where now 
the time shift has been chosen to give the best possible
phase overlap around peak amplitude. What these latter plots
suggest is that the gauge changes by a small amount with time
in the wave extraction regime of the simulation, and the amount
of change is apparently correlated with the amplitude of the GWs being emitted. 

These effects are summarized in Table \ref{tab_speeds},
where we give the average coordinate propagation speeds between the
different extraction radii during the inspiral and coalescence/merger
portion of the signal. Another apparent gauge pathology 
seen in Figs.~\ref{d13_C22_vs_r}--\ref{d19_C22_vs_r}
is that near the peak amplitude the assumed $1/r$
decay in the wave amplitude [factored out in Eq.~(\ref{eq:psi4Ylmdef})] 
does not describe the situation as well as in the
inspiral regime. This could happen (for example)
if the coordinate sphere $r=r_i$ begins to deviate
from a geometric sphere of radius $r_i$; such a change
in gauge could also affect the coordinate velocity
of the wave, and in fact the trend of decreasing
amplitude with extraction radius is consistent
with the decrease in wave speed (though cannot
by itself account for all the change in velocity).

\begin{table}
{
\begin{tabular}[t]{| l || c | c |}
\hline
 {\bf d=13} & $v_{avg,inspiral}$ & $v_{avg,peak}$ \\
\hline
\hline
 r=12.5M to 25M & ---  & 0.84\\
 r=25M to 37.5M & 0.98 & 0.89\\
 r=37.5M to 50M & 0.98 & 0.93\\
\hline
\hline
 {\bf d=16} & $v_{avg,inspiral}$ & $v_{avg,peak}$ \\
\hline
 r=12.5M to 25M & ---  & 0.77\\
 r=25M to 37.5M & 0.99 & 0.85\\
 r=37.5M to 50M & 0.99 & 0.89\\
\hline
\hline
 {\bf d=19} & $v_{avg,inspiral}$ & $v_{avg,peak}$ \\
\hline
 r=12.5M to 25M & ---  & 0.71\\
 r=25M to 37.5M & 1.20 & 0.81\\
 r=37.5M to 50M & 1.00 & 0.86\\
\hline
\end{tabular}
}
\caption{The average GW propagation
speed between various extraction radii for the inspiral
part of the wave and the coalescence/ring-down (``peak'') part---see
Fig.'s \ref{d13_C22_vs_r} to \ref{d19_C22_vs_r} 
\label{tab_speeds}}
\end{table}

\section{Tables}
\label{appendix_tables}
This appendix contains tables from the QNM ring-down analysis presented
in Sec.~\ref{sec4}. Tables \ref{tab:C22Nd13}, \ref{tab:C22Nd13}
and \ref{tab:C22Nd19} are fit parameters for the $_{-\!2}C_{22}$
component of the waveform from the $d=13,16$ and $19$ cases
respectively; similarly Tables \ref{tab:C32Nd13}, \ref{tab:C32Nd13}
and \ref{tab:C32Nd19} are fit parameters for the $_{-\!2}C_{32}$
component of the waveform for the same set of separations.

\begin{table}
\caption{\label{tab:C22Nd13}Fit parameters for $_{-\!2}C_{22}$ for
the $d=13$ case.  $t_r/m$ denotes the time {\em after} the peak in
$|{}_{-\!2}C_{22}|$ at which the fitting started.}
\begin{ruledtabular}
\begin{tabular}{l|c|c|c|c||c|c|c|c}
&\multicolumn{4}{c||}{${\rm Re}[{}_{-\!2}C_{22}]$}
&\multicolumn{4}{c}{${\rm Im}[{}_{-\!2}C_{22}]$} \\
\cline{2-9}
& $N=0$ & $N=1$ & $N=2$ & $N=3$ & $N=0$ & $N=1$ & $N=2$ & $N=3$ \\ 
\hline
$t_r/M$ & 20 & 8.5 & -2.5 & -6.5 & 17 & 4 & -0.5 & -9.5 \\
$a/M_f$ & 0.725 & 0.723 & 0.737 & 0.736 & 0.725 & 0.735 & 0.733 & 0.748 \\
$M_f/M$ & 0.943 & 0.939 & 0.950 & 0.949 & 0.943 & 0.948 & 0.945 & 0.959 \\
\hline
${\cal C}_{2220}$ & -0.143 & -0.146 & -0.140 & -0.140 & -0.145 & -0.142 & -0.144 & -0.138 \\
$\phi_{2220}$ & 1.65 & 1.60 & 1.60 & 1.58 & 1.66 & 1.59 & 1.57 & 1.58 \\
${\cal C}_{2221}$ & - & 0.115 & 0.103 & 0.106 & - & 0.0985 & 0.120 & 0.0946 \\
$\phi_{2221}$ & - & 1.25 & 0.975 & 0.959 & - & 1.07 & 0.979 & 0.946 \\
${\cal C}_{2222}$ &-  & - & -0.0268 & -0.0365 & - & - & -0.0442 & -0.0265 \\
$\phi_{2222}$ & - & - & 0.580 & 0.501 & - & - & 0.688 & 0.445 \\
${\cal C}_{2223}$ & - & - & - & -0.00513 & - & - & - & -0.00265 \\
$\phi_{2223}$ & - & - & - & 3.14 & - & - & - & 3.04 \\
\hline
$M\,\omega_{220}$ & 0.577 & 0.579 & 0.579 & 0.579 & 0.577 & 0.579 & 0.580 & 0.580 \\
$M/\tau_{220}$ & 0.0845 & 0.0850 & 0.0834 & 0.0835 & 0.0847 & 0.0836 & 0.0841 & 0.0820 \\
$M\,\omega_{221}$ & - & 0.567 & 0.568 & 0.568 & - & 0.569 & 0.569 & 0.569 \\
$M/\tau_{221}$ & - & 0.257 & 0.252 & 0.252 & - & 0.253 & 0.254 & 0.248 \\
$M\,\omega_{222}$ & - & - & 0.548 & 0.548 & - & - & 0.548 & 0.550 \\
$M/\tau_{222}$ & - & - & 0.425 & 0.423 & - & - & 0.428 & 0.417 \\
$M\,\omega_{223}$ & - & - & - & 0.520 & - & - & - & 0.522 \\
$M/\tau_{223}$ & - & - & - & 0.602 & - & - & - & 0.591
\end{tabular}
\end{ruledtabular}
\end{table}

\begin{table}
\caption{\label{tab:C22Nd16}Fit parameters for $_{-\!2}C_{22}$ for
the $d=16$ case.  $t_r/m$ denotes the time {\em after} the peak in
$|{}_{-\!2}C_{22}|$ at which the fitting started.}
\begin{ruledtabular}
\begin{tabular}{l|c|c|c|c||c|c|c|c}
&\multicolumn{4}{c||}{${\rm Re}[{}_{-\!2}C_{22}]$}
&\multicolumn{4}{c}{${\rm Im}[{}_{-\!2}C_{22}]$} \\
\cline{2-9}
& $N=0$ & $N=1$ & $N=2$ & $N=3$ & $N=0$ & $N=1$ & $N=2$ & $N=3$ \\ 
\hline
$t_r/M$ & 20.5 & 8.5 & 2.5 & -2.5 & 19 & 10 & 0 & -4 \\
$a/M_f$ & 0.729 & 0.726 & 0.730 & 0.731 & 0.728 & 0.722 & 0.736 & 0.739 \\
$M_f/M$ & 0.947 & 0.942 & 0.944 & 0.945 & 0.947 & 0.939 & 0.949 & 0.951 \\
\hline
${\cal C}_{2220}$ & -0.151 & -0.155 & -0.153 & -0.153 & -0.152 & -0.158 & -0.152 & -0.150 \\
$\phi_{2220}$ & 1.70 & 1.66 & 1.63 & 1.62 & 1.71 & 1.67 & 1.63 & 1.62 \\
${\cal C}_{2221}$ & - & 0.146 & 0.164 & 0.169 & - & 0.151 & 0.147 & 0.153 \\
$\phi_{2221}$ & - & 1.13 & 0.922 & 0.894 & - & 1.26 & 0.920 & 0.811 \\
${\cal C}_{2222}$ & - & - & -0.0971 & -0.126 & - & - & -0.0666 & -0.0959 \\
$\phi_{2222}$ & - & - & 0.636 & 0.548 & - & - & 0.559 & 0.278 \\
${\cal C}_{2223}$ & - & - & - & 0.0444 & - & - & - & 0.0263 \\
$\phi_{2223}$ & - & - & - & 0.325 & - & - & - & -0.133 \\
\hline
$M\,\omega_{220}$ & 0.577 & 0.578 & 0.579 & 0.579 & 0.577 & 0.578 & 0.579 & 0.580 \\
$M/\tau_{220}$ & 0.0840 & 0.0846 & 0.0842 & 0.0842 & 0.0841 & 0.0851 & 0.0836 & 0.0833 \\
$M\,\omega_{221}$ & - & 0.567 & 0.568 & 0.568 & - & 0.566 & 0.568 & 0.569 \\
$M/\tau_{221}$ & - & 0.256 & 0.254 & 0.254 & - & 0.257 & 0.252 & 0.251 \\
$M\,\omega_{222}$ & - & - & 0.547 & 0.547 & - & - & 0.548 & 0.549 \\
$M/\tau_{222}$ & - & - & 0.429 & 0.429 & - & - & 0.426 & 0.424 \\
$M\,\omega_{223}$ & - & - & - & 0.518 & - & - & - & 0.520 \\
$M/\tau_{223}$ & - & - & - & 0.607 & - & - & - & 0.600
\end{tabular}
\end{ruledtabular}
\end{table}

\begin{table}
\caption{\label{tab:C22Nd19}Fit parameters for $_{-\!2}C_{22}$ for
the $d=19$ case.  $t_r/M$ denotes the time {\em after} the peak in
$|{}_{-\!2}C_{22}|$ at which the fitting started.}
\begin{ruledtabular}
\begin{tabular}{l|c|c|c|c||c|c|c|c}
&\multicolumn{4}{c||}{${\rm Re}[{}_{-\!2}C_{22}]$}
&\multicolumn{4}{c}{${\rm Im}[{}_{-\!2}C_{22}]$} \\
\cline{2-9}
& $N=0$ & $N=1$ & $N=2$ & $N=3$ & $N=0$ & $N=1$ & $N=2$ & $N=3$ \\ 
\hline
$t_r/M$ & 21.5 & 7.5 & 1 & -4 & 20.5 & 7 & -1.5 & -5.5 \\
$a/M_f$ & 0.707 & 0.709 & 0.712 & 0.712 & 0.705 & 0.705 & 0.721 & 0.723 \\
$M_f/M$ & 0.948 & 0.947 & 0.948 & 0.947 & 0.946 & 0.943 & 0.955 & 0.955 \\
\hline
${\cal C}_{2220}$ & 0.170 & 0.172 & 0.171 & 0.171 & 0.173 & 0.176 & 0.168 & 0.168 \\
$\phi_{2220}$ & 2.46 & 2.41 & 2.39 & 2.39 & 2.47 & 2.40 & 2.39 & 2.37 \\
${\cal C}_{2221}$ & - & -0.158 & -0.179 & -0.183 & - & -0.177 & -0.160 & -0.167 \\
$\phi_{2221}$ & - & 1.97 & 1.81 & 1.80 & - & 1.97 & 1.79 & 1.71 \\
${\cal C}_{2222}$ & - & - & 0.0859 & 0.107 & - & - & 0.0590 & 0.0824 \\
$\phi_{2222}$ & - & - & 157 & 1.50 & - & - & 1.47 & 1.26 \\
${\cal C}_{2223}$ & - & - & - & -0.0273 & - & - & - & -0.0160 \\
$\phi_{2223}$ & - & - & - & 1.22 & - & - & - & 0.815 \\
\hline
$M\,\omega_{220}$ & 0.566 & 0.567 & 0.568 & 0.568 & 0.565 & 0.567 & 0.568 & 0.569 \\
$M/\tau_{220}$ & 0.0850 & 0.0850 & 0.0848 & 0.0848 & 0.0852 & 0.0855 & 0.0837 & 0.0836 \\
$M\,\omega_{221}$ & - & 0.555 & 0.556 & 0.556 & - & 0.555 & 0.557 & 0.557 \\
$M/\tau_{221}$ & - & 0.257 & 0.256 & 0.256 & - & 0.258 & 0.253 & 0.253 \\
$M\,\omega_{222}$ & - & - & 0.534 & 0.534 & - & - & 0.535 & 0.536 \\
$M/\tau_{222}$ & - & - & 0.432 & 0.432 & - & - & 0.427 & 0.426 \\
$M\,\omega_{223}$ & - & - & - & 0.505 & - & - & - & 0.507 \\
$M/\tau_{223}$ & - & - & - & 0.613 & - & - & - & 0.603 \\
\end{tabular}
\end{ruledtabular}
\end{table}

\begin{table}
\caption{\label{tab:C32Nd13}Fit parameters for $_{-\!2}C_{32}$ for
the $d=13$ case.  $t_r/M$ denotes the time {\em after} the peak in
$|{}_{-\!2}C_{22}|$ at which the fitting started.}
\begin{ruledtabular}
\begin{tabular}{l|c|c|c||c|c|c}
&\multicolumn{3}{c||}{${\rm Re}[{}_{-\!2}C_{32}]$}
&\multicolumn{3}{c}{${\rm Im}[{}_{-\!2}C_{32}]$} \\
\cline{2-7}
& $N=0$ & $N=1$ & $N=2$ & $N=0$ & $N=1$ & $N=2$ \\ 
\hline
$t_r/M$ & 16 & -0.5 & -4.5 & 14.5 & -2 & 10 \\
$a/M_f$ & 0.731 & 0.753 & 0.721 & 0.704 & 0.738 & 0.710 \\
$M_f/M$ & 0.947 & 0.955 & 0.921 & 0.931 & 0.953 & 0.925 \\
\hline
${\cal C}_{3220}$ & -0.00753 & -0.00730 & -0.00868 & -0.00755 & -0.00721 & -0.00804 \\
$\phi_{3220}$ & 1.53 & 1.33 & 1.16 & 0.915 & 0.816 & 0.720 \\
${\cal C}_{3320}$ & 0.00599 & 0.00563 & 0.00628 & 0.00844 & 0.00734 & 0.00811 \\
$\phi_{3320}$ & 2.77 & 2.61 & 2.14 & 2.86 & 2.89 & 2.61 \\
${\cal C}_{3221}$ & - & 0.00822 & 0.0272 & - & 0.00787 & 0.0142 \\
$\phi_{3221}$ & - & 0.749 & 1.17 & - & 0.720 & 0.872 \\
${\cal C}_{3321}$ & - & -0.00592 & -0.0188 & - & -0.007528 & -0.0121 \\
$\phi_{3321}$ & - & 1.56 & 0.298 & - & 2.16 & 1.48 \\
${\cal C}_{3222}$ & - & - & -0.0153 & - & - & -0.00235 \\
$\phi_{3222}$ & - & - & 2.46 & - & - & 1.98 \\
${\cal C}_{3322}$ & - & - & -0.0127 & - & - & -0.00201 \\
$\phi_{3322}$ & - & - & 1.71 & - & - & 3.10 \\
\hline
$M\,\omega_{220}$ & 0.578 & 0.585 & 0.589 & 0.574 & 0.578 & 0.580 \\
$M/\tau_{220}$ & 0.0840 & 0.0821 & 0.0868 & 0.0866 & 0.0831 & 0.0869 \\
$M\,\omega_{320}$ & 0.817 & 0.820 & 0.836 & 0.819 & 0.815 & 0.827 \\
$M/\tau_{320}$ & 0.0874 & 0.0855 & 0.0904 & 0.0902 & 0.0865 & 0.0905 \\
$M\,\omega_{221}$ & - & 0.575 & 0.577 & - & 0.567 & 0.568 \\
$M/\tau_{221}$ & - & 0.248 & 0.262 & - & 0.251 & 0.263 \\
$M\,\omega_{321}$ & - & 0.813 & 0.828 & - & 0.807 & 0.818 \\
$M/\tau_{321}$ & - & 0.258 & 0.273 & - & 0.261 & 0.273 \\
$M\,\omega_{222}$ & - & - & 0.555 & - & - & 0.546 \\
$M/\tau_{222}$ & - & - & 0.443 & - & - & 0.443 \\
$M\,\omega_{322}$ & - & - & 0.812 & - & - & 0.802 \\
$M/\tau_{322}$ & - & - & 0.462 & - & - & 0.462 \\
\end{tabular}
\end{ruledtabular}
\end{table}

\begin{table}
\caption{\label{tab:C32Nd16}Fit parameters for $_{-\!2}C_{32}$ for
the $d=16$ case.  $t_r/m$ denotes the time {\em after} the peak in
$|{}_{-\!2}C_{22}|$ at which the fitting started.}
\begin{ruledtabular}
\begin{tabular}{l|c|c|c||c|c|c}
&\multicolumn{3}{c||}{${\rm Re}[{}_{-\!2}C_{32}]$}
&\multicolumn{3}{c}{${\rm Im}[{}_{-\!2}C_{32}]$} \\
\cline{2-7}
& $N=0$ & $N=1$ & $N=2$ & $N=0$ & $N=1$ & $N=2$ \\ 
\hline
$t_r/M$ & 15.5 & 3.5 & -5 & 14.5 & 0.5 & -5 \\
$a/M_f$ & 0.740 & 0.718 & 0.722 & 0.731 & 0.7.22 & 0.729 \\
$M_f/M$ & 0.952 & 0.936 & 0.922 & 0.944 & 0.928 & 0.918 \\
\hline
${\cal C}_{3220}$ & -0.00903 & -0.00990 & -0.107 & -0.00846 & -0.00932 & -0.0101 \\
$\phi_{3220}$ & 1.57 & 1.61 & 1.25 & 1.32 & 1.05 & 0.828 \\
${\cal C}_{3320}$ & 0.00835 & 0.00919 & 0.00907 & 0.00981 & 0.0102 & 0.00988 \\
$\phi_{3320}$ & 2.88 & 2.75 & 2.28 & 2.88 & 2.49 & 2.20 \\
${\cal C}_{3221}$ & - & 0.00999 & 0.0350 & - & 0.0216 & 0.0413 \\
$\phi_{3221}$ & - & 1.91 & 1.37 & - & 0.911 & 0.918 \\
${\cal C}_{3321}$ & - & -0.0189 & -0.0265 & - & -0.0192 & -0.0273 \\
$\phi_{3321}$ & - & 2.14 & 0.649 & - & 1.34 & 0.404 \\
${\cal C}_{3222}$ & - & - & -0.0179 & - & - & -0.017 \\
$\phi_{3222}$ & - & - & 2.67 & - & - & 2.10 \\
${\cal C}_{3322}$ & - & - & -0.0149 & - & - & -0.0137 \\
$\phi_{3322}$ & - & - & 2.08 & - & - & 1.79 \\
\hline
$M\,\omega_{220}$ & 0.579 & 0.578 & 0.589 & 0.580 & 0.588 & 0.595 \\
$M/\tau_{220}$ & 0.0831 & 0.0856 & 0.0867 & 0.0842 & 0.0859 & 0.0867 \\
$M\,\omega_{320}$ & 0.817 & 0.821 & 0.835 & 0.820 & 0.833 & 0.841 \\
$M/\tau_{320}$ & 0.0865 & 0.0891 & 0.0903 & 0.0877 & 0.0895 & 0.0903 \\
$M\,\omega_{221}$ & - & 0.566 & 0.577 & - & 0.576 & 0.583 \\
$M/\tau_{221}$ & - & 0.259 & 0.262 & - & 0.260 & 0.262 \\
$M\,\omega_{321}$ & - & 0.813 & 0.827 & - & 0.824 & 0.833 \\
$M/\tau_{321}$ & - & 0.269 & 0.273 & - & 0.270 & 0.273 \\
$M\,\omega_{222}$ & - & - & 0.555 & - & - & 0.561 \\
$M/\tau_{222}$ & - & - & 0.442 & - & - & 0.442 \\
$M\,\omega_{322}$ & - & - & 0.811 & - & - & 0.817 \\
$M/\tau_{322}$ & - & - & 0.461 & - & - & 0.461 \\
\end{tabular}
\end{ruledtabular}
\end{table}

\begin{table}
\caption{\label{tab:C32Nd19}Fit parameters for $_{-\!2}C_{32}$ for
the $d=19$ case.  $t_r/M$ denotes the time {\em after} the peak in
$|{}_{-\!2}C_{22}|$ at which the fitting started.}
\begin{ruledtabular}
\begin{tabular}{l|c|c|c||c|c|c}
&\multicolumn{3}{c||}{${\rm Re}[{}_{-\!2}C_{32}]$}
&\multicolumn{3}{c}{${\rm Im}[{}_{-\!2}C_{32}]$} \\
\cline{2-7}
& $N=0$ & $N=1$ & $N=2$ & $N=0$ & $N=1$ & $N=2$ \\ 
\hline
$t_r/M$ & 13.5 & 0 & -5 & 12.5 & -0.5 & -6 \\
$a/M_f$ & 0.729 & 0.723 & 0.706 & 0.730 & 0.713 & 0.704 \\
$M_f/M$ & 0.964 & 0.956 & 0.937 & 0.961 & 0.944 & 0.930 \\
\hline
${\cal C}_{3220}$ & 0.00931 & 0.00978 & 0.0107 & 0.00742 & 0.00823 & 0.00896 \\
$\phi_{3220}$ & 2.40 & 2.34 & 2.23 & 2.10 & 2.02 & 1.91 \\
${\cal C}_{3320}$ & 0.00704 & 0.00729 & 0.00782 & 0.00830 & 0.00880 & 0.00907 \\
$\phi_{3320}$ & 0.471 & 0.326 & 0.493 & 0.431 & 0.218 & -0.0134 \\
${\cal C}_{3221}$ & - & -0.00974 & -0.0210 & - & -0.0148 & -0.0259 \\
$\phi_{3221}$ & - & 2.05 & 2.23 & - & 1.69 & 1.91 \\
${\cal C}_{3321}$ & - & 0.00961 & 0.0187 & - & 0.0137 & 0.0225 \\
$\phi_{3321}$ & - & 2.58 & 1.77 & - & 2.26 & 1.59 \\
${\cal C}_{3222}$ & - & - & -0.00750 & - & - & -0.0853 \\
$\phi_{3222}$ & - & - & 0.442 & - & - & 0.0885 \\
${\cal C}_{3322}$ & - & - & -0.00795 & - & - & -0.00882 \\
$\phi_{3322}$ & - & - & 0.291 & - & - & 0.0739 \\
\hline
$M\,\omega_{220}$ & 0.566 & 0.568 & 0.571 & 0.569 & 0.571 & 0.574 \\
$M/\tau_{220}$ & 0.0826 & 0.0835 & 0.0860 & 0.0828 & 0.0850 & 0.0866 \\
$M\,\omega_{320}$ & 0.802 & 0.805 & 0.815 & 0.805 & 0.812 & 0.819 \\
$M/\tau_{320}$ & 0.0860 & 0.0870 & 0.0896 & 0.0862 & 0.0886 & 0.0902 \\
$M\,\omega_{221}$ & - & 0.557 & 0.559 & - & 0.559 & 0.562 \\
$M/\tau_{221}$ & - & 0.252 & 0.260 & - & 0.227 & 0.262 \\
$M\,\omega_{321}$ & - & 0.797 & 0.806 & - & 0.804 & 0.811 \\
$M/\tau_{321}$ & - & 0.263 & 0.271 & - & 0.268 & 0.273 \\
$M\,\omega_{222}$ & - & - & 0.537 & - & - & 0.539 \\
$M/\tau_{222}$ & - & - & 0.439 & - & - & 0.442 \\
$M\,\omega_{322}$ & - & - & 0.790 & - & - & 0.794 \\
$M/\tau_{322}$ & - & - & 0.458 & - & - & 0.461 \\
\end{tabular}
\end{ruledtabular}
\end{table}

\section{Results for $d=13$}
\label{appendix_d13}

\begin{figure}
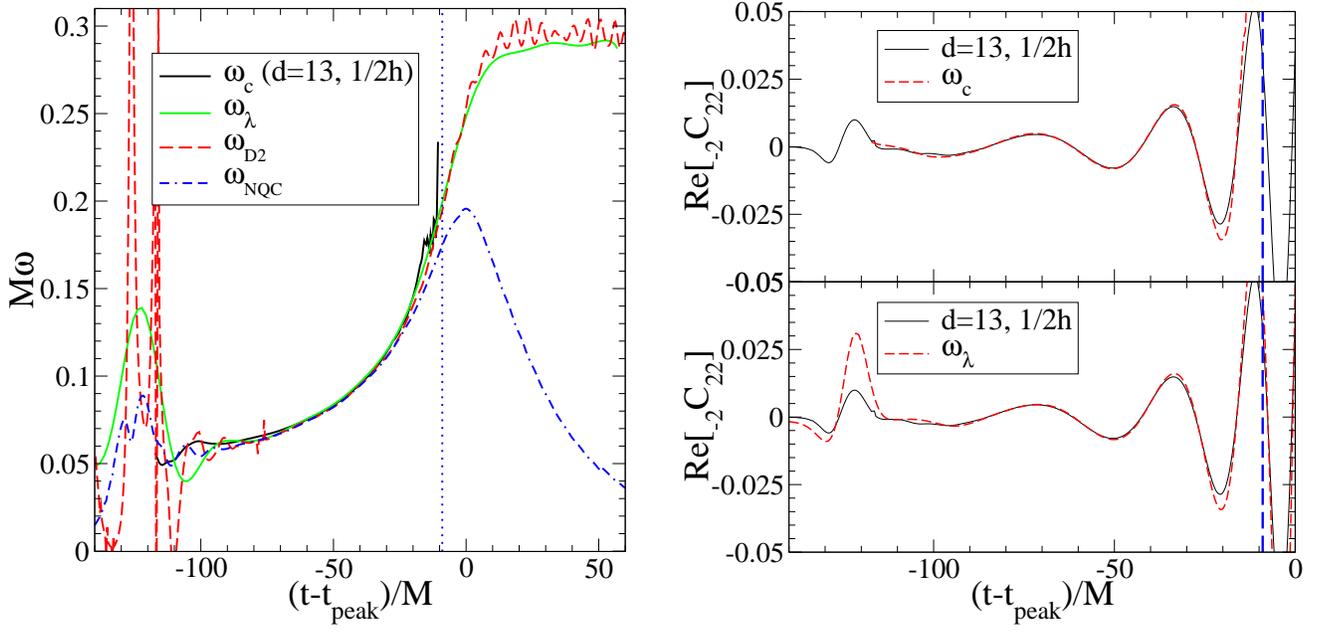

\includegraphics[width=3.25in,clip]{OmegaNQC22d13}
\hspace{0.5cm}
\includegraphics[width=3.25in,clip]{InspiralNQC22d13}
\caption{\label{Fig:OmegaNQC22d13} (left panel) Orbital angular frequencies
for the $d=13$ case.  See Fig.~\ref{Fig:OmegaNQC22d16-19} for
a full description. (right panel) Comparison of numerical and NQC
inspiral waveforms for the $d=13$.  See Fig.~\ref{Fig:InspNQC22d16-19}
for a full description.}
\end{figure}

\begin{figure}
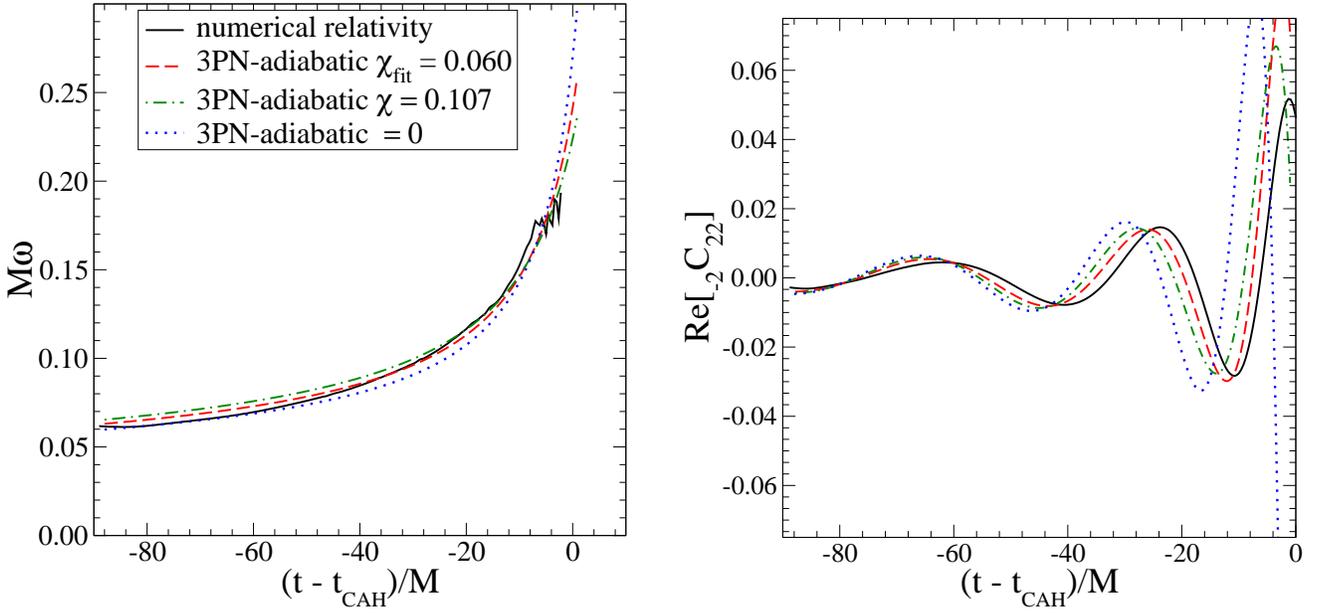

\includegraphics[width=3.25in,clip]{omegaHd13PN}
\hspace{0.5cm}
\includegraphics[width=3.25in,clip]{ReC22Hd13PN}
\caption{\label{d13} (left panel) Comparison of the NR and three analytical orbital frequencies---
see Fig.~\ref{omega1} for a full description. (right panel) Comparison of the NR
and analytical ${\rm Re}[{}_{-\!2}C_{22}]$---see Fig.~\ref{ReC221}
for a full description.}
\end{figure}

\begin{figure}
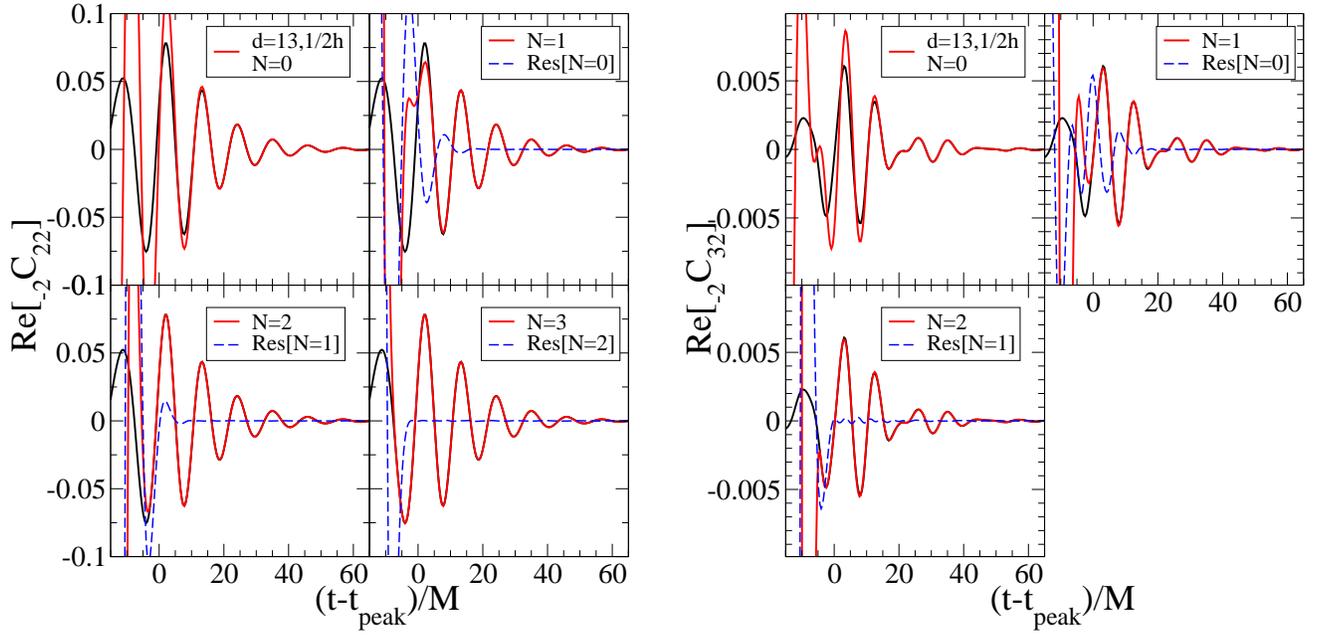

\begin{center}
\includegraphics[width=3.25in,clip]{RDNC22rd13}
\hspace{0.5cm}
\includegraphics[width=3.25in,clip]{RDNC32rd13}
\caption{\label{Fig:RDNC22rd13} The left panel shows a comparison of numerical and QNM
$_{-\!2}C_{22}$ ring-down waveforms for $d=13$---see
Fig.~\ref{Fig:RDNC22rd16-19} for a full description.
The right panel shows a comparison of numerical and QNM
$_{-\!2}C_{32}$ ring-down waveforms for $d=13$---see
Fig.~\ref{Fig:RDNC32rd16-19} for a full description.
}
\end{center}
\end{figure}

\begin{figure}
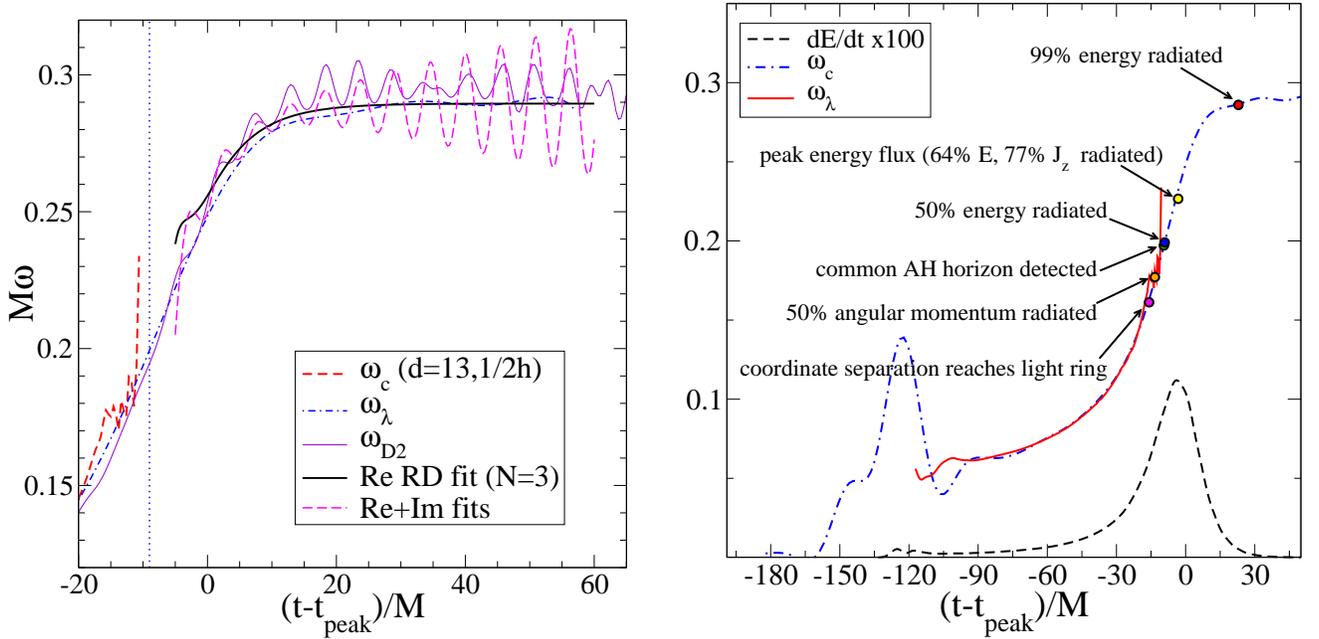

\begin{center}
\includegraphics[width=3.25in,clip]{mOmegad13D}
\hspace{0.5cm}
\includegraphics[width=3.25in,clip]{d13_phases}
\caption{\label{Fig:RDmOC22rd13} (left panel) Dominant frequencies during the ring
down for the $d=13$ case evaluated using several methods.  See
Fig.~\ref{Fig:RDmOC22rd16-19} for a full description. (right panel) 
Features of the $d=13$ merger
phase. See Fig.~\ref{merger16-19} for a full description.}
\end{center}
\end{figure}

This appendix contains some figures showing results from the $d=13$ run
that were excluded from the main text for brevity. Figure~\ref{Fig:OmegaNQC22d13}
shows orbital angular frequencies extracted using several methods (Sec.~\ref{sec:Newt_Quadrupole}),
and a comparison of the numerical and NQC
inspiral waveforms (Sec.~\ref{sec:Newt_Quadrupole}), Fig.~\ref{d13} shows
a comparison of frequencies and waveform components between NR and
various analytical counterparts (Sec.~\ref{sec:adiabatic-PN}), Fig.~\ref{Fig:RDNC22rd13}
shows results from QNM extraction (Sec.~\ref{sec4}), and Fig.~\ref{Fig:RDmOC22rd13}
plots the dominant frequencies during the ring-down phase (Sec.~\ref{sec4}) and 
identifies features of the merger (Sec.~\ref{secmerger}).

\renewcommand{\prd}{\emph{Phys. Rev. D}}

\end{document}